\documentclass[twocolumn,superscriptaddress,pra,letterpaper,showpacs,citesort]{revtex4}
\usepackage{hyperref}
\usepackage{latexsym,amsmath,amssymb,amsfonts,graphicx,color,amsthm}
\usepackage{multirow}
\usepackage{epstopdf}

\def\err{\text{err}}
\def\fnj{f_n^{(j)}}
\def\cD{\mathcal{D}}
\def\cG{\mathcal{G}}
\def\cB{\mathcal{B}}

\def\cQ{\mathcal{Q}}

\def\snj{S_n^{(j)}}

\def\etaDD{\eta_\text{DD}}
\def\Heff{H_\text{eff}}
\def\Gb{\overline G}

\def\Id{\mathbb{I}}
\def\tDD{t_\text{DD}}
\def\DeltaM{\bar\delta}
\def\numpulse{R}

\newtheorem{theorem}{Theorem}
\newtheorem{lemma}[theorem]{Lemma}

\newcommand{\ignore}[1]{}
\newcommand{\blue}[1] {\textcolor{black}{#1}}
\newcommand{\newD}[1] {\textcolor{black}{#1}}

\newcommand{\red}[1]{\textcolor{black}{#1}}
\newcommand{\new}[1]{\textcolor{black}{#1}}

\newcommand{\green}[1] {\textcolor{black}{#1}}

\begin{document}
\title{Combining dynamical decoupling with fault-tolerant quantum computation}
\author{Hui Khoon Ng}
\altaffiliation[Current address: ]{DSO National Laboratories, Applied Physics Lab, Singapore, and Centre for Quantum Technologies, National University of Singapore, Singapore.}
\affiliation{Institute for Quantum Information, California Institute of Technology, Pasadena, CA 91125, USA}
\author{Daniel A. Lidar}
\affiliation{Departments of Electrical Engineering, Chemistry, and Physics, and Center for Quantum Information Science \& Technology, University of Southern California, Los Angeles, California 90089, USA}
\author{John Preskill}
\affiliation{Institute for Quantum Information, California Institute of Technology, Pasadena, CA 91125, USA}

\begin{abstract}
We study how dynamical decoupling (DD) pulse sequences can improve the reliability of quantum computers. We prove upper bounds on
the accuracy of DD-protected quantum gates and derive sufficient conditions for DD-protected gates to outperform unprotected gates. Under suitable conditions, fault-tolerant quantum circuits constructed from DD-protected gates can tolerate stronger noise, and have a lower overhead cost, than fault-tolerant circuits constructed from unprotected gates. Our accuracy estimates depend on the dynamics of the bath that couples to the quantum computer, and can be expressed either in terms of the operator norm of the bath's Hamiltonian or in terms of the power spectrum of bath correlations; we
explain in particular how the performance of recursively generated concatenated pulse sequences can be analyzed from either viewpoint. Our results apply to Hamiltonian noise models with limited spatial correlations. 
\end{abstract}

\maketitle

\section{Introduction}\label{sec:DDFTIntro}

Two well-known methods for protecting quantum systems from noise are
dynamical decoupling (DD) and quantum error correction (QEC). In DD,
pulses are applied to the protected system, chosen so that the damaging effects of the noise nearly average away. In QEC, protected logical qubits are encoded as collective states of many physical qubits, chosen so that damage due to noise can be detected and reversed. 

Each method has advantages and disadvantages. On the plus side, resource requirements for DD are relatively modest \cite{NMRBook,VL98,Zanardi99,Duan99,Vitali99,Viola99,Viola03,BL03,KL05,KL07,Viola05,Kern05,Uhrig}. Only unitary control operations need be applied to the system; there is no need to perform measurements or to replace used ancillary qubits with fresh qubits. Furthermore, a single physical qubit suffices for each protected logical qubit, and protected quantum gates can be implemented using relatively short sequences of pulses. DD pulse sequences are simple enough that experiments on a wide variety of quantum systems have convincingly demonstrated the effectiveness of DD \cite{Berglund01,Fortunato02,Fraval05,Petta05,Morton06,Morton08,Biercuk09,Uys09,Biercuk09a,Damo08,beavan:032308,Sagi09}. On the minus side, DD is effective only against low-frequency noise, slowly varying on the time scale set by the interval between pulses, and its effectiveness is intrinsically limited by imperfections in the timing and shape of the pulses. Furthermore, DD is not an efficient scheme for flushing entropy from the system, if no qubits are replaced or refreshed; thus it seems that DD does not by itself provide a feasible route to scalable quantum computing.

For QEC, on the plus side, the quantum accuracy threshold theorem establishes that QEC, through judicious design of fault-tolerant gadgets acting on code blocks, suffices for accurate simulation of arbitrarily long quantum computations, if the noise is sufficiently weak and reasonably local \cite{Shor96,AB99,Kitaev97,Knill98,TB,AGP,AKP,NP}. QEC can succeed against high-frequency noise, where DD methods fail. On the minus side, though, the resource requirements for QEC are quite daunting. A ready supply of fresh qubits is necessary; furthermore, the number of physical system qubits needed to encode one logical qubit, and the number of physical gates needed to execute one logical gate, can be substantial. Because of the complexity of fault-tolerant quantum computing protocols, and because these protocols work only when the noise is already quite weak, experiments showing that QEC can suppress naturally occurring noise have not yet been performed.

Because of their complementary strengths, DD and QEC used together
should be more effective at protecting quantum computers from noise
than either used by itself. Combining these two methods of error
control is the topic of this paper.
Hybrid schemes combining DD with QEC have been proposed previously \cite{Viola:99a,ByrdLidar:01a,KhodjastehLidar:03}, and even studied experimentally  \cite{Boulant02}. Our new contribution is a systematic investigation of the advantages of hybrid schemes for fault-tolerant quantum computing, including rigorous bounds on performance. 

Our main technical results are analytic expressions for the ``effective noise strength'' of quantum gates implemented using DD pulse sequences. The effective noise strength is (an upper bound on) the deviation in the operator norm of the noisy protected gate from an ideal gate. In the Hamiltonian noise models that we consider, the logarithm of the operator realized by a DD-protected gate can be expanded as a power series (the Magnus expansion) in the noise Hamiltonian; we derive upper bounds on the sum of this series, obtaining formulas for the effective noise strength in terms of parameters in the noise Hamiltonian. We find such bounds both for general DD pulse sequences, and also for pulse sequences that have an approximate time-reversal symmetry; in the latter case the terms of even order in the Magnus expansion are heavily suppressed. 

Armed with our formulas for the effective noise strength, we derive a ``noise-suppression threshold condition'' on the noise parameters. When this condition is satisfied, DD-protected gates are more accurate than unprotected gates. We also compare fault-tolerant quantum circuits composed from DD-protected gates with circuits composed from unprotected gates. In either case, we express the ``accuracy threshold condition'' on the noise parameters. When this condition is satisfied, quantum computation is scalable --- accurate computations of arbitrary size can be performed with a reasonable overhead cost. Typically, improvements in gate accuracy achieved by DD mean that more noise can be tolerated by QEC combined with DD than by QEC alone, and that invoking DD can reduce the overhead cost of QEC.

Our expressions based on the Magnus expansion for the effective noise strength depend on the operator norm of the Hamiltonian that governs the internal dynamics of the quantum computer's environment (the ``bath''), and the results are not useful if this norm is large. But we also describe an alternative method of analysis yielding expressions for the effective noise strength in terms of the frequency spectrum of the bath correlations. Results derived by this method, based on the Dyson expansion,
can be applicable even if the bath Hamiltonian has unbounded norm, as long as the typical bath frequencies are sufficiently small. 

The performance of DD can sometimes be enhanced by using recursively generated ``concatenated'' pulse sequences \cite{KL07}. Adding an extra ``level'' to the recursive hierarchy further suppresses the effective noise Hamiltonian, but at the cost of lengthening the pulse sequence, and the minimal effective noise strength is achieved by choosing the level that optimizes this tradeoff. We analyze concatenated DD sequences and estimate the optimal effective noise strength, using both our bounds on the Magnus expansion and the correlation function viewpoint.

Our analysis of the improvement in 
gate accuracy
that can be achieved by combining DD and QEC applies only to a special class of Hamiltonian noise models. These models satisfy what we call the ``local-bath assumption'' which limits the spatial correlations in the noise. Whether our results can be extended to more general noise models that violate this assumption is an intriguing open question. 

We formulate our noise model in Sec.~\ref{sec:DDFTNoise}. In Sec.~\ref{sec:DDFTTools} and Sec.~\ref{sec:DDFTDDgates} we review and develop some of the tools we need to analyze the performance of DD pulse sequences and fault-tolerant quantum circuits using the Magnus expansion. We state our central results relating the effective noise strength of DD-protected gates to the properties of the noise Hamiltonian, and their implications concerning the noise-suppression threshold and accuracy threshold, in Sec.~\ref{sec:DDFTResults}; then we apply these results to some specific pulse sequences in Sec.~\ref{sec:DDFTExample}. Derivations of these results are contained in Sec.~\ref{sec:DDFTBounds} and the Appendices. We analyze concatenated DD in Sec.~\ref{sec:concatenated}. In Sec.~\ref{sec:scaling}, we emphasize that the effective noise can be related to intensive quantities that are independent of the spatial volume of the bath, and in Sec.~\ref{sec:correlator} we express the noise strength in terms of properties of bath correlations. Sec.~\ref{sec:DDFTConc} contains our conclusions.

\section{Noise model}\label{sec:DDFTNoise}

\subsection{Noise Hamiltonian}

We denote by $S$ the system consisting of all of the qubits in our quantum computer, and we describe the noise acting on  $S$ using a ``noise Hamiltonian'' $H$, which governs the joint evolution of the system and its environment, the bath $B$. During a computation, the Hamiltonian also contains time-dependent terms that realize quantum gates acting on the qubits, but for now consider the case where there are no gates; then $H$ may be expressed as
\begin{equation}
H\equiv H_B + H_\err.
\end{equation}
Here 
\begin{equation}
H_B \equiv\Id_S \otimes B_0
\end{equation} 
describes the ``free'' evolution of the bath (how it would evolve if it were not coupled to the system) while 
\begin{equation}
H_\err\equiv H_S^0+H_{SB}
\end{equation}
includes all the terms in $H$ that act non-trivially on the system. The term
\begin{equation} 
H_S^0\equiv S_0\otimes \Id_B
\end{equation}
describes the unperturbed free evolution of the system; $H_{SB}$ contains terms coupling the system to the bath, and also perhaps other noise terms that act nontrivially only on the system.

Though for some purposes it may seem natural to transform away $H_S^0$ by working in the interaction picture (that is, by considering the motion of the system relative to the rotating frame determined by $H_S^0$), we have included $H_S^0$ in the term $H_\err$ that represents the noise acting on the system. Our reason is that the DD sequences we study are designed to remove the effects of all ``always-on'' terms in the Hamiltonian that act on the system, \emph{i.e.}, not just $H_{SB}$ but also the free evolution term $H_S^0$.  We may by convention choose ${\rm tr}_S(H_{\rm err})=0$, where ${\rm tr}_S$ denotes the system trace, since the trace of $H_{SB}$ may be absorbed into the bath operator $B_0$, and the trace of $S_0$ can be removed by subtracting a term proportional to $\Id_S\otimes \Id_B$, which just shifts the zero point of the energy and has no dynamical effect. 

\begin{figure}[!htb]
\centering
\includegraphics[width=0.25\textwidth]{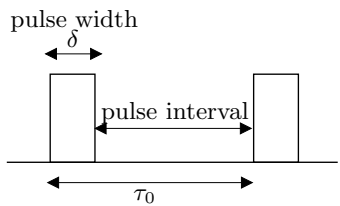}
\caption[Pulse parameters]{\label{fig:PulseTimes} Parameters characterizing a sequence of uniformly spaced rectangular pulses: $\delta$ is the pulse width, and $\tau_0-\delta$ is the interval between the end of one pulse and the beginning of the following pulse.}
\end{figure}

Now consider modeling the noise during a nontrivial quantum
computation. A computation is a circuit containing three types of
operations: qubit state preparations, unitary quantum gates, and qubit
measurements. We model a noisy preparation as an ideal preparation
followed by evolution according to $H$ for a specified time interval,
and we model noisy measurements as ideal measurements preceded by
evolution according to $H$. We assume that quantum gates are executed
using short, hard pulses, where, as in
some experiments, the time interval between consecutive pulses is much longer than the pulse width. Each pulse has its support in a narrow interval of width $\delta$, and we denote by $\tau_0$ the sum of the pulse width and the pulse interval (see Fig. \ref{fig:PulseTimes}), where $\delta\ll \tau_0$. To be concrete, we will sometimes assume that the pulses are perfectly ``rectangular'' --- {\em i.e.}, have vanishing rise-time and fall-time. However, the details of the pulse shape are not really used in our analysis; rather, all that matters is that the pulse is confined to a narrow interval (and even this assumption will be relaxed in Sec.~\ref{subsec:eulerian}). We use the same noise Hamiltonian $H$ to describe the noise both during a pulse and during the interval between pulses. We neglect errors in the timing and strength of the pulses; these are typically small in practice because the pulses are controlled by accurate classical circuitry. 

\subsection{Local-bath assumption}
\begin{figure}[!ht]
\begin{center}
\includegraphics[width=0.35\textwidth]{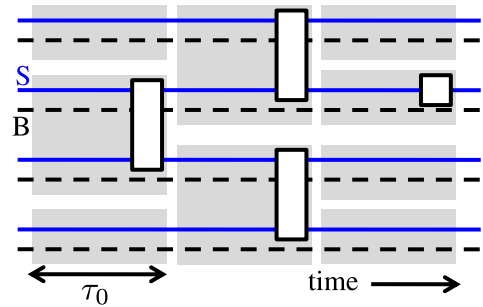}
\end{center}
\caption{\label{fig:localbath} (color online) Illustration of the local-bath assumption. Solid (blue) lines are system qubits, and dashed (black) lines are bath subsystems. Each unfilled rectangle is a quantum gate, and its associated filled (light grey) rectangle represents the joint evolution of system qubits and bath subsystems that are coupled by the noise Hamiltonian. The filled rectangles do not overlap, indicating that 
when two gates act in parallel on distinct sets of system qubits, the associated bath subsystems are also distinct.}
\end{figure}

To further simplify our analysis, we make an additional assumption
about the noise, which we call the \emph{local-bath assumption}
\cite{TB}, illustrated in Fig.~\ref{fig:localbath}.
Let us use the term ``location'' to speak of an operation in a quantum
circuit that is performed in a single time step --- a location may be
a single-qubit or multi-qubit gate, a qubit preparation step, a qubit
measurement, or the identity operation in the case of a qubit that is
idle during the time step. Each time step has duration $t_0$; thus
$t_0=N \tau_0$ if $N$
equally spaced pulses are applied at a particular location. For a specified location labeled by $a$, let $\cQ_a$ denote the set of qubits that participate in the operation applied at that location (for example, a pair of qubits if the operation is a two-qubit gate). Under the local-bath assumption, the noise Hamiltonian can be expressed as
\begin{equation}
  H=\sum_a H_a,
  \label{eq:H-local}
\end{equation}
where the sum is over all locations occurring at a particular time step, and where for any two distinct locations $a$ and $b$ in that time step, $H_a$ and $H_b$ act not only on disjoint sets of system qubits but also on distinct baths. That is, we may write
\begin{equation}
H_a=H_{B,a}+H_{\err,a},
\end{equation}
with
\begin{align}
H_{B,a}&=\Id_{S,a}\otimes B_{0,a},\notag\\
\quad H_{\err,a}&=\sum_\alpha S_{\alpha,a}\otimes B_{\alpha,a},
\end{align}
where the operators $S_{\alpha,a}$ act on $\cQ_a$, and where, for
$a\ne b$, the bath operators $B_{0,a}$ and $B_{\alpha,a}$ associated
with location $a$ commute with the bath operators $B_{0,b}$ and
$B_{\alpha,b}$ associated with location $b$. 
Thus $[H_a,H_b]=0$ for all location pairs $a$ and $b$. Each $H_a$ is assumed to
be time-independent during the duration of location $a$ (this
assumption is helpful because DD pulse sequences are
typically
designed to cope with a time-independent noise Hamiltonian), but Hamiltonians at different locations need not be the same. 

The local-bath assumption allows us express the time evolution
operator for a single time step as a product of unitary operators,
each associated with one particular location, and to analyze the
effectiveness of the DD pulse sequence for each location
separately. Without this assumption, a rigorous analysis of DD-improved fault-tolerant circuits would be far less tractable. We expect  our local-bath model to be a reasonable
approximation to the noise in actual systems, if qubits are well isolated from one another when they are not coupled by quantum gates. However interactions between
qubits (and their associated baths) at different circuit locations are surely
present at some level, and in Sec.~\ref{sec:scaling} we will comment
further on how our analysis is affected when the local-bath assumption
is relaxed.

\subsection{Noise parameters}

To characterize the noise strength, it is useful to introduce the parameters $\beta$, $J$, $\epsilon$ defined by:
\begin{align}
\label{eq:beta-define}\beta&\equiv\max_a\Vert H_{B,a}\Vert,\\
J&\equiv\max_a\Vert H_{\err,a}\Vert,\\
\label{eq:epsilon-define}\epsilon&\equiv\beta+J\geq\max_a\Vert H_a\Vert.
\end{align}
The norm here is the sup operator norm
\begin{equation}
\Vert A\Vert\equiv\sup_{\{|v\rangle\}}\frac{\Vert A|v\rangle\Vert}{\Vert |v\rangle\Vert}, 
\end{equation}
where the vector norm is the Euclidean norm $\Vert |v\rangle\Vert\equiv \sqrt{\langle v|v\rangle}$. Actually, our results concerning the effectiveness of DD pulse sequences apply for any norm that is unitarily-invariant (and therefore also submultiplicative \cite{Bhatia}), but the operator norm will be used to relate these results to the accuracy threshold for fault-tolerant quantum computing \cite{TB,AGP}. We are typically interested in the case where the noise is weak, in the sense that the dimensionless parameter  $\epsilon\tau_0$ is small compared to one (and hence also $J\tau_0\ll 1$ and $\beta\tau_0\ll 1$). We will derive bounds on the performance of DD-protected quantum gates expressed in terms of these small quantities, and also in terms of the dimensionless pulse width $\delta/\tau_0\ll 1$.

For our analysis of fault-tolerant circuits, we will find it convenient to assume that measurements and preparations are at least as fast as pulses, \emph{i.e.}, can be executed in time at most $\delta$. But in Sec.~\ref{subsec:slowMeas} we will discuss how to interpret our results if measurements or preparations take much longer than pulses.

\section{Tools}\label{sec:DDFTTools}

Let us next review some tools for analyzing the noise suppression arising from DD techniques. We focus here on the foundations of our analysis based on the Magnus expansion; further background, needed for our analysis based on bath correlation functions, will be discussed in Sec.~\ref{sec:correlator}. We also provide here a brief discussion of fault tolerance, including the notion of the effective noise strength at a circuit location, a central quantity in our analysis. 

\subsection{Toggling frame}

For now, disregard that we want to do computation, and focus instead
on quantum storage --- the original context for DD methods. In the
absence of any external control, the system and bath evolve under the
time-independent noise Hamiltonian $H$. A DD pulse sequence is
realized via a time-dependent control Hamiltonian $H_c(t)$ acting only
on the system
so that the system and bath evolve according to
$H+H_c(t)$. (In our noise model, we assume that the same noise Hamiltonian $H$ applies during a pulse as between pulses, while recognizing that this assumption is really an idealization.) The DD sequence can be described using either $H_c(t)$
itself or using the time evolution operator $U_c(t)\equiv U_c(t,0)$
generated by $H_c(t)$.

For understanding the effects of the control Hamiltonian, it is convenient to use the interaction picture defined by $H_c(t)$, also known as the \emph{toggling frame} \cite{NMRBook,Viola99,Viola:99a,Viola03,Viola05}. The toggling-frame density operator $\tilde\rho_{SB}(t)$ is related to the Schr{\"o}dinger-picture density operator $\rho_{SB}(t)$ by
\begin{align}
\rho_{SB}(t)&= U(t,0)\rho_{SB}(0)U^\dagger (t,0)\nonumber\\
&\equiv U_c(t)\tilde\rho_{SB}(t)U_c^\dagger(t),
\end{align}
where $U(t,0)$ is the evolution operator generated by the full Hamiltonian $H+H_c(t)$. Therefore the toggling-frame state evolves according to
\begin{equation}
\tilde\rho_{SB}(t)= \tilde U(t,0) \tilde \rho_{SB}(0)\tilde U^\dagger(t,0),
\end{equation}
where the toggling-frame time evolution operator
\begin{equation}
\tilde U(t,0) \equiv U_c^\dagger(t) U(t,0)
\end{equation}
is generated by the \emph{toggling-frame Hamiltonian}
\begin{equation}
\tilde H(t)\equiv U_c^\dagger(t)HU_c(t).
\end{equation}
Since $U_c(t)$ acts nontrivially only on the system, $\tilde H(t)$ can be written as
\begin{equation}
\tilde H(t)=H_B+\tilde H_\err(t),
\label{eq:tildeH}
\end{equation}
where $\tilde H_\err(t)\equiv U_c^\dagger (t)H_\err U_c(t)$ is the toggling-frame version of $H_\err$.  Because the operator norm is unitarily-invariant, we have $\Vert\tilde H(t)\Vert=\Vert H\Vert\leq \epsilon$ and $\Vert\tilde H_\err(t)\Vert=\Vert H_\err\Vert\leq J$.

We consider \emph{cyclic DD}, where
$U_c(t)$ returns to the identity (up to a possible irrelevant overall
phase) at the end of
a cycle taking time $t_{\rm DD}$:
\begin{equation}
U_c(\tDD)=U_c(0)=\Id.
\end{equation}
Therefore, at the end of
the cycle, the toggling-frame and Schr{\"o}dinger-picture states coincide.

\subsection{Finite-width pulses}

In DD, the system is controlled using a sequence of
pulses, where the control Hamiltonian $H_c(t)$ vanishes in between the pulses. The control unitary resulting from a sequence of $R$ pulses can be expressed as
\begin{equation}
\label{eq:UctDD}
U_c=\Id P_\numpulse\Id P_{\numpulse-1}\Id\ldots P_2\Id P_1\Id.
\end{equation}
where $P_k$ is the unitary achieved by the $k$th pulse. We have inserted the identity $\Id$ between successive pulses to indicate the time intervals during which $H_c(t)=0$. For some pulse sequences, including the ones described in Sec.~\ref{sec:DDFTExample}, all pulse intervals have the same duration, but for most of our analysis (excluding some of the discussion of pulse-width effects in Sec.~\ref{subsubsec:Omega1-prime}) we need not assume that the pulses are uniformly spaced.  (It is known that the effectiveness of DD can sometimes be improved by varying the spacing between pulses \cite{Uhrig,Yang,Uys09,Lee08,Uhrig09,Uhrig09a}.)

If the pulses are rectangular with width $\delta$, then we may write
\begin{equation}
P_k\equiv \exp(-i\delta H_{P_k}),
\end{equation}
where $H_{P_k}$ is the time-independent control Hamiltonian that is turned on during the $k$th pulse. If the $k$th pulse begins at time $s_k$, then the control unitary during the pulse ($t\in [s_k,s_k+\delta)$) is
\begin{align}
\label{eq:UcPW}
U_c(t)&=\exp\left(-i\Delta_kH_{P_k}\right)U_c(s_{k})\notag\\
&=\exp\left(-i\Delta_kH_{P_k}\right)P_{k-1}\ldots P_2P_1,
\end{align}
where $\Delta_k=t-s_k$. The toggling-frame Hamiltonian $\tilde H(t)$ can be written as
\begin{align}
\label{eq:Htilde}
&~\quad\tilde H(t)=U_c^\dagger (t)HU_c(t)\\
&=\left\{\begin{array}{l}
\tilde H^{(k-1)}\hspace{3.4cm} \text{for }t\in[s_{k-1} + \delta,s_{k}),\\
e^{i\Delta_k \tilde H_{P_k}^{(k-1)}}\tilde H^{(k-1)}e^{-i\Delta_k \tilde H_{P_k}^{(k-1)}}\text{for }t\in[s_k,s_k+\delta),\\
\end{array}\right.\notag
\end{align}
where
\begin{equation}
\tilde H^{(k-1)} = P_1^\dagger P_2^\dagger \ldots P_{k-1}^\dagger H P_{k-1}\ldots P_2 P_1.
\end{equation}
In the case of cyclic DD, after the last pulse of a complete cycle we have $U_c=\Id$ and $\tilde H= H$. 

\subsection{Magnus expansion}

For a given $\tilde H(t)$, the toggling-frame time evolution operator $\tilde U(\tDD,0)$ can be computed using a \emph{Magnus expansion} \cite{Magnus}. For a unitary time evolution operator $U_M(t,0)$ satisfying the Schr{\"o}dinger equation
\begin{equation}
i\frac{\partial}{\partial t}U_M(t,0)=H_M(t)U_M(t,0),\quad U_M(0,0)=\Id,
\end{equation}
determined by Hamiltonian $H_M(t)$, the Magnus expansion at time $T$ is an operator series 
\begin{equation}
\Omega(T)\equiv\sum_{n=1}^\infty\Omega_n(T)
\end{equation} 
such that
\begin{equation}
U_M(T,0)=\exp\left[\Omega(T)\right],
\end{equation}
and $\Omega_n$ is $n$th order in the Hamiltonian $H_M(t)$. Thus, for the fixed time $T$, time evolution generated by the time-dependent Hamiltonian $H_M(t)$ is equivalent to time evolution generated by the time-independent effective Hamiltonian $H_\text{eff}\equiv \frac{i}{T}\Omega(T)$. 

The leading terms in the Magnus expansion are (see for example, \cite{Blanes08})
\begin{align}
\Omega_1(T)&=-i\int_0^T ds~H_M(s),\\
\label{eq:O2}\Omega_2(T)&=-\frac{1}{2}\int_0^T ds_1\int_0^{s_1} ds_2\left[H_M(s_1),H_M(s_2)\right],\\
\label{eq:O3}\Omega_3(T)&=\frac{i}{6}\int_0^T ds_1\int_0^{s_1} ds_2\int_0^{s_2}ds_3\\
&\hspace{2cm}\big(\left[H_M(s_1),\left[H_M(s_2),H_M(s_3)\right]\right]\notag\\
&\hspace{1.7cm}+\left[H_M(s_3),\left[H_M(s_2),H_M(s_1)\right]\right]\big).\notag
\end{align}
Higher-order terms can be computed using a recursive formula; see Sec.~\ref{sec:DDFTBounds} and Appendix \ref{app:magnus}. In general, $\Omega_n(T)$ is the time integral of a sum of $(n-1)$-nested commutators, each with $n$ factors of $H_M(t)$. The Magnus expansion is thus an infinite series in $H_MT$; a sufficient condition for convergence is \cite{Moan99}
\begin{equation}\label{eq:conv}
\int_0^T dt~\left\Vert H_M(t)\right\Vert<\pi.
\end{equation}

For cyclic DD, we can use the Magnus expansion to compute the toggling-frame time evolution operator $\tilde U(\tDD,0)$ for one complete cycle, where $H_M(t)$ is the toggling-frame Hamiltonian $\tilde H(t)=U_c^\dagger (t)HU_c(t)$ and $U_c(\tDD)=\Id$. In first order we obtain
\begin{align}\label{eq:GrpAvg}
\Omega_1(\tDD)&=-i\int_0^{\tDD}dt~\tilde H(t)\notag\\
&=-i\int_0^{\tDD}dt~U_c^\dagger(t) H U_c(t).
\end{align}
For group-based DD schemes, like the examples we will discuss in Sec.~\ref{sec:DDFTExample}, the integral Eq.~\eqref{eq:GrpAvg} averages $H$ over a finite group $\cG$ if the pulses are ideal, projecting $H$ into the commutant of $\cG$ \cite{Zanardi99}. If $\cG$ acts irreducibly on the system Hilbert space, the commutant contains only the identity operator acting on the system, and therefore $\Omega_1(\tDD)$ acts nontrivially only on the bath. In that case we say that $\Omega_1(\tDD)$ is a ``pure bath'' term.

We say that a DD pulse sequence achieves first-order decoupling if the
first-order term in the Magnus expansion for $U_c(\tDD)$ acts
trivially on the system. More generally, the sequence achieves
$m$th-order decoupling if $\Omega_n(\tDD)$ is a pure bath term for
each $n\leq m$.
In our analysis we will at first
consider pulse sequences that achieve
first-order decoupling for ideal zero-width pulses
(later we will discuss the corrections to first-order decoupling that arise when the pulses have nonzero width, and we will also describe ``Eulerian'' pulse sequences that achieve first-order decoupling even when pulse widths are nonzero \cite{Viola03}).
In particular, these pulse sequences have the property 
\begin{equation}
\label{eq:O1Cond}
\int_0^{\tDD}dt~\tilde H_{\err,0}(t)=0,
\end{equation}
where the subscript ``0'' on $\tilde H_{\err,0}(t)$ indicates that the
 toggling-frame Hamiltonian $\tilde H_\err(t)$ is considered in the
 limit $\delta\to 0$, while holding $\tau_0$
 fixed. (For Eq.~(\ref{eq:O1Cond}) to apply there must be no term in
 $H_{\err,0}$ that acts nontrivially on the system and commutes with
 $H_c(t)$ for all $t\in [0,\tDD]$; if such terms were present they would not be removed by
 the DD sequence described by $H_c(t)$.) For pulse sequences
 satisfying Eq.~(\ref{eq:O1Cond})
 it follows from Eq.~(\ref{eq:tildeH}) that
 the first-order term in the Magnus expansion is
\begin{align}
\Omega_1(\tDD)&=-i\int_0^{\tDD}dt~\tilde H_0(t)\notag\\
&=-iH_B t_{\rm DD}-i\int_0^{\tDD}dt~\tilde H_{\err,0}(t)\notag\\
&=-iH_B\tDD,
\end{align}
a pure bath term, when $\delta=0$. For pulses with nonzero width $\delta$, first-order decoupling is not exact, but the deviation of $\Omega_1(\tDD)$ from a pure bath term is
$O(\delta/\tau_0)$ and thus small when the pulses are sufficiently narrow. For suitably designed pulse sequences the deviation can be improved to a higher power of $\delta/\tau_0$ \cite{Viola03,Pasini08}.

A pulse sequence that achieves first-order decoupling will also achieve second-order decoupling if $\tilde H$ is time-symmetric: $\tilde H(\tDD-t)=\tilde H(t)$ for  $t\in[0,\tDD]$. This condition is satisfied provided
\begin{equation}
U_c(\tDD-t)=V_tU_c(t),
\end{equation}
where $V_t$ is unitary and commutes with $H$ (for example, if $V_t=e^{i\phi_t}\Id$ is a phase). In fact, when $\tilde H$ is time-symmetric, not just the second-order term, but all even terms in the Magnus expansion vanish \cite{Wang72}, as we show in Appendix \ref{app:Sym}.

\subsection{Quantum accuracy threshold theorem}
\label{subsec:ToolsFT}

The quantum accuracy threshold theorem establishes that a noisy quantum computer can operate reliably if the noise is sufficiently weak. Under the local-bath assumption formulated in Sec.~\ref{sec:DDFTNoise}, the operation applied at location $a$ in the noisy circuit is a unitary transformation $\Gb_a$ acting on the system and bath, which can be expressed as
\begin{equation}
\label{eq:good-plus-bad}
\Gb_a= \cG_a+\cB_a.
\end{equation}
Here $\cG_a$ is the ``good part'' of the operation; it can be expressed as $G_a\otimes B_a$, where $G_a$ is the ideal operation that would be applied to the system in the absence of noise, and $B_a$ is a unitary transformation acting on the bath. The operator $\cB_a$ is the ``bad part,'' the deviation of $\Gb_a$ from the ideal operation, which acts jointly on system and bath. (Recall that we model a noisy qubit preparation as an ideal preparation followed by a noisy unitary transformation, and a noisy qubit measurement as a noisy unitary transformation followed by an ideal measurement; for preparation or measurement locations, $\Gb_a$ denotes the noisy transformation that follows or precedes the ideal preparation or measurement.) In this noise model, we may characterize the noise strength by
\begin{equation}
\label{eq:eta-def}
\bar\eta\equiv \max_a\Vert\cB_a\Vert,
\end{equation}
the maximum value of the operator norm of the bad part, where the maximum is with respect to all locations in the noisy circuit. The threshold theorem asserts that an ideal circuit of arbitrary size can be simulated accurately if $\bar\eta$ is less than a critical value $\eta_0$, the accuracy threshold. The threshold theorem proved in \cite{AGP} actually applies to a broader class of noise models that do not necessarily satisfy the local-bath assumption, but this class includes the noise model of Sec.~\ref{sec:DDFTNoise} as a special case. The analysis in \cite{AC} established a lower bound on the accuracy threshold, $\eta_0\gtrsim 10^{-4}$. \red{If $\bar\eta < \eta_0$, then we say the noise is below the accuracy threshold, meaning that scalable quantum computing is possible.}

In this paper we will relate the noise strength $\bar\eta$ defined by Eq.~(\ref{eq:good-plus-bad}) and Eq.~(\ref{eq:eta-def}) to the parameters that characterize the noise model defined in Sec.~\ref{sec:DDFTNoise}. We denote by $\etaDD$ the value of $\bar\eta$ that can be achieved using dynamical decoupling, and we denote by $\eta$ the value of $\bar\eta$ achieved without using dynamical decoupling. If $\etaDD < \eta$, then we say that the noise is below the \emph{noise suppression threshold}, meaning that dynamical decoupling reduces the effective noise strength. 

In Sec.~\ref{sec:DDFTResults} we express $\etaDD$ in terms of the parameters $J$ and $\epsilon$ defined in Eq.~(\ref{eq:beta-define})-(\ref{eq:epsilon-define}). In Sec.~\ref{sec:correlator} we express $\etaDD$ in terms of properties of bath correlation functions, using a different formula than Eq.~(\ref{eq:eta-def})\blue {, based on the Dyson expansion.}

\section{DD-protected gates}\label{sec:DDFTDDgates}

\subsection{Including the gate pulse}

So far we have described how to reduce the noise in a quantum memory
using cyclic DD.
Now we want to estimate the effective noise strength achieved
by DD for operations other than the identity, so we must explain how
DD is used to suppress the noise in these nontrivial operations, We
will describe nontrivial quantum gates, postponing discussion of
preparations and measurements until later. 

We refer to one cycle of the DD pulse sequence for the identity gate
as the ``memory'' sequence. To perform a DD-protected nontrivial gate
$G_a$, we must modify the memory sequence accordingly. In fact our DD
pulse sequence for the gate is exactly the same as the memory
sequence, except for the very last pulse. If the memory sequence
of $R$ pulses
ends with a period of trivial evolution, then we \emph{append} a pulse
implementing $G_a$ to the end the memory sequence. Thus, if the memory
sequence lasts time $\tDD$ and the pulse width is $\delta$, then the
$G_a$ pulse sequence lasts time $t_0=\tDD + \delta$
and uses $N=R+1$ pulses.
If on the other hand the
$R$-pulse
memory sequence ends with a nontrivial pulse implementing
$P$, then we combine this pulse and the gate pulse into a single pulse
implementing $G_aP$. Again, we denote the total time for the $G_a$
pulse sequence by $t_0$,
and the total number of pulses by $N(=R)$.

While we assume for simplicity that every pulse has the same width $\delta$, we recognize that in some cases different types of pulses may have different time scales. For example, in recent experiments with quantum dot qubits, $X$ gates are implemented using (fast) exchange couplings and $Z$ gates are implemented using (slow) magnetic field gradients \cite{Petta05}.
One may interpret $\delta$ as the duration of the longest pulse used, or one could easily refine our analysis by allowing different pulses to have different widths.

In a DD-protected circuit, each $G_a$ gate is replaced by the corresponding DD-protected gate; under the local-bath assumption, the noisy DD-protected gate is a unitary transformation denoted $\Gb_a$ acting jointly on the \red{system qubits involved in the gate and the associated bath subsystem. }Though the duration $t_0$ of a DD-protected gate is longer than the duration $\tau_0$ of an unprotected gate, the DD-protected gate may be more accurate than the unprotected gate, if the noise is weak enough.

At the end of the complete 
$G_a$
pulse sequence, the unitary operator $U_{c,a}(t_0)$ generated by the control Hamiltonian $H_c(t)$ (which now includes the gate pulse) is
\begin{equation}
U_{c,a}(t_0)=G_aU_c(\tDD)=G_a,
\end{equation}
because the cyclic memory sequence satisfies $U_c(\tDD)=\Id$ (up to a possible phase). 
Therefore the noisy DD-protected gate at location $a$ is
\begin{equation}
\label{eq:DD-protected}
\Gb_a\equiv U_a(t_0,0)=U_{c,a}(t_0)\tilde U_a(t_0,0)=G_a\tilde U_a(t_0,0),
\end{equation}
where $\tilde U_a(t_0,0)$ is the toggling-frame time evolution operator.
The corresponding toggling-frame Hamiltonian is similar to the toggling-frame Hamiltonian Eq.~\eqref{eq:Htilde} for the memory sequence, except for the appended gate pulse:
\begin{align}\label{eq:GHtilde}
&\quad\tilde H_a(t)=U_{c,a}^\dagger (t)H_aU_{c,a}(t)\\
=&\left\{\begin{array}{l}
e^{i\Delta_k \tilde H_{P_k}^{(k-1)}}\tilde H_a^{(k-1)}e^{-i\Delta_k \tilde H_{P_k}^{(k-1)}}\text{for }t\in[s_k,s_k+\delta),\\
\tilde H_a^{(k)}\hspace{3.6cm}\text{for }t\in[s_k+\delta,s_{k+1}),\\
e^{i\Delta_{\numpulse+1} \tilde H_{G_a}^{(\numpulse)}}\tilde H_a^{(\numpulse)}e^{-i\Delta_{\numpulse+1} \tilde H_{G_a}^{(\numpulse)}}\text{for }t\in[s_{\numpulse+1},s_{\numpulse+1}+\delta ),\notag\\
G_a^\dagger H_a G_a \hspace{2.95cm}\text{for }t=t_0.
\end{array}\right.
\end{align}
Eq.~(\ref{eq:GHtilde}) applies to the case where the gate pulse is appended to the end of the memory sequence; the memory sequence contains $R$ equally spaced pulses labeled by $k=1,2,\ldots, \numpulse$, and the gate pulse begins at time $s_{\numpulse+1}$. 

The DD-protected qubit measurement is the memory pulse sequence followed by an ideal measurement. We assume that the measurement takes time $\delta$, the same as the pulse width, so that the duration $t_0$ of the protected measurement matches the duration of the DD-protected gate. Similarly, the DD-protected qubit preparation is an ideal preparation lasting time $\delta$ followed by the memory pulse sequence. See Sec.~\ref{subsec:slowMeas} for discussion of how our analysis is modified when preparations and measurements are slow compared to other operations.

\subsection{Effective noise strength}

To define the effective noise strength for the DD-protected gate, we divide the noisy gate into a good part and a bad part as in Eq.~(\ref{eq:good-plus-bad}), obtaining 
\begin{eqnarray}
\Gb_a&=\underbrace{G_aU_{B,a}(t_0)}_{\equiv \cG_a}+\underbrace{\Gb_a-G_aU_{B,a}(t_0)}_{\equiv \cB_a}
\label{eq:Gb}
\end{eqnarray}
The good part $\cG_a$ describes the ideal evolution in the absence of noise ($H_\err=0$) --- the ideal gate $G_a$ is applied to the system, while the bath evolves according to its unperturbed Hamiltonian $H_{B,a}$. The bad part $\cB_a$ describes the effects of noise, as modified by the DD pulse sequence. 

\blue {As we discuss in more detail in} \green{Sec.~\ref{subsec:gen-case}, we may choose a different way of separating the pure bath dynamics into a good and bad part than the choice made in Eq.~(\ref{eq:Gb}). Incorporating $U_{B,a}(t_0)$ into $\cG_a$ is convenient when we use the Magnus expansion to analyze the performance of DD-protected gates, but another choice is more convenient for the analysis based on the Dyson expansion in Sec~\ref{subsec:gen-case}.}

Using Eq.~(\ref{eq:DD-protected}) and the unitary invariance of the operator norm, we obtain an expression for the noise strength of the DD-protected circuit:
\begin{align}\label{eq:maxBa}
\etaDD \equiv \max_a\left\Vert\cB_a\right\Vert
&=\max_a\left\Vert \tilde U_a(t_0,0)-U_{B,a}(t_0)\right\Vert;
\end{align}
this is just the norm of the bad part expressed in the toggling frame. In what follows, we will sometimes drop the subscript $a$ when context makes the intended meaning clear.

We can now estimate $\etaDD$ using the Magnus expansion. We write 
\begin{equation}\label{eq:UHeff}
\tilde U(t_0,0)=\exp[\Omega(t_0)]\equiv \exp[-it_0\Heff],
\end{equation}
where $\Heff\equiv \frac{i}{t_0}\Omega(t_0)$, and $\Omega(t_0)$ can be computed using the (gate-appended) toggling-frame Hamiltonian $\tilde H(t)$ in Eq.~\eqref{eq:GHtilde}. To bound the quantity $\Vert \tilde U(t_0,0)-U_B(t_0)\Vert$, 
we make use of Lemma \ref{lem:error} in Appendix \ref{app:error}, which gives:
 \ignore{
 we use a special case \cite{Nakamoto} of Lemma \ref{lem:error} in Appendix \ref{app:error}:
 For any two bounded  Hermitian operators $A$ and $B$,
 \begin{equation}
 \label{eq:DiffUnitary}
 \Vert e^{-iA}-e^{-iB}\Vert\leq \Vert A-B\Vert .
 \end{equation}
Taking $A\equiv t_0\Heff$ and $B\equiv t_0H_B$, Eq.~(\ref{eq:DiffUnitary}) implies
}
\begin{align}\label{eq:Udiff}
\Vert \tilde U(t_0,0)-U_B(t_0)\Vert 
&\leq t_0\Vert\Heff-H_B\Vert.
\end{align}
Inserting the Magnus expansion $-it_0\Heff=\Omega(t_0)=\sum_{n=1}^\infty\Omega_n(t_0)$ we find
\begin{align}
\etaDD &\leq t_0\max_a\Vert H_{\text{eff},a}-H_{B,a}\Vert\notag\\
&=\max_a\Vert\sum_{n=1}^\infty \Omega_{n,a}(t_0)+it_0H_{B,a}\Vert\notag\\
\label{eq:BBound}&\leq \max_a\Big(\Vert\Omega_{1,a}'(t_0)\Vert+\sum_{n=2}^\infty\Vert\Omega_{n,a}(t_0)\Vert\Big),
\end{align}
where $\Omega_{1,a}'(t)\equiv \Omega_{1,a}(t)+itH_{B,a}$. For a pulse sequence that achieves first-order decoupling with ideal zero-width pulses, $\Omega_{1,a}'(t_0)$ vanishes in the limit $\delta\to 0$. To derive a useful upper bound on the effective noise strength $\etaDD$, we will need good bounds on the other terms in Eq.~\eqref{eq:BBound}.

\subsection{The effective noise strength for a time-symmetric sequence}\label{sec:ens-sym}

We say that the memory pulse sequence is time-symmetric (or ``palindromic'') if $\tilde H(\tDD-t)=\tilde H(t)$ for $t\in [0,\tDD]$. We will show in Appendix \ref{app:Sym} that a time-symmetric pulse sequence that achieves first-order decoupling also achieves second-order decoupling. However, the time symmetry is broken if we construct the DD-protected gate by appending the gate pulse to the memory sequence, even if the memory sequence by itself is time-symmetric.

\begin{figure*}[!t]
\begin{center}
\includegraphics[width=\textwidth]{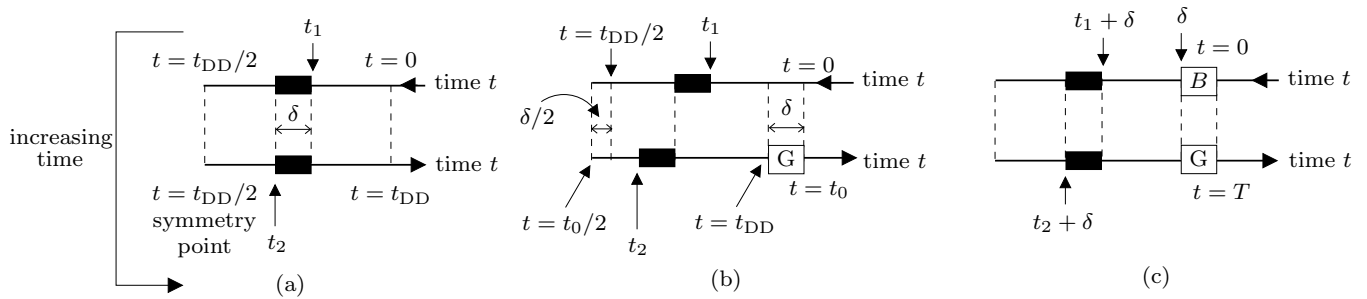}
\end{center}
\caption [Figure explaining $H_M(t)$ for a time-symmetric DD sequence]{\label{fig:TimeSym} Schematic representation of $H_M(t)$ for time-symmetric DD pulse sequences. The time axis is bent in half, with time flowing counterclockwise from the upper right to the lower left, so that times aligned on the upper and lower branches are related by time symmetry. (a) Two pulses (marked as black boxes) in a time-symmetric memory sequence with $\tilde H(\tDD-t)=\tilde H(t)$; the pulse on the bottom branch is the time-reversed version of the pulse on the top branch. (b) Appending the gate pulse (box $G$) to the memory sequence spoils the time symmetry; the black pulses on the upper and lower branches are no longer aligned. (c) Appending fictitious time evolution governed by $-H_B$ during $t\in[0,\delta]$ (box $B$) restores the time symmetry of the memory sequence for $t\in[\delta,T-\delta]$, where $T=t_0+\delta$.}
\end{figure*}

For the purpose of estimating the effective noise strength, we can nearly restore the time symmetry of the DD-protected gate by a simple trick (see Fig. \ref{fig:TimeSym}). Recalling that our goal is to derive an upper bound on $\Vert \tilde U(t_0,0)-U_B(t_0)\Vert$, we observe that the unitary invariance of the operator norm implies
\begin{equation}\label{eq:BTimeSym}
\Vert \tilde U(t_0,0)-U_B(t_0)\Vert=\Vert \tilde U(t_0,0)U_B^\dagger(\delta)-U_B(t_0-\delta)\Vert,
\end{equation}
where $\delta$ is the width of the gate pulse, and $t_0=\tDD +\delta$. Furthermore, we may regard $\tilde U(t_0,0)U_B^\dagger(\delta)$ as the time evolution operator generated by the Hamiltonian
\begin{equation}\label{eq:HMTimeSym}
H_M(t)\equiv\left\{\begin{array}{ll}-H_B&t\in[0,\delta),\\\tilde H(t-\delta)&t\in[\delta,T], \end{array}\right.
\end{equation}
where $T=t_0 + \delta$. If the memory sequence is time-symmetric, then $H_M(T-t)=H_M(t)$ for $t\in [\delta,T-\delta]$. Thus $H_M(t)$ is ``nearly time-symmetric'' in the interval $[0,T]$; the symmetry is broken only in the small intervals $[0,\delta]$ and $[T-\delta,T]$ at the beginning and end of $[0,T]$.

The unitary operator $\tilde U(t_0,0)U_B^\dagger(\delta)\equiv \exp[\Omega(T)]$ can be computed using the Magnus expansion for Hamiltonian $H_M(t)$. Viewing $\tilde U(t_0,0)U_B^\dagger(\delta)$ as being generated by the time-independent Hamiltonian $i\Omega(T)/(t_0-\delta)$ for time $t_0-\delta$, and again using Lemma \ref{lem:error} in Appendix \ref{app:error}, we obtain instead of Eq. \eqref{eq:BBound},
\begin{align}
\etaDD&=\max_a\Vert \tilde U_a(t_0,0)U_{B,a}^\dagger(\delta)-U_{B,a}(t_0-\delta)\Vert\notag\\
&\leq \max_a\Vert \Omega_a(T)+i(T-2\delta)H_{B,a}\Vert\notag\\
\label{eq:TimeSymBound}&=\max_a\Big(\Vert\Omega_{1,a}'(T)\Vert+\sum_{n=2}^\infty\Vert\Omega_{n,a}(T)\Vert\Big),
\end{align}
where $\Omega_{1,a}'(T)$ is now defined as $\Omega'_{1,a}(T)\equiv \Omega_{1,a}(T)+i(T-2\delta)H_{B,a}$.

More generally, we say that the Hamiltonian $H_M(t)$ is nearly time-symmetric in $[0,T]$ if the time symmetry holds everywhere except in a small interval or the disjoint union of several small intervals. We denote by $\Delta$ the region in which the time symmetry is violated; thus
\begin{equation}
\left\{\begin{array}{ll}
H_M(T-t)=H_M(t)&\text{for }t\notin\Delta\\
H_M(T-t)\neq H_M(t)&\text{for }t\in\Delta.
\end{array}\right.
\end{equation}
We also use the same symbol $\Delta(\ll T)$ to denote the total length of this region. Thus $\Delta=0$ for a perfectly time-symmetric sequence. In what follows, we will sometimes say that the pulse sequence realizing a DD-protected gate is ``time-symmetric'' if the corresponding memory sequence is time-symmetric, even though the time symmetry may be broken by the gate pulse appended to the memory sequence. We say that the memory sequence and the DD-protected gates are ``general'' if the memory sequence has no special time symmetry properties. 

\section{Effective noise strength and threshold conditions}\label{sec:DDFTResults}

In this section, we state some of our conclusions concerning the effective noise strength $\etaDD$ achieved by dynamical decoupling, and the implications for fault-tolerant quantum computing. Derivations will be postponed until Sec.~\ref{sec:DDFTBounds}. Here we focus on results derived using the Magnus expansion; results relating $\etaDD$ to properties of bath correlation functions 
\blue {derived using the Dyson expansion}
are discussed in Sec.~\ref{sec:correlator}.

\subsection{Bounds on \blue {the} Magnus expansion}
\label{subsec:BoundsMagnus}

Combining Eq.~(\ref{eq:BBound}) and Eq.~(\ref{eq:TimeSymBound}), we can state our upper bound on the effective noise strength $\etaDD$ as
\begin{equation}\label{eq:etaDD-magnus}
\etaDD \le \Vert\Omega_1'(T)\Vert+\sum_{n=2}^\infty\Vert\Omega_n(T)\Vert,
\end{equation}
where $\Omega_1'(T)\equiv \Omega_1(T)+i(T-2\Gamma)H_B$, $T\equiv t_0+\Gamma$, and the maximum over all locations is understood. The Magnus expansion $\Omega(T)=\sum_{n=1}^\infty\Omega_n(T)$ is computed using the Hamiltonian 
\begin{equation}\label{eq:HMGen}
H_M(t)\equiv\left\{\begin{array}{ll}-H_B&t\in[0,\Gamma),\\\tilde H(t-\Gamma)=H_B+\tilde H_\err(t-\Gamma)&t\in[\Gamma,T]. \end{array}\right.
\end{equation}
For the general case, in which we are not trying to exploit the time symmetry of the memory sequence, we choose $\Gamma=0$. For the nearly-time-symmetric case we choose $\Gamma=\delta$, and $H_M$ is time-symmetric in the interval $[\delta,T-\delta]$.

If the memory sequence achieves first-order decoupling in the limit $\delta\to 0$, then $\Vert \Omega'_1(T)\Vert$ vanishes apart from finite-width corrections. The $n$th-order Magnus term $\Omega_n(T)$ for $n\ge 2$ satisfies $\left\Vert\Omega_n(T)\right\Vert=O\left((\epsilon T)^n\right)$, because $\Vert H_M(t)\Vert \le \epsilon$, and the integral $\Omega_n(T)$ can be bounded by the volume of the integration region times an upper bound on the integrand. In fact, this estimate can be improved to $\left\Vert\Omega_n(T)\right\Vert=O\left((JT)(\epsilon T)^{n-1}\right)$, because $H_M(t)$ has the form $\pm H_B +H'(t)$ where $H'(t)$ is either $0$ or $\tilde H_\err(t)$; therefore $\Vert [H_M(t_1),H_M(t_2)]\Vert = O(J\epsilon)$, since $\Vert H'(t)\Vert\leq J$ and $H_B$ commutes with itself.

We anticipate, then, that at any location $a$, the terms in the Magnus expansion can be bounded as
\begin{subequations}\label{eq:Gen}
\begin{align}
\left\Vert\Omega_1'(T)\right\Vert&\leq C_1(JT);\\
\left\Vert\Omega_n(T)\right\Vert&\leq C_n(JT)(\epsilon T)^{n-1},~~ n=2,3,4;\\
\label{eq:Gen-5th}\sum_{n=5}^\infty\left\Vert\Omega_n(T)\right\Vert&\leq C_5 (JT)(\epsilon T)^4,
\end{align}
\end{subequations}
where $C_1, C_2, C_3, C_4, C_5$ are constants. Note that the last of these results bounds the sum of all high-order Magnus terms with $n\geq 5$. Combining Eq.~(\ref{eq:etaDD-magnus}) and Eq.~(\ref{eq:Gen}) we find 
\begin{equation}
\label{eq:etaDD-Cn}
\etaDD\le  (JT)\sum_{n=1}^5C_n(\epsilon T)^{n-1}.
\end{equation}
The constants $C_n$, derived in Sec.~\ref{sec:DDFTBounds}, are listed in Table \ref{tab:Coeffs} for both general and time-symmetric pulse sequences. Our value of $C_5$, obtained by bounding an infinite series, holds only for $\epsilon T\leq 0.54$, a condition likely to be satisfied when DD works effectively. If desired, tighter bounds can be derived on the $n$th order terms with $n\ge 5$ using results from Sec.~\ref{sec:DDFTBounds}. However, we judge Eq.~(\ref{eq:Gen-5th}) to be good enough for our purposes, since this bound on the sum of higher-order terms is already rather small for $\epsilon T\ll 1$, as in typical cases of interest. Also listed in the last column of Table \ref{tab:Coeffs} are values of $\{C_n\}$ derived in Sec.~\ref{sec:DDFTBounds} using the Dyson expansion rather than the Magnus expansion, also under the assumption $\epsilon T\leq 0.54$. These upper bounds are weaker for $n=2, 3$ but stronger for $n=4, 5$, and hence provide a tighter estimate of the effective noise strength for pulse sequences that achieve third-order decoupling. 

\begin{table}
\begin{ruledtabular}
\begin{tabular}{c|c|c|c}
&{\bf General}&{\bf Nearly time symmetric}&{\bf Dyson \red{(General)}}\\
\hline
\hline
\multirow{2}{*}{$C_1$}&\multicolumn{3}{@{\hspace{0.6cm}}l @{\hspace{0.2cm}}}{$2N\delta/T$ in general,}\\
 &\multicolumn{3}{@{\hspace{0.6cm}}l @{\hspace{0.2cm}}}{$N\delta/T$ if pulses are regularly spaced in time}\\
\hline
$C_2$ &$1/2$&$2\left(\frac{\Delta}{T}\right)\left(1-\frac{1}{2}\frac{\Delta}{T}\right)$&$1$\\
\hline
 $C_3$ &$2/9$&$2/9$&
$~1/2$\\
\hline
 $C_4$ &$11/9$ &$56\left(\frac{\Delta}{T}\right)\left(1-\frac{1}{2}\frac{\Delta}{T}\right)$ & $1/6$\\
\hline
$C_5$ & 9.43 & 9.43 & .0466\\

\end{tabular}
\end{ruledtabular}
\caption{\label{tab:Coeffs} Constants $\{C_n\}$ appearing in the upper bound Eq.~(\ref{eq:etaDD-Cn}) on the effective noise strength $\eta_{\rm DD}$, derived from the Magnus expansion in the general case and the nearly-time-symmetric case, and from the Dyson expansion in the general case. $N$ denotes the total number of pulses in the DD-protected gate, $\delta$ is the pulse width, $T=t_0$ in the general case, and $T=t_0 +\delta$ in the nearly-time-symmetric case. For the nearly-time-symmetric case, $\Delta$ is the size of the small region in which the time symmetry is violated. The value of $C_5$ applies assuming $\epsilon T\leq 0.54$. }
\end{table}

Our bounds on $\Omega_n(T)$ for $n$ odd is not improved by invoking time symmetry, but for $n=2, 4$, the bounds listed in Table \ref{tab:Coeffs} are tighter for the time-symmetric case than the general case, assuming $\Delta/T\ll 1$. In fact, $C_2$ and $C_4$ vanish in the limit $\Delta/ T\to 0$, reflecting the property that all even-order terms in the Magnus expansion vanish when the time symmetry is exact. For the time-symmetric case, we derive bounds on $C_n$ for even $n\ge 6$ in Appendix \ref{app:Even}, but these results were not used in our estimate of $C_5$.

\subsection{Noise suppression threshold}

DD-protected gates will outperform unprotected gates if the noise is
weak enough. In a circuit of unprotected gates, each gate is realized
by a single pulse, where the pulses are separated in time by the pulse
interval $\tau_0$. For the noise model of Sec.~\ref{sec:DDFTNoise},
the effective noise strength for this computation may be expressed as
\cite{TB,AGP}
\begin{equation}\label{eq:AGP}
\eta=\left(\max_a \Vert H_{SB,a}\Vert\right)\tau_0.
\end{equation}
Eq.~(\ref{eq:AGP}) is not derived using the Magnus expansion; rather it follows directly from Lemma \ref{lem:error} in Appendix \ref{app:error}.
%
\green{
The noise strength $\eta$ does not depend on the pulse shape; all that matters is the strength of the noise Hamiltonian $H_{SB,a}$ and the time $\tau_0$ allotted for executing the gate.
}
If we assume that $H_S^0=0$, Eq.~(\ref{eq:AGP}) becomes
\begin{equation}\label{eq:eta-unprotected}
\eta = J\tau_0.
\end{equation}
We say that the noise model satisfies the \emph{noise suppression threshold condition} if the effective noise strength can be reduced by using DD-protected gates instead, \emph{i.e.}, if
\begin{equation}
\etaDD<\eta.
\end{equation}
In our noise model, this condition can be expressed in terms of the parameters $\epsilon\tau_0$, $\delta/\tau_0$ and $\tau_0/t_0$.

For example, continuing to assume that $H_S^0=0$, suppose in addition that $\delta/\tau_0$ is negligible and $\epsilon T$ is small enough so that the Magnus expansion is well-approximated by the lowest-order nonzero term. Then, in the general (non-time-symmetric) case, using $C_1= 0$ and $C_2=1/2$, we can approximate $\etaDD$ by
\begin{equation}
\label{eq:etaDD-general-lowest}
\etaDD\simeq \frac{1}{2}(JT)(\epsilon T)=\frac{1}{2}\left(\frac{J\tau_0}{\tau_0/t_0}\right)\left(\frac{\epsilon \tau_0}{\tau_0/t_0}\right);
\end{equation}
we use the $\simeq$ symbol to emphasize that higher order corrections in $\delta/\tau_0$ and $\epsilon t_0$ are neglected. The noise suppression threshold condition $\etaDD<\eta=J\tau_0$ is satisfied for
\begin{equation}\label{eq:eptau0}
\epsilon \tau_0 \lesssim 2\left(\frac{\tau_0}{t_0}\right)^2,
\end{equation}
or 
\begin{equation}
\epsilon \tau_0 \lesssim 2 N^{-2}
\end{equation}
for a sequence of $N$ equally spaced pulses. As the pulse sequence
grows, the duration $t_0$ of DD-protected gates increases relative to
the duration $\tau_0$ of unprotected gates, and Eq.~\eqref{eq:eptau0}
imposes a stronger restriction on $\epsilon$.

Note that $\etaDD$ depends on the norm of the bath Hamiltonian 
$\beta$
(which contributes to $\epsilon$), while $\eta$ does not. 
\blue{Technically, this difference comes about because the second order Magnus term exhibited in Eq.~(\ref{eq:etaDD-general-lowest}) contains a contribution from the non-vanishing commutator between $H_{SB}$ and $H_B$, while $\eta$ is computed directly as a difference between the ideal and noisy Hamiltonians, differing only by $H_{SB}$ (see Appendix \ref{app:error}). }
\green{
Physically, $\etaDD$ depends on $\beta$ because dynamical decoupling works effectively only if the bath dynamics is sufficiently slow. Alternatively, we could estimate $\etaDD$ in terms of parameters other than $\beta$ that characterize the speed of the bath dynamics; for example, we will derive in Sec.~\ref{sec:correlator} an expression for $\etaDD$ involving the bath's frequency spectrum rather than the operator norm $\beta$.
}

In the limit of zero-width pulses, a time-symmetric pulse sequence that achieves first-order decoupling achieves second-order decoupling as well, so that $C_1=C_2=0$. Imposing time symmetry may lengthen the pulse sequence; we denote the duration of a DD-protected time-symmetric gate by $t_0'$, to contrast with the duration $t_0$ of the gate when the pulse sequence is not time-symmetric. In the time-symmetric case, the effective noise strength becomes (assuming $\delta=0$ and thus $T=t_0$, and using $C_3=2/9$)
\begin{equation}\label{eq:etaDD_symmetric-lowest}
\etaDD \simeq\frac{2}{9}(JT)(\epsilon T)^2=\frac{2}{9}\left(\frac{J\tau_0}{\tau_0/t_0'}\right)\left(\frac{\epsilon \tau_0}{\tau_0/t_0'}\right)^2.
\end{equation}
Therefore the noise suppression threshold condition $\etaDD < J\tau_0$ is satisfied if
\begin{equation}\label{eq:eptau0-prime}
\epsilon \tau_0 \lesssim \frac{3}{\sqrt 2}\left(\frac{\tau_0}{t_0'}\right)^{3/2},
\end{equation}
or 
\begin{equation}
\epsilon \tau_0 \lesssim \frac{3}{\sqrt 2}\left(N'\right)^{-3/2}
\end{equation}
for the case of 
$N'$
equally spaced pulses. Even though $t_0' > t_0$, Eq.~(\ref{eq:eptau0-prime}) places a less stringent condition on $\epsilon$ than Eq.~(\ref{eq:eptau0}), provided $t_0'/t_0 \lesssim(9t_0/2\tau_0)^{1/3}$. We emphasize again that Eq.~(\ref{eq:eptau0}) and Eq.~(\ref{eq:eptau0-prime}) are derived using lowest-order approximations in an expansion in $\delta/\tau_0$ and $\epsilon$.

The expression Eq.~(\ref{eq:etaDD-general-lowest}) for $\etaDD$ indicates that to achieve effective noise suppression we should favor short DD pulse sequences (with $t_0/\tau_0$ not too large) to minimize the exposure to noise during the DD-protected gate. On the other hand Eq.~(\ref{eq:etaDD_symmetric-lowest}) illustrates that a longer pulse sequence can pay off if it allows us to achieve higher-order decoupling. These results exemplify a more general tradeoff between shorter sequences and better decoupling that must be optimized to design DD-protected gates with the smallest possible effective noise strength. The tradeoff is also manifested by the analysis in Sec.~\ref{sec:concatenated} of concatenated DD pulse sequences.

\subsection{Accuracy threshold and overhead cost}

\begin{figure}[!ht]
\begin{center}
\includegraphics[width=0.5\textwidth]{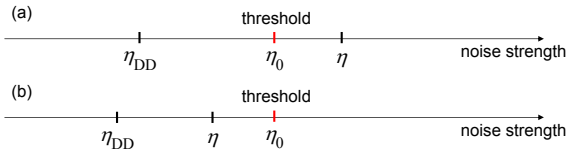}
\end{center}
\caption{\label{fig:th} Two scenarios where DD-protected gates
  outperform unprotected gates. (a) Quantum computing is scalable with DD-protected gates, but not with unprotected gates. (b) Quantum computing is scalable with either DD-protected gates or with unprotected gates, but DD reduces the overhead cost of fault tolerance.}
\end{figure}

A quantum computation unprotected by DD is scalable if the noise
strength of unprotected gates is below the accuracy threshold, $\eta <
\eta_0$. For DD-protected gates, the accuracy threshold condition
becomes $\etaDD < \eta_0$. If the noise suppression threshold
condition is satisfied, so that $\etaDD < \eta$, it may be that $\eta
> \eta_0$ and $\etaDD < \eta_0$; in that case, arbitrarily large
quantum circuits can be simulated accurately with DD-protected gates,
but not with unprotected gates.
This is illustrated in Fig.~\ref{fig:th}(a).

Even when $\eta < \eta_0$, DD may reduce the overhead cost of
fault-tolerant quantum computing if $\etaDD < \eta$
[Fig.~\ref{fig:th}(b)].
Suppose that we wish to simulate an ideal quantum circuit containing
$L$ gates. If our noisy gates have noise strength $\bar\eta$, which is
below the threshold value $\eta_0$, the simulation is possible using
$L^*$ noisy gates where \cite{AGP}
\begin{eqnarray}
\frac{L^*}{L} = \left(\frac{\log (c\eta_0 L/\theta)}{\log (\eta_0/\bar\eta)}\right)^a;
\end{eqnarray}
here $c$ and $a$ are constants, and $\theta$ is the ``error'' in the
simulation (the $L^1$ distance between the ideal probability
distribution of outcomes and the simulated distribution). Denote by
$L^*_{\rm un}$ the number of pulses in the fault-tolerant circuit
built from unprotected gates, and by $L^*_{\rm DD}$ the number of
pulses in the fault-tolerant circuit built from DD-protected gates,
and suppose that each DD-protected gate uses
$N$ pulses, while each unprotected gate uses a single pulse. Then the ratio
\begin{eqnarray}\label{eq:NumGates}
\frac{L^*_{\rm DD}}{L^*_{\rm un}} = N \left(\frac{\log (\eta_0/\eta)}{\log (\eta_0/\etaDD)}\right)^a
\end{eqnarray}
is independent of the size $L$ of the simulated circuit. If using DD
substantially improves the effective noise strength, ${L^*_{\rm
    DD}}/{L^*_{\rm un}}$ may be small, especially if $\eta$ is only
slightly below the threshold value $\eta_0$. Even though each
DD-protected gates requires multiple pulses, the total number of
pulses used in the simulation may be reduced, because DD improves the
gate accuracy.

Of course, we have reached this conclusion using the local-bath assumption, which allows us to assign a well-defined effective noise strength to the DD-protected gate. Furthermore our results are useful only if the Hamiltonian of the local bath has finite norm (so that $\epsilon < \infty$). However, we will see that the correlation function analysis in Sec.~\ref{sec:correlator} can provide useful upper bounds on $\etaDD$ even if $\epsilon$ is infinite.

\subsection{Slow preparations and measurements}\label{subsec:slowMeas}

Another drawback of this analysis is that our model of qubit preparations and measurements may be unrealistic in some physical situations. In our estimates of the effective noise strength in a DD-protected quantum computation, we have treated preparations and measurements like gates. We have assumed that each preparation and measurement location in the circuit, like each gate location, has duration $t_0$. A DD-protected preparation location consists of a preparation taking time $\delta$ followed by a DD memory sequence, and a DD-protected measurement location consists of a DD memory sequence followed by a measurement taking time $\delta$. Thus we have assumed that the preparations and measurements are just as fast as the pulses. In some physical systems, however, preparations and measurements are relatively slow; in solid-state devices, for example, the measurement time can be orders of magnitude longer than the gate time. 

If the actual time $\DeltaM$ required for a preparation or measurement is longer than the pulse width $\delta$ but still short compared to the pulse interval $\tau_0$, then we could still try to improve measurements and preparations using DD sequences. If it makes sense to model the noise during a preparation or measurement as we have modeled the noise in the pulses, then we could modify our analysis by using the measurement width $\DeltaM$ in estimating the effective noise strength $\etaDD$ at preparation and measurement locations, while continuing to use the pulse width $\delta$ in estimating $\etaDD$ at gate locations. But if $\DeltaM\gtrsim \tau_0$, or more generally if the noise in preparations and measurements is modeled much differently than the noise in gates, then it may be more appropriate to consider the preparation/measurement noise strength to be a separate parameter in the analysis, not necessarily related to the parameters $J$ and $\epsilon$ that characterize the noise Hamiltonian described in Sec.~\ref{sec:DDFTNoise} and appear in the estimate of $\etaDD$ at gate locations. 

Measurement locations might be much noisier than gate locations because gates can be improved using sequences of fast DD pulses, while slow measurements cannot be improved by DD. Or measurements might be noisier than gates for other quite different reasons. Previous work has shown that scalable fault-tolerant quantum computing is still possible, and that the accuracy threshold is not much affected, when measurements are much \emph{slower} than gates \cite{DiVincenzo}. What deserves further study, though, is how fault-tolerant circuit design can be optimized when measurements are much \emph{noisier} than gates. 

\section{Examples}\label{sec:DDFTExample}

Now we will analyze the effectiveness of several different DD pulse sequences, applying the results from Sec.~\ref{subsec:BoundsMagnus}. We adopt a noise model that includes only single-qubit errors acting on the system; thus the noise Hamiltonian is 
\begin{equation}
H=H_B+\sum_{i,\alpha} \sigma^{(i)}_\alpha \otimes B^{(i)}_\alpha,
\end{equation}
where $i$ labels the qubits, $\sigma_\alpha^{(i)}$ for $\alpha=x,y,z$ are the Pauli operators acting on qubit $i$, and 
\begin{equation}
H_B=\Id_S\otimes B_0.
\end{equation} 
In some realistic situations,
such as electron-spin qubits interacting
with a nuclear spin bath \cite{Petta05,Morton06,Morton08,Lee08},
such single-qubit errors are the dominant noise in the system.

In principle, $H_\err$ could also contain errors that act collectively on several qubits at once; for example, errors acting jointly on two qubits might be expected to occur during the execution of a two-qubit gate. Efficient DD pulse sequences can be constructed that suppress multi-qubit errors \cite{Wocjan,RW06}, but in this Section we limit our attention to single-qubit noise and pulse sequences that combat it. The more general results in Sec.~\ref{subsec:BoundsMagnus} can also be applied to other models that include multi-qubit noise and to the corresponding pulse sequences that achieve first-order decoupling for such noise.

We will discuss three different DD pulse sequences that can suppress the single-qubit noise. The first is the simplest DD scheme that protects against arbitrary single-qubit errors. The second is a time-symmetric sequence that achieves second-order decoupling in the limit of zero-width pulses. The third is the Eulerian DD scheme \cite{Viola03}, which is more robust against pulse errors than the other schemes.

\subsection{Universal decoupling sequence}\label{subsec:univ}

The shortest pulse sequence that suppresses arbitrary single-qubit
errors is called the  ``universal decoupling sequence''
\cite{Viola99,KL07},
or ``XY-4'' in the NMR literature \cite{Gullion:1990:479}.
For this sequence, the unitary operator generated by the control Hamiltonian, acting on a single qubit, can be expressed as
\begin{equation}\label{eq:univ}
U_c(\tDD)=Z\Id X\Id Z\Id X\Id.
\end{equation}
The notation in Eq.~(\ref{eq:univ}) is meant to convey that one complete cycle of the memory sequence contains four equally spaced pulses (each of width $\delta$) that successively apply the Pauli operators $X$, $Z$, $X$, $Z$, where $X=\sigma_x$ and $Z=\sigma_z$; therefore the product of the four Pauli operators is $-\Id$. Each $\Id$ in Eq.~(\ref{eq:univ}) represents trivial evolution during the pulse interval of width $\tau_0-\delta$. The total duration of the pulse sequence is $\tDD=4\tau_0$.

This sequence achieves first-order decoupling. In the limit of zero-width pulses, the toggling frame Hamiltonian is
\begin{widetext}
\begin{align}
\tilde H(t)&=U_c^\dagger(t) H U_c(t)=\left\{\begin{array}{lll}
~\Id H\Id~~=H_B+X \otimes B_X +Y\otimes B_Y+Z\otimes B_Z&~~\text{for }t\in[0,\tau_0),\\
XHX=H_B+X\otimes B_X-Y\otimes B_Y-Z\otimes B_Z&~~\text{for }t\in[\tau_0,2\tau_0),\\
YHY=H_B-X\otimes B_X+Y\otimes B_Y-Z\otimes B_Z&~~\text{for }t\in[2\tau_0,3\tau_0),\\
ZHZ=H_B-X\otimes B_X-Y\otimes B_Y+Z\otimes B_Z&~~\text{for }t\in[3\tau_0,4\tau_0),
\end{array}\right.
\end{align}
\end{widetext}
and we find
\begin{align}
\Omega_1(\tDD)&=-i\int_0^{\tDD} dt \tilde H(t)\nonumber\\
&=-i\tau_0\left(\Id H\Id+XHX+YHY+ZHZ\right) \nonumber\\
\label{eq:group-ave}&=-i\tDD H_B,
\end{align}
a pure-bath term. The first-order Magnus term (up to the factor $-i\tDD$) is the Pauli-group average of the noise Hamiltonian $H$, which commutes with any nontrivial Pauli operator acting on the system qubit.

In a DD-protected gate, the final pulse in the universal decoupling sequence is modified by combining with the gate pulse. For a single-qubit gate, the pulse sequence realizing the gate $G$ is
\begin{equation}
U_c(t_0)=(G Z)\Id X\Id Z\Id X\Id,
\end{equation}
where now $GZ$ represents a single pulse with duration $\delta$ and $t_0=\tDD$. In a two-qubit gate, the universal pulse sequence is applied in parallel to both qubits, except that the final pulse $Z\otimes Z$ in the memory sequence is replaced by the two-qubit pulse $G(Z\otimes Z)$.

To estimate the effective noise strength $\etaDD$, we note that the
total number of pulses is $N=4$, and that $\tau_0/t_0=1/4$. From the bounds in Eq.~\eqref{eq:Gen} and Table
\ref{tab:Coeffs} (for the case where the sequence is not time
symmetric) we obtain
\begin{align}
\etaDD&=(4J\tau_0)\left[\frac{\delta}{\tau_0}+\frac{1}{2}(4\epsilon \tau_0)+\frac{2}{9}(4\epsilon \tau_0)^2\right.\notag\\
\label{eq:eU}&\hspace{2.1cm}\left.+\frac{11}{9}(4\epsilon\tau_0)^3+9.43(4\epsilon\tau_0)^4\right],
\end{align}
where we have used $C_1=N\delta/t_0$ because the pulses are regularly spaced in time. Note that the parameters $J$ and $\epsilon$ include sums over all qubits in the set $\cQ_a$ that participate in the gate at location $a$ in the circuit. 

\begin{figure}[!ht]
\begin{center}
\includegraphics[width=0.48\textwidth]{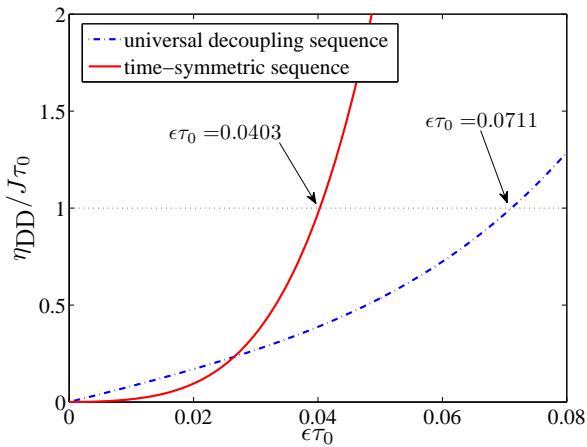}
\end{center}
\caption{\label{fig:plotDDFT} (color online) Plot of $\etaDD/\eta$ versus $\epsilon\tau_0$ for the universal decoupling sequence (Eq.~\ref{eq:eU}) and for the time-symmetric sequence (Eq.~\ref{eq:eU-sym}), assuming zero-width pulses. The noise strength for the DD-protected gate is weaker than the noise strength for the unprotected gate for $\epsilon\tau_0<0.0711$ in the case of the universal decoupling sequence, and for $\epsilon\tau_0<0.0403$ in the case of the time-symmetric sequence. For $\epsilon\tau_0$ sufficiently small, using the time-symmetric sequence reduces the noise strength further than the universal decoupling sequence.}
\end{figure}

In Fig. \ref{fig:plotDDFT}, $\etaDD/\eta$ (where $\eta=J\tau_0$) is plotted as a function of $\epsilon\tau_0$, in the limit $\delta/\tau_0\rightarrow 0$. The noise suppression threshold condition $\etaDD< \eta$ is satisfied when
\begin{equation}
\epsilon\tau_0<0.0711.
\end{equation}
In the limit $\epsilon\tau_0\rightarrow 0$, the noise suppression threshold condition is satisfied for
\begin{equation}\label{eq:1-4}
\frac{\delta}{\tau_0}<\frac{1}{4}.
\end{equation}

\subsection{Time-symmetric sequence}\label{subsec:timeSym}

We can construct a time-symmetric DD sequence by composing two copies of the universal decoupling sequence --- first we perform the sequence in the forward direction, and then run it backwards in time. For zero-width pulses, using the same notation as in Eq.~(\ref{eq:univ}), in which $\Id$ represents trivial evolution for time $\tau_0$ between successive pulses, this sequence can be expressed as
\begin{equation}\label{eq:TSseq}
U_c(\tDD)=\Id X\Id Z\Id X\Id \Id X\Id Z\Id X\Id,
\end{equation}
where we have combined the two $Z$ operators in the middle into a
zero-width identity ``pulse'' [not shown in Eq.~(\ref{eq:TSseq})]. The
total duration of the pulse sequence is $\tDD=8\tau_0$, twice as long
as the universal decoupling sequence. Like the universal decoupling
sequence, this sequence achieves first-order decoupling. In addition,
it satisfies the time-symmetry property $U_c(\tDD-t)=U_c(t)$, so that
the toggling-frame Hamiltonian obeys $\tilde H(\tDD-t)=\tilde H(t)$,
and thus this pulse sequence achieves second-order decoupling as well.
This pulse sequence is known in the NMR literature as ``XY-8'' \cite{Lizak91}.

For finite-width pulses, we modify our notation to emphasize that the second half of the sequence is the time reverse of the first half. We write 
\begin{equation}\label{eq:timesymseq}
U_c(\tDD)=\Id X^{(-)}\Id Z^{(-)}\Id X^{(-)}\Id\Id_{\delta} \Id X^{(+)}\Id Z^{(+)}\Id X^{(+)}\Id.
\end{equation}
Now, each $\Id$ represents trivial evolution for time $(\tau_0-\delta)$. The $\Id_{\delta}$ operator in the middle represents trivial evolution for time $\delta$, arising from combining two $Z$ pulses. It might seem more natural to use $\Id_{2\delta}$ instead, matching the total duration of two $Z$ pulses each with width $\delta$, but we choose the sequence Eq.~\eqref{eq:timesymseq} so that our upper bound on $\Omega_3(T)$, the dominant Magnus term when $\delta/\tau_0$ is negligible, will have a simple form. Since $\delta/\tDD$ is small anyway, it does not matter much which of these sequences we choose. $X^{(\pm)}$ and $Z^{(\pm)}$ represent finite-width pulses implementing $X$ and $Z$. Before the midpoint of the sequence at $t=\tDD/2$, the $X$ pulses are executed using the constant Hamiltonian $H_{P_X}$ such that $X=\exp(-i\delta H_{P_X})$ and the $Z$ pulse is executed using $H_{P_Z}$ such that $Z=\exp(-i\delta H_{P_Z})$, assuming the pulses are perfectly rectangular. After the midpoint, the universal decoupling sequence runs backwards; $X$ is executed using $-H_{P_X}$ and $Z$ using $-H_{P_Z}$. Thus, $U_c(\tDD-t)=U_c(t)$. 

Appending the gate pulse to this memory sequence breaks the time symmetry, which can be nearly restored using the trick explained in Sec.~\ref{sec:ens-sym}. The region $\Delta$ in which the time symmetry is violated is the union of two intervals: the duration of the gate pulse, and its image under time reversal, during which evolution is governed by the Hamiltonian $-H_B$. Thus $\Delta=2\delta$ (recall that we use $\Delta$ to denote both the region and its size). The DD-protected gate contains $N=8$ pulses (seven pulses in the memory sequence, including the identity pulse in the middle, plus the gate pulse) and has duration $t_0=8\tau_0$, so that $\tau_0/t_0=1/8$ and $T=t_0+\delta$. From the bounds in Eq.~\eqref{eq:Gen} and Table \ref{tab:Coeffs} (for the case where the sequence is nearly time-symmetric) we obtain an estimate of the effective noise strength $\etaDD$ of the DD-protected gates; we may use $C_1=N\delta/T\le\delta/\tau_0$ because the pulses are regularly spaced in time. 

In the limit of zero-width pulses ($\delta/\tau_0\rightarrow 0$), the effective noise strength becomes
\begin{equation}\label{eq:eU-sym}
\etaDD=(8J\tau_0)\left[\frac{2}{9}(8\epsilon\tau_0)^2+9.43(8\epsilon\tau_0)^4\right]; 
\end{equation}
$\etaDD/\eta$ is plotted in Fig. \ref{fig:plotDDFT}. The noise suppression threshold condition $\etaDD< \eta$ is satisfied when
\begin{equation}
\epsilon\tau_0<0.0403
\end{equation}
This condition is more stringent than for the universal decoupling sequence, which is not surprising since the time-symmetric sequence is twice as long. As Fig. \ref{fig:plotDDFT} illustrates, the time-symmetric sequence becomes more advantageous when $\epsilon\tau_0$ is small, as is likely to be the case when $\etaDD$ is below the accuracy threshold $\eta_0$. In the limit $\epsilon\tau_0\rightarrow 0$, only $C_1$ survives, and we find $\etaDD \le 8\eta(\delta/\tau_0)$; thus the noise suppression threshold condition is satisfied for
\begin{equation}
\frac{\delta}{\tau_0}<\frac{1}{8}.
\end{equation}
The largest permissible pulse width is half as large as in the case of the universal decoupling sequence [Eq.~(\ref{eq:1-4})] because the time-symmetric sequence is twice as long.

\begin{figure}[!htbp]
\begin{center}
\includegraphics[width=0.48\textwidth]{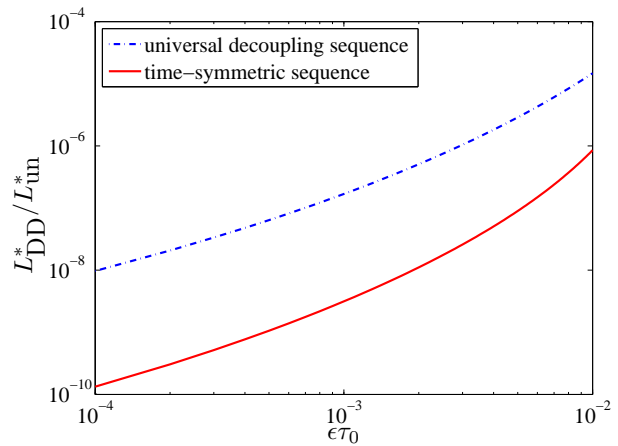}
\end{center}
\caption{\label{fig:NumGates} (color online) Plot of $L^*_\text{DD}/L^*_\text{un}$ (Eq.~\eqref{eq:NumGates}) versus $\epsilon\tau_0$ for the universal decoupling sequence and the time-symmetric sequence, in the limit $\delta/\tau_0\rightarrow 0$. We have assumed $\eta_0/\eta=2$, and have used the value $a=\log_2(291)\approx 8.18$ appropriate for the fault-tolerant gadget constructions in \cite{AGP}.} 
\end{figure}

Using Eq.~\eqref{eq:NumGates} and the expressions for $\etaDD$ in Eq.~(\ref{eq:eU}) (with $\delta/\tau_0=0$) and Eq.~(\ref{eq:eU-sym}), we plot in Fig. \ref{fig:NumGates} the ratio $L^*_\text{DD}/L^*_\text{un}$ versus $\epsilon\tau_0$ for both the universal decoupling sequence and the time-symmetric sequence. Here, just to illustrate the idea that DD can drastically reduce the overhead requirements for fault-tolerant quantum computing, we have assumed $\eta_0/\eta=2$, and we have taken the value $a=\log_2(291)\approx 8.18$ from \cite{AGP} ($291$ is the number of locations, including measurements and preparations, contained in the fault-tolerant {\sc cnot} gadget constructed in \cite{AGP}). Because the noise strength for the unprotected gate is close to the threshold value, and because $\epsilon\tau_0$ is well below the noise suppression threshold for each DD sequence in the range plotted, the reduction in the number of pulses achieved by using DD-protected gates is substantial. Furthermore, although the time-symmetric sequence is longer than the universal decoupling sequence, the time-symmetric sequence reduces the total number of pulses more effectively than the universal sequence, by more than 
\red{an order }of magnitude for $\epsilon\tau_0< 10^{-2}$.

In brief, the overhead improvement achieved by DD, illustrated by Fig. \ref{fig:NumGates}, arises as follows. The accuracy threshold analysis and overhead estimate in \cite{AGP} is based on concatenated coding, a hierarchy of codes within codes. The number of coding levels $k$ needed to simulate accurately a circuit of fixed size varies with the effective noise strength $\bar\eta$ according to
\begin{equation}
2^{k} \propto \frac{1}{\log\left( \eta_0 /\bar\eta\right)},
\end{equation}
and the number of noisy gates used in the fault-tolerant simulation grows like $2^{ak}$. By improving the effective noise strength, DD reduces the number of levels needed, substantially reducing the overhead cost. This savings in the number of gates more than compensates for the additional pulses used to achieve the DD improvement of each gate.

\begin{figure}[!htbp]
\begin{center}
\includegraphics[width=0.48\textwidth]{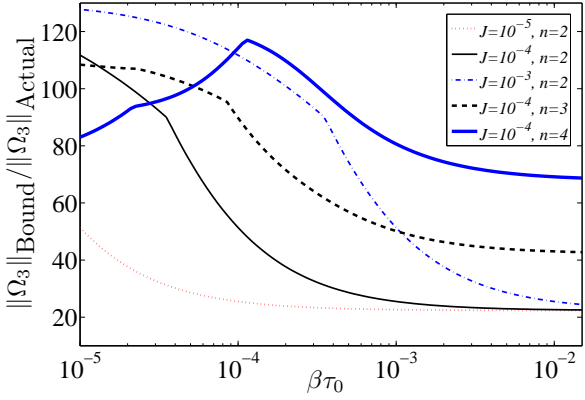}
\end{center}
\caption[Plot of $\Vert\Omega_3\Vert_\text{Bound}/\Vert\Omega_3\Vert_\text{Actual}$ versus $\beta\tau_0$ for various values of $J\tau_0$ and bath sizes.]{\label{fig:O3} (color online)  Plot of $\Vert\Omega_3\Vert_\text{Bound}/\Vert\Omega_3\Vert_\text{Actual}$ versus $\beta\tau_0$ for the time-symmetric DD sequence Eq.~\eqref{eq:TSseq}. The noise Hamiltonian is $H=H_B+H_{SB}$ ($H_S^0=0$), where $H_B=(\beta/2)\sum_i\sigma^z_i$ and $H_{SB}=(J/4)\sum_{\alpha=x,y,z}\sigma^\alpha_S\otimes \left(\sum_{i=1}^n\sigma^\alpha_i\right)$; the index $i$ labels the bath spins. Here $\Vert\Omega_3\Vert_\text{Bound}$ is computed using Eq.~(\ref{eq:Gen}) and  Table \ref{tab:Coeffs} (where $\delta=0$ and $T=8\tau_0$), while $\Vert\Omega_3\Vert_\text{Actual}$ is computed by evaluating the integral in Eq.~(\ref{eq:O3}) exactly. The kinks in the plots arise because the operator norm can have a discontinuous first derivative when eigenvalues cross.}
\end{figure}

For some noise models, the value of $\etaDD$ derived by our general arguments may be overly pessimistic. For example, using the time-symmetric sequence Eq.~\eqref{eq:TSseq}, we computed $\Omega_3(T)$ for a single-qubit system coupled to an $n$-spin bath in an external magnetic field, assuming an isotropic Heisenberg interaction between the system qubit and each bath spin. The ratio of the bound from Eq.~(\ref{eq:Gen}) and Table \ref{tab:Coeffs} to the actual value of $\Vert\Omega_3(T)\Vert$ for this model is plotted in Fig. \ref{fig:O3} as a function of $\beta\tau_0$, for $H_S^0=0$ and $\delta=0$.  The bound is larger than the actual value by at least a factor of 20.

\subsection{Eulerian decoupling sequences}
\label{subsec:eulerian}

The effects of finite pulse width and other pulse imperfections can be
suppressed by using an ``Eulerian'' memory sequence
\cite{Viola03}. In Eulerian decoupling, the operator applied by each
pulse is the generator of a finite group, and $U_c(t)$ traverses an
Euler cycle in the Cayley graph of this group. As a result, the
error Hamiltonian is group averaged and first-order decoupling is maintained even when the
pulses have (reproducible) imperfections. We will describe a simple
Eulerian memory sequence here; see \cite{Viola03} for a more general
discussion.

A simple Eulerian memory sequence protecting against single-qubit noise is \cite{Viola03}
\begin{equation}\label{eq:euler-sequence}
U_c(\tDD)=X\Id Z\Id X\Id Z\Id Z\Id X\Id Z\Id X\Id.
\end{equation}
Here the pulses are equally spaced in time; each $\Id$ 
operator
represents the same time interval, and the spacing between the start of two consecutive pulses is $\tau_0$. This sequence looks like two repetitions of the universal decoupling sequence, except that the $X$ and $Z$ pulses are swapped in the second repetition. In contrast to the time-symmetric sequence Eq.~(\ref{eq:timesymseq}), we use the same Hamiltonian $H_{P_X}$ to execute each $X$ pulse, rather than reversing the sign of the Hamiltonian during the second half of the sequence; similarly we use the same Hamiltonian $H_{P_Z}$ to execute each $Z$ pulse.

Without making any assumption about the pulse widths or shapes (except for assuming that all $X$ pulses are alike and that all $Z$ pulses are alike), we may express the unitary evolution operator over the pulse interval of duration $\tau_0$ as $u_X(t)$ for an $X$ pulse and $u_Z(t)$ for a $Z$ pulse.  Then, for $t\in [0,\tDD\equiv 8\tau_0]$, $U_c(t)$ becomes
\begin{align}
U_c(t)=\left\{\begin{array}{ll}
u_X(t)\Id&t\in[0,\tau_0)\\
u_Z(t-\tau_0)X&t\in[\tau_0,2\tau_0)\\
u_X(t-2\tau_0)(iY)&t\in[2\tau_0,3\tau_0)\\
u_Z(t-3\tau_0)(-Z)&t\in[3\tau_0,4\tau_0)\\
u_Z(t-4\tau_0)(-\Id)&t\in[4\tau_0,5\tau_0)\\
u_X(t-5\tau_0)(-Z)&t\in[5\tau_0,6\tau_0)\\
u_Z(t-6\tau_0)(iY)&t\in[6\tau_0,7\tau_0)\\
u_X(t-7\tau_0)X&t\in[7\tau_0,8\tau_0)
\end{array}\right..
\end{align}
The first-order Magnus term $\Omega_1(\tDD)$ can be expressed in terms of effective Hamiltonians $H_X$ and $H_Z$, obtained by averaging the Hamiltonian over an $X$ or $Z$ pulse respectively: 
\begin{eqnarray}
\tau_0 H_X \equiv \int_0^{\tau_0} dt ~u^\dagger_X(t) H u_X(t),\nonumber\\
\tau_0 H_Z \equiv\int_0^{\tau_0} dt~ u^\dagger_Z(t) H u_Z(t).
\end{eqnarray} 
Since $u_{X}$ and $u_Z$ act only on the system, they commute with the bath Hamiltonian $H_B$; while averaging over the pulse alters $H_{\rm err}$, it has no effect on $H_B$. Therefore we find that 
\begin{align}\label{Omega_1-Euler}
\quad~\Omega_1(\tDD)&=\int_0^{\tDD} dt~\tilde H(t)=\int_0^{\tDD} dt~ U_c^\dagger(t) HU_c(t)\nonumber\\
&= \tau_0\left(H_X+XH_XX+YH_XY+ZH_XZ\right) \nonumber\\
&+\tau_0\left(H_Z+XH_ZX+YH_ZY+ZH_ZZ\right)\nonumber\\
&=8H_B\tau_0;
\end{align}
thus $\Omega_1(\tDD)$ is a pure bath term. To derive the last line of Eq.~(\ref{Omega_1-Euler}), we have used the property $H+XHX+YHY+ZHZ=4H_B$ [as in Eq.~(\ref{eq:group-ave})]. We conclude that first-order decoupling is perfectly attained irrespective of the pulse shape, as long as the \emph{same} $u_{X(Z)}(t)$ is applied for every $X(Z)$ pulse, and the \emph{integrated} pulses are exactly right. 

\begin{figure}[!htbp]
\centering
\includegraphics[width=0.52\textwidth]{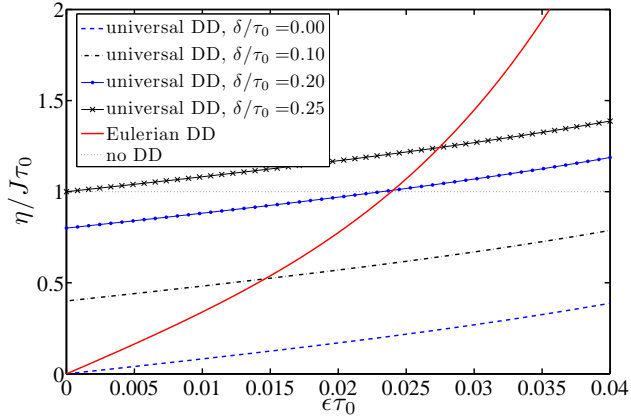}
\caption{\label{fig:EDD1} (color online) Comparison of effective noise strengths \newD{$\eta_{\rm DD}$ and $\eta_{\rm EDD}$} for the universal decoupling sequence \newD{given in Eq.~\eqref{eq:univ}} (for different pulse-widths) and the Eulerian decoupling sequence \newD{given in Eq.~\eqref{eq:euler-sequence}, respectively. The universal decoupling sequence is always worse than no decoupling ($\eta=J\tau_0$) for $\delta/\tau_0 \geq 1/4$, and Eulerian decoupling is worse than no decoupling for $\epsilon\tau_0 \geq 0.0239$. The Eulerian sequence is always better than universal DD for $\delta/\tau_0 \geq 0.1983$. For smaller values of $\delta/\tau_0$,} as the pulse-width increases, the Eulerian 
sequence outperforms the universal 
sequence for small values of $\epsilon\tau_0$. However, because of its longer length, the Eulerian sequence 
\newD{offers no} advantage over the universal sequence or no decoupling when $\epsilon\tau_0$ 
\newD{is} too large.} 
\end{figure}

\begin{figure}[!htbp]
\centering
\includegraphics[width=0.52\textwidth]{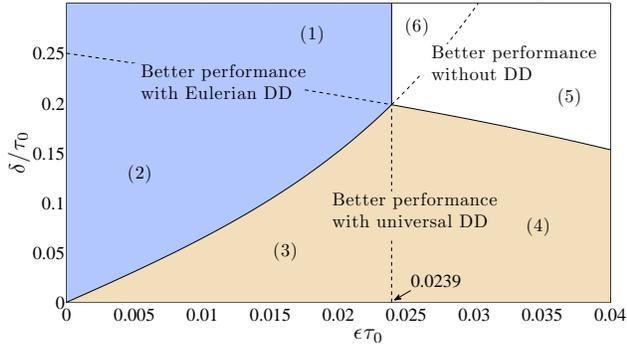}
\caption{\label{fig:EDD2} (color online) Illustration of the parameter regions in which no dynamical decoupling, the universal decoupling sequence 
\newD{(DD), Eq.~\eqref{eq:univ},} or the Eulerian decoupling sequence (EDD)\newD{, Eq.~\eqref{eq:euler-sequence},} emerges as the best strategy. Different regions 
\newD{indicated}
correspond to the following inequalities: 
(1) $\eta_\textrm{EDD}<\eta_\textrm{noDD}<\eta_\textrm{DD}$; (2) $\eta_\textrm{EDD}<\eta_\textrm{DD}< \eta_\textrm{noDD}$; (3) $\eta_\textrm{DD}<\eta_\textrm{EDD}<\eta_\textrm{noDD}$; (4) $\eta_\textrm{DD}<\eta_\textrm{noDD}<\eta_\textrm{EDD}$; (5) $\eta_\textrm{noDD}<\eta_\textrm{DD}<\eta_\textrm{EDD}$; (6) $\eta_\textrm{noDD}<\eta_\textrm{EDD}<\eta_\textrm{DD}$. \newD{The noise strengths are given by $\eta_\textrm{noDD} = J\tau_0$ and Eqs.~\eqref{eq:eU}, \eqref{eq:etaEDD}}.} 
\end{figure}

\new{
To demonstrate the advantage of using an Eulerian memory sequence, let us compare it with the universal decoupling sequence, taking into account finite pulse-width effects. The effective noise strength of the universal decoupling sequence is given in Eq.~\eqref{eq:eU}. For the Eulerian decoupling sequence described in Eq.~\eqref{eq:euler-sequence}, the effective noise strength is given by a similar expression, but with $4\tau_0$ replaced by $8\tau_0$ to account for the longer Eulerian sequence ($N=8$). Furthermore, 
\newD{in this case} we can drop the first-order\newD{, pulse-width dependent} term 
$\delta/\tau_0$, which gives
\begin{align}
\eta_\text{EDD}&=(8J\tau_0)\left[\frac{1}{2}(8\epsilon \tau_0)+\frac{2}{9}(8\epsilon \tau_0)^2\right.\notag\\
\label{eq:etaEDD}&\hspace{2.1cm}\left.+\frac{11}{9}(8\epsilon\tau_0)^3+9.43(8\epsilon\tau_0)^4\right].
\end{align}
The comparison between the universal decoupling sequence and the Eulerian decoupling sequence is illustrated in Figs. \ref{fig:EDD1} and \ref{fig:EDD2},} \newD{with numerical values delineating different regions easily deduced by solving the corresponding inequalities comparing Eqs.~\eqref{eq:eU} and \eqref{eq:etaEDD}.}

Adding a gate pulse $G$, by combining $G$ with the final $X$ pulse of the Eulerian memory sequence, introduces an error depending on the width of the final pulse. However, because this nonvanishing contribution to $\Omega_1(T)$ arises only from the final pulse, it does not depend on the length of the memory sequence. Other contributions to $\Omega(T)$ that depend on pulse shapes, in the second order of the Magnus expansion and beyond, are suppressed by additional factors of $\epsilon\tau_0$.

The contributions that depend on the pulse shape can be further suppressed by making the Eulerian memory sequence time-symmetric. Consider, for example, the sequence
\begin{align}
&U_c(\tDD)\notag\\
&= X^{(-)}\Id Z^{(-)}\Id X^{(-)}\Id Z^{(-)}\Id Z^{(-)}\Id X^{(-)}\Id Z^{(-)}\Id X^{(-)}\Id\notag\\
\label{eq:eulerian-time-sym}&\times\Id X^{(+)}\Id Z^{(+)}\Id X^{(+)}\Id Z^{(+)}\Id Z^{(+)}\Id X^{(+)}\Id Z^{(+)}\Id X^{(+)},
\end{align}
where the control Hamiltonian is chosen such that $u_{X^{(-)}}(\tDD -t)= u_{X^{(+)}}(t)$ and $u_{Z^{(-)}}(\tDD -t)= u_{Z^{(+)}}(t)$. Because this sequence obeys the time symmetry condition $U_c(\tDD-t)= U_c(t)$, the even-order Magnus terms vanish. Furthermore, because Eq.~(\ref{eq:eulerian-time-sym}) is just two copies of the Eulerian sequence Eq.~(\ref{eq:euler-sequence}), the first running backward in time and the second running forward, the sequence achieves first-order decoupling for any pulse shape. Corrections depending on the pulse shape enter only in third order and beyond. Of course, making the Eulerian sequence time-symmetric (or making the time-symmetric sequence Eulerian) lengthens the pulse sequence and so increases the time $T$ appearing in the Magnus expansion. Whether using this longer sequence actually improves the noise suppression depends on the values of the parameters $\epsilon\tau_0$, $\delta\tau_0$ and $\tau_0/t_0$, but it could pay off if the pulse width is relatively large, \newD{as suggested by Figs. \ref{fig:EDD1} and \ref{fig:EDD2}}. Adding a gate pulse to the time-symmetric Eulerian memory sequence spoils the first-order decoupling and breaks the time symmetry, but the resulting contributions to $\Omega_1$ and $\Omega_2$ depend only on the width of the final pulse, not on the length of the pulse sequence. 

Eulerian DD-protected gates that achieve exact first-order decoupling for nonzero-width pulses can be devised using the dynamically corrected gates recently introduced in
\cite{KV09,KV09a}. This scheme is based on the idea that distinct gates can have related errors, so that the errors cancel for a suitably constructed pulse sequence. The errors in distinct gates can be similar if the gates are constructed from control unitaries that traverse similar time-dependent paths, differing only by rescaling or reversing the time along the path. Arbitrary-order decoupling for nonzero-width pulses can be achieved by concatenating dynamically corrected gates \cite{KLV09}.

\section{Derivations}\label{sec:DDFTBounds}

In this section, we derive the coefficients for the bounds on the Magnus expansion listed in Table \ref{tab:Coeffs}. The Magnus expansion is computed for the Hamiltonian $H_M(t)$ given in Eq.~\eqref{eq:HMGen}; at any time $t$, $H_M(t)=\pm H_B+H'(t)$, where $H'(t)$ is either 0 or $\tilde H_\err(t)$. The two terms in $H_M(t)$ are bounded as $\Vert H_B\Vert\leq \beta$ and $\Vert H'(t)\Vert\leq J$; thus $\Vert H_M(t) \Vert\leq\beta+J=\epsilon$. The Magnus terms can be computed from $H_M(t)$ using the following recursive formulas \cite{Klarsfeld}, derived in Appendix \ref{app:magnus}:
\begin{subequations}\label{eq:recurv}
\begin{align}
\label{eq:A} A(t)&=-iH_M(t);\\
\label{eq:O1} \Omega_1(T)&=\int_0^{T}dt A(t);\\
\label{eq:On} \Omega_n(T)&=\sum_{j=1}^{n-1}\frac{B_j}{j!}\int_0^{T} dt\snj(t),~ n\geq 2;\\
\label{eq:sn1} S_n^{(1)}(t)&=\left[\Omega_{n-1}(t),A(t)\right];\\
\label{eq:snj} \snj(t)&=\sum_{m=1}^{n-j}\left[\Omega_m(t),S_{n-m}^{(j-1)}(t)\right],2\leq j\leq n-1,
\end{align}
\end{subequations}
where $\{B_j\}$ are the Bernoulli numbers. Explicit formulas for $\Omega_2(T)$ and $\Omega_3(T)$ were given in Eqs.~\eqref{eq:O2} and \eqref{eq:O3}. 

\subsection{General case: Magnus expansion}\label{subsec:magnus-derive}
For the general ({\em i.e.}, not time-symmetric) case, Table \ref{tab:Coeffs} gives $C_1=N\delta/T$ for regularly spaced pulses or $2N\delta/T$ in general, $C_2=1/2$, $C_3=2/9$, $C_4=11/9$ and $C_5= 9.43$. Now we derive these coefficients.

\subsubsection{Bound for \texorpdfstring{$\Omega_1'$}{Omega' 1}}
\label{subsubsec:Omega1-prime}
We assume that first-order decoupling is attained, so that in the limit of zero-width pulses $\tilde H(t)$ for the memory sequence satisfies Eq.~\eqref{eq:O1Cond}: $\int_0^{\tDD}dt\tilde H_{\err,0}(t)=0$. Recall that the subscript ``0" on $\tilde H_\err$ means we are to take $\delta$ to zero in $\tilde H_\err(t)$ while holding $\tau_0$ fixed. If a zero-width gate pulse is appended to the memory sequence, then $\tilde H_{\err,0}(t)$ in the DD-protected gate differs from $\tilde H_{\err,0}(t)$ in the memory sequence only during the final instantaneous pulse, and therefore still integrates to zero. Hence the DD-protected gate as well as the memory sequence satisfies  $\Omega_1(T) = -iTH_B$ and $\Omega_1'(T)=0$.

When the pulses have finite width ($\delta>0$), $\Omega_1'$ picks up corrections that depend on $\delta$. Noting that $\tilde H(t)$ differs from $\tilde H_0(t)$ only during the pulses, we write
\begin{align}
\Omega_1'(T)&=-i\int_0^{t_0} dt \tilde H(t)+it_0H_B\nonumber\\
&=-i\int_0^{t_0} dt \tilde H_0(t)+it_0H_B\notag\\
&\quad+i\int_0^{t_0}dt_{\text{PW}} \tilde H_0(t)-i\int_0^{t_0}dt_{\text{PW}}\tilde H(t)\nonumber\\
\label{eq:C1eq1}&=i\int_0^{t_0}dt_{\text{PW}} \tilde H_0(t)-i\int_0^{t_0}dt_{\text{PW}}\tilde H(t).
\end{align}
Here, $dt_\text{PW}$ indicates integration only over times within the pulses. Now, $\tilde H_0(t)=H_B+\tilde H_{\err,0}(t)$, so for a sequence with $N$ pulses (including the gate pulse), we have $i\int_0^{t_0}dt_{\text{PW}}\tilde H_0(t)=iN\delta H_B+i\int_0^{t_0}dt_\text{PW}\tilde H_{\err,0}(t)$ and $-i\int_0^{t_0}dt_{\text{PW}}\tilde H(t)=-iN\delta H_B- i\int_0^{t_0}dt_\text{PW}\tilde H_\err(t)$. The two $iN\delta H_B$ terms cancel, and we are left with
\begin{equation}\label{eq:C1eq}
\Omega_1'(T)=i\int_0^{t_0}dt_\text{PW}\tilde H_{\err,0}(t)-i\int_0^{t_0}dt_\text{PW}\tilde H_\err(t).
\end{equation}
The second term can be upper bounded by $N\delta J$. For the first term, Eq.~\eqref{eq:GHtilde} tells us that for $\delta=0$, $\tilde H_0(t)=\tilde H^{(k)}=H_B+\tilde H_\err^{(k)}$ for $t\in[s_k,s_{k+1})$. Hence, we have that $i\int_0^{t_0}dt_\text{PW}\tilde H_{\err,0}(t)= i\delta\sum_k \tilde H_\err^{(k)}$. Now, the first-order decoupling condition can be written as
\begin{align}\label{eq:C1eq2}
\int_0^{t_0}dt\tilde H_{\err,0}&=\sum_k(s_{k+1}-s_k)\tilde H_\err^{(k)}=0.
\end{align}
If all the pulses are regularly spaced in time, so that $s_{k+1}-s_k$ are all equal for all $k$, this condition implies that $\sum_k \tilde H_\err^{(k)}=0$. In this case, the first term of the right-hand side of Eq.~\eqref{eq:C1eq} vanishes and
$\Omega_1'(T)$ is bounded by the norm of the second term only:
\begin{equation}
\left\Vert\Omega_1'(T)\right\Vert\leq N\delta J=\frac{N\delta}{T}(JT).
\end{equation}
Hence, $C_1=N\delta/T$ if pulses are regularly spaced in time. Even if the pulses are not regularly spaced in time, this value of $C_1$ works whenever $\sum_k \tilde H_\err^{(k)}=0$. Otherwise, we can still upper bound the first term in Eq.~\eqref{eq:C1eq} by $N\delta J$, so that $\Vert\Omega_1'(T)\Vert\leq 2N\delta J=(2N\delta/T)(JT)$. This gives $C_1=2N\delta/T$ in general.

\subsubsection{Bound for \texorpdfstring{$\Omega_2$}{Omega 2}}
We will derive an upper bound on 
\begin{align}
\Omega_2(T)&=-\frac{1}{2}\int_0^Tds_1\int_0^{s_1}ds_2[H_M(s_1),H_M(s_2)]\label{eq:O2bound}
\end{align}
where $H_M(t)=H_B + H'(t)$. The term quadratic in $H_B$ vanishes, because $H_B$ is time independent and $[H_B,H_B]=0$. The term of linear order in $H_B$ can be expressed as 
\begin{eqnarray} -\frac{1}{2}\int_0^T ds_1\int_0^{s_1}ds_2  [H_B,H'(s_2)-H'(s_1)].
\end{eqnarray}
We note that 
\begin{equation}
\int_0^T ds_1\int_0^{s_1}ds_2 = \int_0^T ds_2\int_{s_2}^{T}ds_1;
\end{equation}
either way we are integrating over the triangle with $s_2\le s_1\le T$. Therefore,
\begin{align}
&\int_0^T ds_1\int_0^{s_1}ds_2 ~[H_B,H'(s_1)] =\int_0^T ds~s[H_B,H'(s)],\notag\\
&\int_0^T ds_2\int_{s_2}^{T}ds_1 ~[H_B, H'(s_2)] \notag\\
&\hspace{2cm}= \int_0^T ds~(T-s)[H_B,H'(s)].
\end{align}
Combining terms we find
\begin{eqnarray}
&&\int_0^T ds_1\int_0^{s_1}ds_2~ [H_B, H'(s_1) - H'(s_2)]
\nonumber\\
&&= \int_0^T (2s - T)[H_B,H'(s)],
\end{eqnarray}
and hence
\begin{align}
&\left \|\frac{1}{2} \int_0^T ds_1\int_0^{s_1}ds_2 ~[H_B, H'(s_1) - H'(s_2)]\right \| \notag\\
&\hspace{1cm}\le J \beta \int_0^T ds \left| 2s-T \right| = \frac{1}{2} J \beta T^2.
\end{align}
This bound on the sum of two terms is better by a factor of two than we would have found by bounding the two terms separately, because of a partial cancellation between the two terms. 

We bound the term in $\Omega_2(T)$ of zeroth order in $H_B$ using
\begin{eqnarray}
\left\| [H'(s_1),H'(s_2)]\right\| \le 2J^2,
\end{eqnarray}
and therefore
\begin{align}
&\left\|\frac{1}{2}\int_0^Tds_1\int_0^{s_1}ds_2[H'(s_1),H'(s_2)]\right\| \notag\\
&\hspace{1cm}\le \frac{1}{2}(2J^2)(T^2/2) = \frac{1}{2}J^2T^2.
\end{align}
Combining with the terms linear order in $H_B$ we obtain
\begin{eqnarray}
\left\| \Omega_2(T)\right \| \le \frac{1}{2} J\beta T^2 + \frac{1}{2} J^2 T^2= \frac{1}{2}(\epsilon T)(JT),
\end{eqnarray}
where $\epsilon = \beta + J$; hence $C_2 = 1/2$.

\subsubsection{Bound for \texorpdfstring{$\Omega_3$}{Omega 3}}

The integrand in the expression Eq.~\eqref{eq:O3} for $\Omega_3(T)$ is 
\begin{align}
&\frac{i}{6}\big([H_M(s_1),[H_M(s_2),H_M(s_3)]] \notag\\
& \hspace{.2cm}
+ [H_M(s_3),[H_M(s_2),H_M(s_1)]]\big),
\end{align}
where $H_M(s) = H_B + H'(s)$; because $[H_B,H_B]=0$, the term cubic in $H_B$ vanishes, and the terms quadratic in $H_B$ can be written in the form 
\begin{eqnarray}
\frac{i}{6}[H_B,[H_B,H'(s_1) +H'(s_3)-2H'(s_2)]].
\end{eqnarray}
The time-ordered integration 
\begin{eqnarray}
\int_0^T ds_1\int_0^{s_1} ds_2\int_0^{s_2}ds_3
\end{eqnarray}
can be expressed as $\int_0^T ds_1 \left(s_1^2/2\right)$ for a function independent of $s_2,s_3$, as $\int_0^T ds_3\left(\left(T-s_3\right)^2/2\right)$ for a function independent of $s_1,s_2$, and as $\int_0^T ds_2 ~s_2\left(T-s_2\right)$ for a function independent of $s_1,s_3$. Therefore, the contribution to $\Omega_3(T)$ quadratic in $H_B$ is
\begin{align}\label{eq:omega-three-quadratic}
\left[\Omega_3(T)\right]_{\rm quadratic} &=\frac{i}{6}\int_0^T ds [H_B,[H_B,H'(s)]]\notag\\
&\times \left( \frac{1}{2} s^2 + \frac{1}{2} (T{-}s)^2 {-} 2 s(T{-}s)\right);
\end{align}
using
\begin{eqnarray}
\left\| [H_B,[H_B,H'(s]]\right\| \le 4\beta^2 J,
\end{eqnarray}
it can be bounded as
\begin{eqnarray}\label{eq:omega-three-quadratic-absolute}
&&\left\| \left[\Omega_3(T)\right]_{\rm quadratic} \right\|\nonumber\\
&\le& \frac{2}{3}\beta^2 J \int_0^T ds \left| \frac{1}{2} s^2 + \frac{1}{2} (T-s)^2 - 2 s(T-s)\right|\nonumber\\
&=& \frac{2}{3} \beta^2 J \frac{T^3}{3\sqrt{3}} = \frac{2}{9\sqrt{3}}(\beta T)^2(JT).
\end{eqnarray}
(The integrand has zeros at $s_\pm = \frac{1}{2} \pm \frac{1}{2\sqrt{3}}$; it is positive in $[0,s_-]$ and $[s_+,T]$, negative in $[s_-,s_+]$. The integrals over these three intervals are respectively $\frac{T^3}{12\sqrt{3}}$, $-\frac{T^3}{6\sqrt{3}}$, $\frac{T^3}{12\sqrt{3}}$, and the integral of the absolute value is $\frac{T^3}{3\sqrt{3}}$.)

Now consider the terms linear in $H_B$, with integrand
\begin{eqnarray}\label{eq:omega-three-terms}
\frac{i}{6}
\big( [B23] + [B21] + [1B3] + [3B1] + [12B] + [32B] \big)\nonumber\\
\end{eqnarray}
where 
\begin{eqnarray}
[B23]\equiv [H_B,[H'(s_2),H'(s_3)]],
\end{eqnarray} etc. We note that 
\begin{align}
&\int_{T\ge s_1\ge s_2\ge s_3\ge 0}ds_1ds_2ds_3 \big( [1B3] + [12B] \big) \notag\\
=& \int_{T\ge s_1\ge s_3\ge 0}ds_1ds_3 ~(s_1-s_3)[1B3] \notag\\
&- \int_{T\ge s_1\ge s_2\ge 0}ds_1ds_2 ~s_2[1B2]\notag\\
=& \int_{T\ge s_1\ge s_2\ge 0}ds_1ds_2 ~\left(s_1-2s_2\right)[1B2],
\end{align}
and hence
\begin{align}
&\left| \int_{T\ge s_1\ge s_2\ge s_3\ge 0}ds_1ds_2ds_3 \big( [1B3] + [12B] \big) \right|\notag\\
&\le \|[1B2]\|_{\rm max} 
 \int_{T\ge s_1\ge s_2\ge 0}ds_1ds_2 ~\left|s_1-2s_2\right|\notag\\
&= 4\beta J^2 \frac{T^3}{6} = \frac{2}{3} \beta J^2T^3.
\end{align}
Similarly,
\begin{align}
&\int_{T\ge s_1\ge s_2\ge s_3\ge 0}ds_1ds_2ds_3 \big( [3B1] + [32B] \big) \notag\\
=& \int_{T\ge s_1\ge s_3\ge 0}ds_1ds_3 ~(s_1-s_3)[3B1] \notag\\
&- \int_{T\ge s_2\ge s_3\ge 0}ds_2ds_3 ~(T-s_2)[3B2]\notag\\
=& \int_{T\ge s_1\ge s_3\ge 0}ds_1ds_3 ~\left(2s_1 - s_3 -T\right)[3B1],
\end{align}
and hence
\begin{align}
&\left| \int_{T\ge s_1\ge s_2\ge s_3\ge 0}ds_1ds_2ds_3 \big( [3B1] + [32B] \big) \right|\notag\\
&\le \|[3B1]\|_{\rm max} \notag\\
&\hspace{.1cm}\times\int_{T\ge T-s_3\ge T-s_1\ge 0}ds_1ds_3 ~\left|(T-s_3) - 2(T-s_1)\right|\notag\\
&= 4\beta J^2 \frac{T^3}{6} = \frac{2}{3} \beta J^2T^3.
\end{align}
Also, 
\begin{align}
&\int_{T\ge s_1\ge s_2\ge s_3\ge 0}ds_1ds_2ds_3 \big( [B23] + [B21] \big) \notag\\
= &\int_{T\ge s_2\ge s_3\ge 0}ds_2ds_3 ~(T-s_2)[B23] \notag\\
&- \int_{T\ge s_1\ge s_2\ge 0}ds_1ds_2 ~s_2[B12]\notag\\
=& \int_{T\ge s_2\ge s_3\ge 0}ds_2ds_3 ~\left(T-s_2-s_3\right)[B23],
\end{align}
and hence
\begin{align}
&\left| \int_{T\ge s_1\ge s_2\ge s_3\ge 0}ds_1ds_2ds_3 \big( [B23] + [B21] \big) \right|\notag\\
&\le \|[B23]\|_{\rm max} \int_{T\ge s_2\ge s_3\ge 0}ds_2ds_3 ~\left|T-s_2-s_3\right|\notag\\
&= 4\beta J^2 \frac{T^3}{6} = \frac{2}{3} \beta J^2T^3.
\end{align}
Combining these three bounds, we obtain an upper bound on the terms in $\Omega_3(T)$ linear in $H_B$:
\begin{align}
\left\| \left[\Omega_3(T)\right]_{\rm linear} \right\|\le \frac{1}{6}\times 3\times \left(\frac{2}{3} \beta J^2T^3\right)= \frac{1}{3}\beta J^2 T^3.
\end{align}

For the term in $\Omega_3(T)$ independent of $H_B$, we have
\begin{align}
&\left\| \left[\Omega_3(T)\right]_{\rm zeroth-order} \right\|\notag\\
&\le \frac{1}{6} \left(\frac{T^3}{6}\right) (2) \|~[H'(s_1),[H'(s_2),H'(s_3)]]~\|_{\rm max}\notag\\
&= \frac{2}{9} (JT)^3.
\end{align}
Putting together the bounds on the terms of second, first, and zeroth order in $H_B$, we find
\begin{align}
&\left\| \Omega_3(T) \right\| \le\frac{2}{9\sqrt{3}}(\beta T)^2(JT)+ \frac{1}{3} (\beta T)(JT)^2+\frac{2}{9} (JT)^3 \notag\\
&\hspace{,5cm}= \frac{2}{9\sqrt{3}}(\epsilon T)^2 (JT) +\frac{1}{3}\left(1- \frac{4}{3\sqrt{3}}\right)(\epsilon T)(JT)^2 \notag\\
&\hspace{1.5cm} + \frac{1}{9}\left(\frac{2}{\sqrt{3}}-1\right) (JT)^3.
\end{align}Using $J \le \epsilon$, we obtain a weaker but simpler bound:
\begin{eqnarray}
\left\| \Omega_3(T) \right\| \le  \frac{2}{9}(\epsilon T)^2 (JT).
\end{eqnarray}
Hence $\red{C_3} = 2/9$.

\subsubsection{Bounds for \texorpdfstring{$\Omega_{n\geq 4}$}{Omega n}}
To bound the Magnus terms for $n\geq 4$, we use the recursive formulas Eq.~\eqref{eq:recurv} and ideas from \cite{Blanes98, Moan}. In Appendix \ref{app:snj}, we show that the $\snj$ operators satisfy:
\begin{equation}
\Vert \snj(t)\Vert\leq \fnj J\left(2\epsilon t\right)^{n-1},
\end{equation}
for all $n\geq 2$, $1\leq j\leq n-1$, where the coefficients $\fnj$ are given in Eq.~\eqref{eq:fnj}. Using this, we can write down bounds for $\Omega_{n\geq 4}$ as follows:
\begin{align}
\left\Vert\Omega_n(T)\right\Vert&\leq \sum_{j=1}^{n-1}\frac{\vert B_j\vert}{j!}\int_0^T ds\Vert\snj(s)\Vert\notag\\
&\leq\frac{1}{n}\sum_{j=1}^{n-1}\frac{\vert B_j\vert}{j!}f_n^{(j)}(JT)(2\epsilon T)^{n-1}\notag\\
\label{eq:Olarge}&= f_n(JT)(4\epsilon T)^{n-1},
\end{align}
where the coefficients $f_n$ are defined as
\begin{equation}\label{eq:fn}
f_n=\frac{1}{n2^{n-1}}\sum_{j=1}^{n-1}\frac{\vert B_j\vert}{j!}f_n^{(j)}.
\end{equation}

Using Eq.~\eqref{eq:fn} and the recursive formula  for $\fnj$ from Eq.~\eqref{eq:fnj}, one can show that $f_4=11/576$. Then, $\Omega_4(T)$ can be bounded as
\begin{equation}
\left\Vert\Omega_4(T)\right\Vert\leq \frac{11}{576} (JT)(4\epsilon T)^3,
\end{equation}
so $C_4= 4^3(11/576)=11/9$.

The bounds for $\Omega_n$ for $n\geq 5$ can be gathered together into a single bound by writing
\begin{equation}\label{eq:O5}
\sum_{n=5}^\infty\left\Vert\Omega_n(T)\right\Vert\leq (JT)(4\epsilon T)^4\left[\sum_{n=5}^\infty f_n(4\epsilon T)^{n-5}\right].
\end{equation}
In \cite{Moan}, the $\{f_n\}$ were shown to be coefficients in the power series expansion of $G^{-1}(y)=\sum_{n=1}^\infty f_n y^n$; $G^{-1}(y)$ is the inverse function of
\begin{equation}
y=G(s)=\int_0^s dx\left[2+\frac{x}{2}\left(1-\cot\frac{x}{2}\right)\right]^{-1},
\end{equation}
defined for domain $-2\pi\leq s\leq 2\pi$, the interval over which $G(s)$ is monotonically increasing. A self-contained proof of this fact is provided in Appendix \ref{app:fn}. We want to relate the expression in the brackets in Eq.~\eqref{eq:O5} to $G^{-1}$. Define $\zeta$ as
\begin{equation}
\zeta=G(2\pi)=2.17374\ldots,\quad G^{-1}(\zeta)=2\pi,
\end{equation}
and assume that $\epsilon T\leq 0.54$ so that $4\epsilon T\leq \zeta$. Then, $G^{-1}(4\epsilon T)\leq 2\pi$ since $G(s)$ is monotonically increasing over its domain, and therefore,
\begin{align}\label{eq:sum-from-5}
\left[\sum_{n=5}^\infty f_n(4\epsilon T)^{n-5}\right]&\leq \sum_{n=5}^\infty f_n\zeta^{n-5}\notag\\
&= \frac{1}{\zeta^5}\left[G^{-1}(\zeta)-\sum_{n=1}^4 f_n\zeta^n\right].
\end{align}
Using $f_1=1$, $f_2=\frac{1}{4}$, $f_3=\frac{5}{72}$ and $f_4=\frac{11}{576}$, which can be derived from Eq.~\eqref{eq:fn} and  Eq.~\eqref{eq:fnj}, Eq.~(\ref{eq:sum-from-5}) implies
\begin{equation}
\left[\sum_{n=5}^\infty f_n(4\epsilon T)^{n-5}\right]\leq 0.03685\ldots \equiv C'.
\end{equation}
Then,
\begin{equation}
\sum_{n=5}^\infty\left\Vert\Omega_n(T)\right\Vert\leq C'(JT)(4\epsilon T)^4.
\end{equation}
Therefore, $C_5\equiv 4^4\times C'\simeq 9.43$.

Note that the condition $\epsilon T\leq 0.54$ is more stringent than the sufficient condition for convergence of the Magnus expansion given in Eq.~\eqref{eq:conv}, which requires $\epsilon T<\pi$. If $0.54 < \epsilon T < \pi$, we need to use a different method to find an upper bound on the sum of the high-order Magnus terms.

\subsection{General case: Dyson expansion}
\label{subsec:gen-case}

In Sec.~\ref{subsec:magnus-derive} we used the Magnus expansion to obtain bounds on the noise strength of DD-improved quantum gates. Here we derive bounds on the noise strength by a different method based on time-ordered perturbation theory in the toggling frame. These new bounds are easier to derive than those in Sec.~\ref{subsec:magnus-derive}, and they apply without any upper bound imposed on the expansion parameter $\epsilon T$; furthermore, in the case of a pulse sequence that achieves third-order decoupling, they are actually tighter than the previous bounds. Unfortunately, in the case of first-order or second-order decoupling, they are not as tight. In this derivation, we will assume pulses have zero width, and we will consider only the general case (without time symmetry).

In the local-bath model, we consider the toggling-frame system-bath Hamiltonian 
\begin{eqnarray}
\lambda \tilde H(t) = \lambda \left(H_B + \tilde H_{\rm err}(t)\right),
\end{eqnarray}
which describes the noise at a particular circuit location. Here $\lambda H_B$ is the Hamiltonian of the local bath (acting trivially on the system) and $\lambda \tilde H_{\rm err}$ is the Hamiltonian responsible for the noise (acting jointly on system and bath). We have introduced the coupling parameter $\lambda$ here for convenience, to keep track of terms in the Dyson and Magnus expansions, and we will set $\lambda=1$ momentarily. 

Consider the toggling-frame time-evolution operator $\tilde U(T)$ obtained by integrating the Schr\"odinger equation with Hamiltonian $\lambda \tilde H(t)$ for time $T$ (if the control unitary $U_c(T)$ for this time interval is the identity --- i.e. if the control sequence is {\em cyclic} --- then the toggling-frame and Schr\"odinger picture evolution operators coincide). The Dyson expansion is the expansion of $\tilde U(T)$ in powers of $\lambda$:
\begin{eqnarray}\label{eq:lambda-expanding-U}
\tilde U(T) = \sum_{n=0}^\infty \frac{\lambda^n}{n!} \int_0^T dt_1 \cdots \int_0^T dt_n~ 
\blue{\mathcal{T}}
\left(\tilde H(t_1) \cdots \tilde H(t_n)\right),\nonumber\\
\end{eqnarray}
where 
\blue{$\mathcal{T}$}
denotes time ordering. The Magnus expansion is the expansion of the {\em logarithm} of $\tilde U(T)$ in powers of $\lambda$:
\begin{eqnarray}\label{eq:lambda-expanding-logarithm}
\tilde U(T) = \exp \left( \sum_{n=1}^\infty \lambda^n \Omega_n(T)\right). 
\end{eqnarray}
We say that the control sequence achieves $n$th order decoupling if the first $n$ terms in the Magnus expansion are pure-bath terms, acting trivially on the system. By expanding the exponential in Eq.~(\ref{eq:lambda-expanding-logarithm}) and comparing with Eq.~(\ref{eq:lambda-expanding-U}), we see that for a control sequence that achieves $n$th order decoupling, the terms of order $\lambda^m$ for $m\le n$ in the Dyson expansion are all pure bath terms [and that the $(n{+}1)$st-order term in the Dyson expansion is $\lambda^{n+1} \left(\Omega_{n+1}(T) + \cdots\right)$, where the ellipsis represents a pure bath term.]

In Sec.~\ref{sec:DDFTDDgates}, we defined the effective noise strength $\eta_{\rm DD}$ as an upper bound on the deviation of the noisy operation $\tilde U(T)$ from a pure-bath unitary operator $U_B(T)$:
\begin{eqnarray}
\eta_{\rm DD} = \max_{\blue{a}} \| \tilde U(T) - U_B(T) \|.
\end{eqnarray}
This definition was convenient because each order in the Magnus expansion is anti-Hermitian, so that in the case where $n$th-order decoupling is achieved, the exponential of the sum of the first $n$ terms in the Magnus expansion is a pure-bath unitary. However, when we express the noisy unitary as the sum of good and bad parts (where the good part acts trivially on the system), it is not necessary for the good part to be unitary --- the criterion for scalable quantum computing is $\eta \le \eta_0$ (where $\eta$ is the operator norm of the bad part) whether the action of the good part on the bath is unitary or not. Therefore, to estimate the noise strength, we can separate the terms in the Dyson expansion into pure bath terms (whose sum is not necessarily unitary) and remaining terms that may act nontrivially on the system. Then the noise strength $\eta$ is an upper bound on the operator norm of the sum of these remaining terms. 

The operator norm of the $n$th order term in the Dyson expansion Eq.~(\ref{eq:lambda-expanding-U}) can be bounded above by $\frac{1}{n!}(\epsilon T)^{n}$ (with $\lambda$ now set equal to 1). This is simply an upper bound on the norm of the integrand times the volume of the integration region. But we can also do a double expansion of the $n$th order term in $H_B$ and $\tilde H_{\rm err}$, bounding each term separately \cite{Lidar-Uhrig}. In this double expansion, the terms that are zeroth order in $\tilde H_{\rm err}$ are of course pure bath terms, and their sum has operator norm bounded above by $\frac{1}{n!}(\beta T)^{n}$.  Thus, the upper bound on the sum of all the order $n$ terms in the Dyson expansion that are not zeroth order in $J$ is
\begin{eqnarray}\label{eq:n-bound}
\frac{T^{n}}{n!}\left((\beta + J)^{n} - \beta^{n}\right)
\end{eqnarray}
To express this bound in terms of $\epsilon$, 
{\blue{we note that $f(\beta)=\beta^{n}$ is a convex function for $n\ge 1$, so that $f(\beta) \ge f(\beta+J) - J f'(J+\beta)$; thus,}
\begin{eqnarray}
(\beta + J)^{n} - \beta^{n} \le nJ (\beta+J)^{n-1} = nJ\epsilon^{n-1},
\end{eqnarray}
and the upper bound in 
Eq.~(\ref{eq:n-bound}) becomes
\begin{eqnarray}
\frac{1}{(n-1)!}(JT)(\epsilon T)^{n-1}.
\label{eq:Dyson-n}
\end{eqnarray}

Now consider a cyclic control sequence that achieves $n$th order decoupling, so that all terms up to $n$th order in the Dyson expansion are pure bath terms. We estimate the effective noise strength using an upper bound on the non-pure-bath parts of all higher order terms, finding
\begin{eqnarray}
\eta_{\rm DD} = \sum_{m=n+1}^\infty \frac{1}{(m-1)!}(JT)(\epsilon T)^{m-1}
\end{eqnarray}
Thus, by using the Dyson expansion rather than the Magnus expansion we have found%
\blue{, we read off $C_n = 1/(n-1)!$ for $n=2,3,4$ from Eq.~(\ref{eq:Dyson-n}), i.e.,}
\begin{align}
&C_2 = 1, \quad C_3 = 1/2, \quad C_4 = 1/6, 
\end{align}
and
\begin{align}
 C_5 &= \sum_{m=5}^\infty \frac{1}{(m-1)!}(\epsilon T)^{m-5}\notag\\
&= \frac{e^x -1 -x -x^2/\blue{2} - x^3/6}{x^4}\Big|_{x=\epsilon T}\notag\\
&=\frac{1}{24} + \frac{\epsilon T}{120} + \frac{(\epsilon T)^2}{720} +\cdots \approx \red{0}.0466,
\end{align}
where the numerical value of $C_5$ was obtained by setting $\epsilon T= \red{0}.54$
\blue{in order to have a meaningful comparison with the $C_5$ value we obtained from the Magnus expansion}. 
Thus, comparing with the bounds derived in Sec.~\ref{subsec:magnus-derive}, we have improved the values of $C_4$ and $C_5$ substantially, but not the values of $C_2$ and $C_3$. This means that for a (not time-symmetric) cyclic control sequence achieving third order decoupling, we get a smaller value for $\eta_{\rm DD}$ using the Dyson expansion rather than the Magnus expansion. 

\subsection{Time-symmetric case}\label{app:TimeSym}

Now we derive bounds on the Magnus terms that apply when the pulse sequence is time-symmetric except inside a small region $\Delta\subseteq [0,T]$; as before, we use $\Delta$ to denote both this region and its size. As in Eq.~(\ref{eq:HMGen}), we are interested in the Hamiltonian $H_M(t)$ describing evolution for time $\Gamma$ governed by the Hamiltonian $-H_B$, followed by evolution for time $T-\Gamma$ governed by the toggling-frame Hamiltonian of a DD-protected gate. But our analysis in this Section applies to any Hamiltonian $H_M(t)$ that is time-symmetric outside region $\Delta$.

Even Magnus terms vanish when $\Delta=0$, and we will derive explicit $\Delta$-dependent bounds on $\Omega_2(T)$ and $\Omega_4(T)$, which are linear in $\Delta$ when $\Delta$ is small. 
We could also exploit the time symmetry to derive improved bounds on the higher-order Magnus terms ($\Omega_{n\geq 5}(T)$); however we will not bother to do so. Instead we use the same upper bounds on these terms that apply in the general case, with the expectation that these bounds are already quite small in typical cases of interest. 

To obtain a bound on $\Omega_2(T)$ for a nearly time-symmetric sequence, we 
observe that the double time integral in Eq.~(\ref{eq:O2bound}) can be split into four cases: (i) $s_1,s_2\notin\Delta$, (ii) $s_1\in\Delta,s_2\notin\Delta$, (iii) $s_1\notin\Delta,s_2\in\Delta$ and (iv) $s_1,s_2\in\Delta$. The contribution from case (i) vanishes, because $H_M(t)$ is time-symmetric in this region. The contribution from the remaining three cases can be bounded by 
\begin{eqnarray}
\left\| \Omega_2(T)\right\| &\le& \frac{1}{2} \left\| [H_M(s_1),H_M(s_2)]\right\|_{\rm max}\cdot {\rm Volume}\nonumber\\
&\le& 2 J\epsilon \cdot {\rm Volume},
\end{eqnarray}
where ``Volume'' means the total volume of integration regions (ii), (iii), and (iv) combined. 

 
We recall that the integral is time ordered, so that $s_1\ge s_2$. The region $\Delta$ is the union of a disjoint set of intervals $\{\Delta_i\}$. We assume these intervals are labeled consecutively, so that $\Delta_j > \Delta_i$ for $j > i$. For case (ii), if $s_1\in \Delta_i$, then $s_2$ lies in the part of $T \backslash \Delta$ less then $\Delta_i$. Call this region $T_{< i}$. Similarly, for case (iii), if $s_2\in \Delta_i$, then $s_1$ lies in the part of $T \backslash \Delta$ greater then $\Delta_i$. Call this region $T_{> i}$. Adopting the convention in which the same symbol is used to represent both a region and its length, the total integration region for cases (ii) and (iii) combined has volume 
\begin{align}
&\sum_i \Delta_i \left(T_{< i}\right) + \Delta_i \left(T_{> i}\right) = \sum_i \Delta_i \left(T_{< i} + T_{>i}\right)\notag\\
&=\sum_i \Delta_i (T-\Delta) = \Delta(T-\Delta),
\end{align} 
with the first contribution coming from case (ii) and the second from case (iii). 

For case (iv), if $s_1$ and $s_2$ are in the same interval $\Delta_i$, the integration region has volume $\frac{1}{2}\Delta_i^2$. If $s_1\in \Delta_i$ and $s_2\not\in \Delta_i$, then $s_2\in \Delta_j$ for $j<i$. Summing the volumes of all regions with $s_1,s_2\in\Delta$ gives
\begin{eqnarray}
\sum_i \frac{1}{2} \Delta_i^2 + \sum_{i<j} \Delta_i \Delta_j = \frac{1}{2}\left(\sum_i \Delta_i\right)^2 = \frac{1}{2}\Delta^2.
\end{eqnarray}
Adding the contributions from cases (ii), (iii), and (iv), we find that the total volume is $\Delta T - \frac{1}{2}\Delta^2$, and conclude that \cite{Litsch}
\begin{eqnarray}\label{eq:omega-2-even-bound}
\left\| \Omega_2(T)\right\| &\le& 2 J\epsilon \cdot {\rm Volume} \nonumber\\
&=& 2\frac{\Delta}{T}\left(1-\frac{\Delta}{2T}\right)(JT)\epsilon T).
\end{eqnarray}
Hence, $C_2 = 2(\Delta/T)\left(1-{\Delta}/{2T}\right)$.

Since each even Magnus term vanishes in the time-symmetric case, there are upper bound on all even Magnus terms that depend linearly on $\Delta/T$ to lowest order. Such bounds are derived in Appendix \ref{app:Even}. For $\Omega_4(T)$,  Eq.~\eqref{eq:Even} yields
\begin{align}
\left\Vert\Omega_4(T)\right\Vert&\leq 14(JT)(\epsilon T)^3\left[1-(1-\frac{\Delta}{T})^4\right]\nonumber\\
&=14(JT)(\epsilon T)^3\left[4\left(\frac{\Delta}{T}\right)-6\left(\frac{\Delta}{T}\right)^2\right.\notag\\
&\hspace{2.3cm}\left.+4\left(\frac{\Delta}{T}\right)^3-\left(\frac{\Delta}{T}\right)^4\right].
\end{align}
Since $4(\Delta/T)^3\leq 4(\Delta/T)^2$ and $(\Delta/T)^4\geq 0$, we can rewrite this as
\begin{align}\label{eq:omega-4-even-bound}
\left\Vert\Omega_4(T)\right\Vert
&\leq 14(JT)(\epsilon T)^3\left[4\frac{\Delta}{T}-2\left(\frac{\Delta}{T}\right)^2\right]\nonumber\\
&= 56\frac{\Delta}{T}\left(1-\frac{\Delta}{2T}\right)(JT)(\epsilon T)^3.
\end{align}
Hence, $C_4=56(\Delta/T)\left(1-\Delta/2T\right)$. (For $\Omega_2(T)$ the bound Eq.~\eqref{eq:Even} is actually weaker by a factor of 2 than Eq.~\eqref{eq:omega-2-even-bound}, because a looser estimate of the integration volume is used to derive Eq.~\eqref{eq:Even}.)

\section{Concatenated dynamical decoupling} 
\label{sec:concatenated}

A concatenated DD pulse sequence is a recursively generated sequence with a self-similar structure \cite{KL05,KL07}. For example, from the ``level-1'' universal pulse sequence
\begin{equation}
p_1= Z\Id X\Id Z\Id X\Id
\end{equation}
we obtain the corresponding  ``level-2'' sequence by replacing each pulse interval $\Id$ in the level-1 sequence by the complete level-1 sequence $p_1$, obtaining
\begin{equation}
p_2= Z p_1 X p_1 Z p_1 X p_1;
\end{equation}
similarly, the level-$k$ sequence is
\begin{equation}
p_k= Z p_{k-1} X p_{k-1} Z p_{k-1} X p_{k-1}.
\end{equation}
If the duration of a single pulse is $\tau_0$ and $p_1$ is an $R$-pulse sequence that achieves first-order decoupling, then the corresponding level-$k$ sequence $p_k$ has duration $T^{(k)}=R^k\tau_0$ and achieves $k$th-order decoupling; {\em i.e.}, has effective noise strength $O(J\epsilon^k)$. 

\blue{The advantages of concatenated DD over standard
periodic pulse sequences (such as cycles of the universal decoupling, or XY-4 sequence) have been documented numerically \cite{KL07,Witzel:07a,PhysRevB.75.201302,Zhang:08,West:10} and
confirmed in a number of recent experimental studies \cite{Alvarez:10,2010arXiv1011.1903T,2010arXiv1011.6417W,2010arXiv1007.4255B}.
Concatenated pulse sequences are substantially less efficient than ``optimal'' sequences with nonuniform pulse intervals that achieve $k$th-order decoupling with exponentially fewer pulses \cite{Uhrig,Yang,Uhrig09,WFL:09,Mukhtar:10,NUDD}, but nevertheless have some nice properties. 
For one thing, concatenated pulse sequences are relatively robust against pulse imperfections, because pulse errors arising at each level get suppressed at higher levels. 
Experimental evidence for this robustness was provided in a recent NMR study of a qubit in a rapidly fluctuating spin bath, where pulse imperfections played a role, and concatenated DD sequences outperformed a variety of other sequences, including ``optimal'' ones with nonuniform pulse intervals, in preserving an unknown quantum state \cite{Alvarez:10}.}

We will analyze the performance of concatenated pulse sequences in two ways, first using the Magnus expansion, and then in Sec.~\ref{subsec:dyson-concatenated} using the Dyson expansion and bath correlation functions. 

\ignore{
Concatenated pulse sequences are less efficient than sequences with nonuniform pulse intervals that achieve $k$th-order decoupling with fewer pulses, but nevertheless have some nice properties. For one thing, the concatenation method can be applied to enhance any pulse sequence that achieves first-order (or higher-order) decoupling, while ``optimal'' pulse sequences \cite{Uhrig,Yang,Uhrig09,WFL:09} are known only for qubits. Furthermore, concatenated pulse sequences are relatively robust against pulse imperfections, because pulse errors arising at each level get suppressed at higher levels. We will analyze the performance of concatenated pulse sequences in two ways, first using the Magnus expansion, and then in Sec.~\ref{subsec:dyson-concatenated} using the Dyson expansion and bath correlation functions.
}
%

\blue{
Before presenting the analysis, we briefly state our main results. For ideal, zero-width pulses, 
}
\green{
we find that if an $R$-pulse sequence is concatenated $k$ times, then the effective noise strength is
} 
\blue{
\begin{eqnarray}
\eta_{\rm DD}^{(k)} = R^{k(k+3)/2}\left(\bar c\epsilon\tau_0\right)^k (J\tau_0)
\label{eq:eta-level-k-temp}
\end{eqnarray}
}
\green{
where $\tau_0$ is the pulse interval and ${\bar c}$ is a constant of order one. Increasing the concatenation level produces higher order decoupling, reflected in the $k$-dependent power of $\epsilon\tau_0$ in Eq.~(\ref{eq:eta-level-k-temp}), but also lengthens the pulse sequence, reflected in the $k$-dependent power of $R$. Thus there is an optimal concatenation level $k$, given by
\begin{equation}
k_{\max}=\lfloor\log_R(1/\bar c\epsilon\tau_0)-1\rfloor,
\label{eq:kmax}
\end{equation}
where $\lfloor \cdot \rfloor$ denotes the ``floor'' function. Using this optimal value of $k$, we find that the optimal effective noise strength satisfies the bound 
\begin{equation}
\label{eq:eta-optimal}
\eta_{\rm DD}^{(\rm opt)}/(J\tau_0)\le R^{-1} \left(\bar c\epsilon\tau_0\right)^{\frac{1}{2}\log_R(1/\bar c \epsilon \tau_0)-\frac{3}{2}}.
\end{equation}
If a time-symmetric $R$-pulse sequence is concatenated $k$ times, then the effective noise strength is 
\begin{eqnarray}
\eta_{\rm DD}^{(k)}=R^{k(k+2)}\left(\bar c\epsilon\tau_0\right)^{2k} (J\tau_0).
\label{eq:eta-level-k-pali}
\end{eqnarray}
which yields
\begin{equation}
\label{eq:eta-optimal-time-sym}
\eta_{\rm DD}^{(\rm opt)}/(J\tau_0)\le R^{-3/4} \left(\bar c\epsilon\tau_0\right)^{\log_R(1/\bar c \epsilon \tau_0)-2}
\end{equation}
after choosing the optimal value of $k$.
}

\blue{
\indent Optimal noise strengths for the universal and time-symmetric sequences,
plotted in Fig.~\ref{fig:Concatenated},
are orders of magnitude lower than the noise strengths
achievable without concatenation, shown in Fig.~\ref{fig:plotDDFT}.
Though longer, the time-symmetric sequence performs much better when
$\bar{c}\epsilon\tau_0$
is sufficiently small.
\newline
\indent When the pulses have a finite width $\delta$ and consequently experience systematic errors that arise from the time-independent noise Hamiltonian that is ÒonÓ during the pulses, there is a floor on the effective noise strength, namely
\begin{eqnarray}
\eta_{\rm DD}^{(k)} \ge 4 R\delta J.
\end{eqnarray}
As the level $k$ increases, $\eta_{\rm DD}^{(k)}$ falls as in Eq.~(\ref{eq:eta-level-k-temp}) or Eq.~(\ref{eq:eta-level-k-pali}) as long as it remains well above the floor, but reaches a plateau as the floor is approached. Such behavior was observed in the numerical simulations reported in \cite{KL07}. This floor might be substantially suppressed by using Eulerian pulse sequences as in Sec.~\ref{subsec:eulerian}\green{.}
\newline
\begin{figure}[!htbp]
\begin{center}
\includegraphics[width=0.48\textwidth]{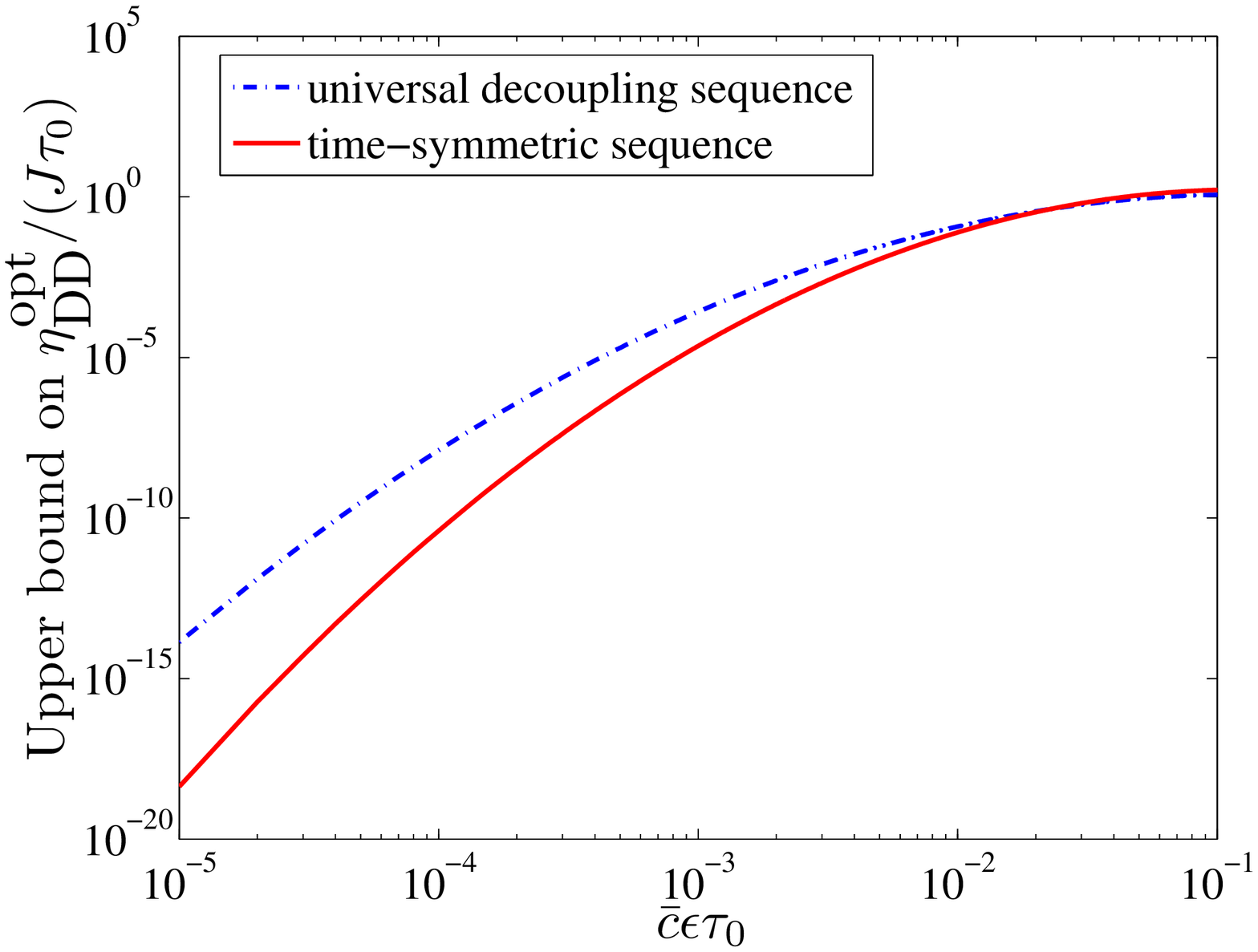}
\end{center}
\caption{\label{fig:Concatenated}
(color online) Upper bounds on the
effective noise strengths achieved by the concatenated universal DD
pulse sequence (blue dashed line, based on Eq.~(\ref{eq:eta-optimal}) with $R=4$) 
and by the
concatenated time-symmetric DD pulse sequence
(red solid line, based on Eq.~(\ref{eq:eta-optimal-time-sym}) with $R=8$),
as a function of $\bar c\epsilon\tau_0$, where $\bar c$ is defined in the text.} 
\end{figure}
}

\subsection{Magnus expansion analysis}
\label{subsec:magnus-concatenated}

The noise Hamiltonian has an unambiguous decomposition into two parts: $H=H_B+H_{\rm err}$, where $H_B=I\otimes B_0$, $H_{\rm err}=\sum_\alpha S_\alpha\otimes B_\alpha$, and $\{S_\alpha\}$ is a basis for the {\em traceless} operators acting on the system. For a level-1 pulse sequence with duration $T$, the toggling-frame time evolution operator is $\tilde U(T)=\exp (\Omega(T))$; writing $\Omega(T)= -iH^{(1)} T$, we may regard $H^{(1)}$ as the level-1 ``effective Hamiltonian.'' Like $H$, $H^{(1)}$ has an unambiguous decomposition into two parts,
\begin{equation}
H^{(1)}= H_B^{(1)}+H_{\rm err}^{(1)}\equiv I\otimes B_0^{(1)} + \sum_\alpha S_\alpha\otimes B_\alpha^{(1)},
\end{equation}
and we may define parameters that characterize the effective noise at level 1:
\begin{equation}\label{eq:level-1-noise-parameters}
\|H_B^{(1)}\|\le \beta^{(1)},\quad \|H_{\rm err}^{(1)}\|\le J^{(1)}, \quad \epsilon^{(1)}= \beta^{(1)}+J^{(1)}.
\end{equation}
Now, we can analyze the level-2 pulse sequence just as we did the level-1 sequence, but with the level-0 noise Hamiltonian $H=H^{(0)}$ replaced by the level-1 effective Hamiltonian $H^{(1)}$. Proceeding in this way, we can estimate properties of the toggling-frame time evolution operator $\tilde U^{(k)}$ for the level-$k$ pulse sequence using the level-$(k{-}1)$ Hamiltonian $H^{(k{-}1)}$. At each level, we can define noise parameters $\beta^{(k)}$, $J^{(k)}$, and $\epsilon^{(k)}$ as in Eq.~(\ref{eq:level-1-noise-parameters}), and derive recursion relations that relate the level-$k$ noise parameters to level-$(k{-}1)$ noise parameters.

To understand how first-order decoupling is achieved by the level-1
sequence, we assumed that the toggling-frame Hamiltonian is constant
in the interval between pulses. For the concatenated sequence at level
2 and above, this assumption is not true, since the interval in
between the level-$k$ pulses contains a complex level-$(k{-}1)$ pulse
sequence. However the unitary operator describing the evolution from
the end of one level-$k$ pulse to the beginning of the next level-$k$
pulse is equivalent to the evolution operator that would have been
derived from the constant Hamiltonian $H^{(k{-}1)}$ during the pulse
interval. Thus for the purpose of understanding the time evolution in
the toggling frame resulting from the level-$k$ sequence, it does no
harm to imagine that the Hamiltonian is constant between pulses and do
the analysis just as for the level-1 sequence.

For a sequence that achieves first-order decoupling, $\Omega_1$ at each level is a pure bath term
\begin{equation}\label{eq:level-k-omega1}
\Omega^{(k)}_1(T^{(k)}) = -iT^{(k)}H_B^{(k{-}1)}, 
\end{equation}
where $T^{(k)} = R^k\tau_0$ is the duration of the level-$k$ sequence,
constructed by concatenating $k$ times a sequence with $R$
pulses. Suppose we consider a pulse sequence such that each pulse
either commutes or anticommutes with each of the traceless operators
in the set $\{S_\alpha\}$ (the argument below can be easily adapted to
more general pulse sequences). Under this assumption, the second-order
term $\Omega_2$ in the Magnus expansion has no pure-bath component
(see Appendix \ref{app:omega2-bath}) and thus contributes only to
$H_{\rm err}^{(1)}$. Therefore $H_B^{(k)}$ arises from
$\Omega_1^{(k)}$ and the pure bath component of $\Omega_{\ge
  3}^{(k)}=\sum_{n=3}^\infty \Omega_n^{(k)}$.
As shown in Appendix \ref{app:CDD}, the norm of the pure-bath component of $\Omega^{(k)}_j$ is no larger than $\|\Omega^{(k)}_j\|$; it follows that we may choose $\beta^{(k)}$ such that $\beta^{(k)}T^{(k)}$ is an upper bound on 
\begin{equation}
\|\Omega^{(k)}_1(T)\|+ \|\Omega^{(k)}_{\ge 3}(T)\|.
\end{equation}
From Eq.~(\ref{eq:Gen}) and Table~\ref{tab:Coeffs} we see that 
\begin{equation}
\|\Omega^{(k)}_{\ge 3}\|\le c^{(k)}_3\left(J^{(k{-}1)}T^{(k)}\right)\left(\epsilon^{(k{-}1)} T^{(k)}\right)^2, 
\end{equation}
where the ``constant'' $c_3^{(k)}$ actually depends on the value of $\epsilon^{(k-1)} T^{(k)}$:
\begin{equation}
c_3^{(k)}= \frac{2}{9} + \frac{11}{9}\left(\epsilon^{(k-1)} T^{(k)}\right) + 9.43 \left(\epsilon^{(k-1)} T^{(k)}\right)^2, 
\label{eq:c_3-estimate}
\end{equation}
assuming $\epsilon^{(k-1)} T^{(k)} \le \red{0}.54$ ({\em e.g.}, $c_3^{(k)} = 0.44$ for $\epsilon^{(k-1)} T^{(k)} = 0.1$ and $c_3^{(k)}= 0.24$ for $\epsilon^{(k-1)} T^{(k)}= 0.01$). 
Recalling Eq.~(\ref{eq:level-k-omega1}), we conclude that
\begin{eqnarray}\label{eq:beta-level-k}
\beta^{(k)}= \beta^{(k{-}1)} + c^{(k)}_3 J^{(k{-}1)}\left(\epsilon^{(k{-}1)} T^{(k)}\right)^2.
\end{eqnarray}
Though Eq.~(\ref{eq:beta-level-k}) has been expressed as an equality, the right-hand side is actually an upper bound on $\|H_B^{(k)}\|$.

The level-$k$ error Hamiltonian $H^{(k)}_{\rm err}$ arises from $\Omega^{(k)}_2$ and the traceless component of $\Omega^{(k)}_{\ge 3}$. It is shown in Appendix \ref{app:CDD} that the norm of the traceless component of $\Omega^{(k)}_j$ is no larger than $2\|\Omega^{(k)}_j\|$; therefore we may choose $J^{(k)}$ such that $J^{(k)}T^{(k)}$ is an upper bound on 
\begin{equation}
\|\Omega^{(k)}_2(T)\|+ 2\|\Omega^{(k)}_{\ge 3}(T)\|.
\end{equation}
(If the system is a single qubit, then the norm of the traceless component of $\Omega^{(k)}_j$ is no larger than $\|\Omega^{(k)}_j\|$, and thus the factor of $2$ in the second term
can be omitted.) Therefore, again using Eq.~(\ref{eq:Gen}) and Table~\ref{tab:Coeffs} we find
\begin{eqnarray}\label{eq:J-level-k}
J^{(k)}= c^{(k)}_2 J^{(k{-}1)}\left(\epsilon^{(k{-}1)} T^{(k)}\right),
\end{eqnarray}
where
\begin{eqnarray}
\label{eq:c_2-estimate}
c_2^{(k)}&=& \frac{1}{2} + 2\left[\frac{2}{9}\big(\epsilon^{(k-1)} T^{(k)}\big) + \frac{11}{9}\big(\epsilon^{(k-1)} T^{(k)}\big)^2 \right.\notag\\
&&\hspace{1cm}\left.+ 9.43 \big(\epsilon^{(k-1)} T^{(k)}\big)^3\right],
\end{eqnarray}
assuming $\epsilon^{(k-1)} T^{(k)} \le \red{0}.54$ ({\em e.g.}, $c_2^{(k)} = \red{0}.588$ for $\epsilon^{(k-1)} T^{(k)} = 0.1$ and $c_2^{(k)}= \red{0}.505$ for $\epsilon^{(k-1)} T^{(k)}= 0.01$).

Eq.~(\ref{eq:beta-level-k}) can be rewritten as 
\begin{equation}
\beta^{(k)} = \beta^{(k{-}1)} + K^{(k)}.
\end{equation}
where
\begin{equation}\label{eq:K-level-k}
K^{(k)}= c^{(k)}_3 J^{(k{-}1)}\left(\epsilon^{(k{-}1)} T^{(k)}\right)^2,
\end{equation}
and iterating this equation yields
\begin{equation}
\beta^{(k)} = \beta +K^{(1)} + K^{(2)} + \cdots + K^{(k)}
\end{equation}
and
\begin{eqnarray}
\label{eq:epsilon-level-k}
\epsilon^{(k)} &=& \beta^{(k)} +J^{(k)}\nonumber\\
&=&\beta +K^{(1)} + K^{(2)} + \cdots + K^{(k)}+J^{(k)}\notag.\\
\end{eqnarray}
The solution to the recursion relations Eq.~(\ref{eq:J-level-k}), (\ref{eq:K-level-k}), (\ref{eq:epsilon-level-k}) cannot be expressed easily in closed form, but the properties of the solution can be grasped if we assume that
\begin{equation}\label{eq:epsilon-k-bound}
c_2^{(k)}\epsilon^{(k{-}1)}\le \bar c \epsilon
\end{equation}
for each $k$, where $\bar c$ is a constant.
That is, if we iterate the recursion relations to estimate $J^{(\ell)}$, our assumption is that Eq.~(\ref{eq:epsilon-k-bound}) is satisfied for all $k\le\ell$.
Then using $T^{(k)}=R^k\tau_0$, we can replace Eq.~(\ref{eq:J-level-k}) by
\begin{equation}
J^{(k)}= \left(\bar c \epsilon \tau_0\right)R^k J^{(k{-}1)},
\end{equation}
which has the solution
\begin{equation}\label{J-level-k-solution}
J^{(k)}= \left(\bar c \epsilon \tau_0\right)^k R^{k(k+1)/2} J
\end{equation}
where $J^{(0)}=J$ \cite{prior-CDD-comment}.

The effective noise strength for the level-$k$ sequence is
\begin{eqnarray}
\label{eq:eta-level-k}
\eta_{\rm DD}^{(k)} &=& \|H_{\rm err}^{(k)}\| T^{(k)} = J^{(k)}R^k\tau_0\nonumber\\
&=& R^{k(k+3)/2}\left(\bar c\epsilon\tau_0\right)^k (J\tau_0),
\end{eqnarray}
so that
\begin{equation}
\eta_{\rm DD}^{(k)} =R^{k+1}(\bar c\epsilon\tau_0)\eta_{\rm DD}^{(k{-}1)} ;
\end{equation}
therefore the optimal suppression of the noise strength is achieved by choosing the level $k$ to be the largest integer such that $R^{k+1}(\bar c\epsilon\tau_0 )<1$, or equivalently,
$k_{\max}=\lfloor\log_R(1/\bar c\epsilon\tau_0)-1\rfloor$ [Eq.~(\ref{eq:kmax})].

For example, if $\bar c\epsilon \tau_0 = 10^{-3}$ and $R=4$, we choose $k_{\max}=3$ ({\em i.e.}, a sequence with duration $T^{(3)}=64\tau_0$) and obtain $\eta_{\rm DD}^{(k_{\rm max})}/(J\tau_0)= 2.6 \times 10^{-4}$, an improvement by a factor of 60 over the noise strength $\eta_{\rm DD}^{(1)}$ achieved by the level-1 sequence.

The expression for $\eta_{\rm DD}^{(k)}$ in Eq.~(\ref{eq:eta-level-k})
is the exponential of a quadratic function of $k$, minimized at
$k=\log_R(1/\bar c\epsilon\tau_0)-3/2$. The nearest integer differs from
this optimal value by at most $1/2$; substituting $k+1=\log_R(1/\bar
c\epsilon\tau_0)$ into Eq.~(\ref{eq:eta-level-k}), we conclude that the
optimal effective noise strength satisfies
$\eta_{\rm DD}^{(\rm opt)}/(J\tau_0)\le R^{-1} \left(\bar c\epsilon\tau_0\right)^{\frac{1}{2}\log_R(1/\bar c \epsilon \tau_0)-\frac{3}{2}}$ [Eq.~(\ref{eq:eta-optimal})].

The condition Eq.~(\ref{eq:epsilon-k-bound}), used in the derivation of Eq.~(\ref{eq:eta-optimal}), can be justified for $\bar c = O(1)$. Suppose for example that $J$ is small compared to $\beta$. In that case, $\epsilon^{(k{-}1)}$ grows slowly with $k$, and it is a good approximation to assume $\epsilon^{(k{-}1)}\simeq \epsilon$. The optimal value of $k$ is chosen such that $R^{k+1}(\bar c\epsilon\tau_0)< 1$ and hence
\begin{equation}
\epsilon^{(k{-}1)}T^{(k)} \simeq \epsilon \tau_0 R^k < 1/(\bar cR).
\label{eq:1/cR}
\end{equation}
Using Eq.~(\ref{eq:c_2-estimate}) we see that Eq.~(\ref{eq:epsilon-k-bound}) applies for $k \le k_{\rm max}$ provided that
\begin{eqnarray}
\frac{1}{2} + 2\left[ \frac{2}{9}\left(\bar cR\right)^{-1}+\frac{11}{9}\left(\bar cR\right)^{-2}+9.43 \left(\bar cR\right)^{-3}\right] \le \bar c,\nonumber\\
\end{eqnarray}
which for $R=4$ is satisfied by
\begin{equation}
\bar c= 1.027.
\end{equation}
For consistency, we note that 
\ignore{
\begin{equation}
\epsilon^{(k{-}1)}T^{(k)} \simeq \epsilon \tau_0 R^k < 1/(R\bar c) = .244 < .54
\end{equation}
}
\blue{with these values Eq.~(\ref{eq:1/cR}) yields $\epsilon^{(k{-}1)}T^{(k)} < \red{0}.244 < \red{0}.54$,}
as assumed in the derivation of Eq.~(\ref{eq:c_2-estimate}). 

We can also check the self consistency of the approximation $\epsilon^{(k{-}1)}\simeq \epsilon$. Using this approximation together with Eq.~(\ref{eq:epsilon-k-bound}) and Eq.~(\ref{J-level-k-solution}) we find 
\begin{eqnarray}
K^{(k)} &=& c_3^{(k)} J^{(k-1)}\left(\epsilon^{(k-1)}T^{(k)}\right)^2\nonumber\\
&\le& \left[c_3^{(k)}/\left(c_2^{(k)}\right)^2 \right] \left((\bar c\epsilon\tau_0)^{k{-}1}R^{k(k-1)/2}J\right)\left(\bar c\epsilon\tau_0 R^k\right)^2\nonumber\\
&=& \left[c_3^{(k)}/\left(c_2^{(k)}\right)^2 \right] R^{k(k+3)/2}(\bar c\epsilon\tau_0)^{k}(\bar c \epsilon) \left(J\tau_0\right),
\end{eqnarray}
and hence%
\blue{, using Eq.~(\ref{eq:eta-level-k})},
\begin{eqnarray}
{K^{(k)}}/{\epsilon}
&\le&  \left[\bar c c_3^{(k)}/\left(c_2^{(k)}\right)^2 \right]~ \eta_{\rm DD}^{(k)}.
\end{eqnarray}
Since $\eta_{\rm DD}^{(k)} \ll 1$ for $1\le k \le k_{\rm max}$, 
\blue{and Eqs.~(\ref{eq:c_3-estimate}) and (\ref{eq:c_2-estimate}) yield $c_3^{(k)} \lesssim c_2^{(k)}$,}
we conclude that $K^{(k)} \ll \epsilon$ for each $k$. Thus for $J \ll \beta$ we have $\epsilon\simeq \beta$ and $\epsilon^{(k)} \simeq \beta^{(k)}\simeq \beta\simeq \epsilon$ for each $k$%
\blue{, where we have used Eq.~(\ref{eq:epsilon-level-k})}.

Numerical iteration of the recursion relations confirms that the approximation $\epsilon^{(k{-}1)}\simeq \epsilon$ works well for $J/\beta < \red{0}.3$, and that our estimate of $\eta_{\rm DD}^{(\rm opt)}$ is reasonably tight in that case \cite{Litsch}.  For $J \gg \beta$, though, $\epsilon \simeq J$ and $\epsilon^{(k)} \ll \epsilon$ for $1\le k\le k_{\rm max}$; we may still use Eq.~(\ref{eq:epsilon-k-bound}) to derive an upper bound on $\eta_{\rm DD}^{(\rm opt)}$in that case, but our estimate Eq.~(\ref{eq:eta-optimal}) becomes overly pessimistic \cite{Litsch}. 
\blue{Indeed, the case $J \gg \beta$ is favorable for DD, since the bath dynamics is relatively slow and the system-bath coupling, which DD suppresses, is larger to begin with. For an analysis of concatenated DD in this case see Ref.~\cite{KL07}.}

If we concatenate a time-symmetric pulse sequence, which achieves second-order decoupling, then we may replace Eq.~(\ref{eq:J-level-k}) by 
\begin{eqnarray}\label{eq:J-level-k-palindrome}
J^{(k)}= 2c^{(k)}_3 J^{(k{-}1)}\left(\epsilon^{(k{-}1)} T^{(k)}\right)^2
\end{eqnarray}
(the factor of 2 can be omitted if the system is a qubit), and we can also improve the estimate of $c_3$ to
\begin{equation}\label{eq:c3-estimate}
c_3^{(k)}= \frac{2}{9} + 9.43 \left(\epsilon^{(k-1)} T^{(k)}\right)^2,
\end{equation}
where $\epsilon^{(k-1)} T^{(k)} \le .54$. 
Defining $\bar c$ for a time-symmetric sequence by 
\begin{equation}\label{eq:c-bar-time-symmetric}
2c_3^{(k)}\left(\epsilon^{(k{-}1)}\right)^2\le (\bar c \epsilon)^2, 
\end{equation}
Eq.~(\ref{eq:J-level-k-palindrome}) becomes
\begin{eqnarray}\label{eq:J-level-k-palindrome-again}
J^{(k)}= \left(\bar c \epsilon \tau_0\right)^2R^{2k} J^{(k{-}1)},
\end{eqnarray}
which has the solution
\begin{equation}
J^{(k)}= \left(\bar c \epsilon \tau_0\right)^{2k} R^{k(k+1)} J,
\end{equation}
and thus
\begin{eqnarray}
\eta_{\rm DD}^{(k)}=R^{k(k+2)}\left(\bar c\epsilon\tau_0\right)^{2k} (J\tau_0).
\end{eqnarray}
The noise strength is optimized by choosing the largest integer $k$ such that 
$k+\frac{1}{2} < \log_R(1/\bar c\epsilon\tau_0)$. For example, if $R=8$ and 
$\bar c\epsilon\tau_0=10^{-3}$, we choose $k_{\rm max}=2$ ({\em i.e.}, a sequence with duration $T^{(2)}=64\tau_0$) and obtain $\eta_{\rm DD}^{(k_{\rm max})}/(J\tau_0)= 1.7 \times 10^{-5}$, an improvement by a factor of 30 over the noise strength $\eta_{\rm DD}^{(1)}$ achieved by the level-1 sequence. The optimal noise strength satisfies
$\eta_{\rm DD}^{(\rm opt)}/(J\tau_0)\le R^{-3/4} \left(\bar c\epsilon\tau_0\right)^{\log_R(1/\bar c \epsilon \tau_0)-2}$ [Eq.~(\ref{eq:eta-optimal-time-sym})]. 

If we make the approximation $\epsilon^{(k{-}1)}\simeq \epsilon$ , then, because $R^{k{+}\frac{1}{2}}(\bar c\epsilon\tau_0)< 1$ for the optimal value of $k$, we have
\begin{equation}
\epsilon^{(k{-}1)} T^{(k)}\simeq \epsilon \tau_0 R^k < 1/\left(\bar c\sqrt{R}\right).
\end{equation}
Using Eq.~(\ref{eq:c3-estimate}) we see that Eq.~(\ref{eq:c-bar-time-symmetric}) applies for $k \le k_{\rm max}$ provided that
\begin{eqnarray}
2\left[\frac{2}{9} + 9.43 (\bar c^2 R)^{-1}\right] < \bar c^2,
\end{eqnarray}
which for $R=8$ is satisfied by 
\begin{equation}
\bar c = 1.332.
\end{equation}
As in our analysis for the non-time-symmetric case, the approximation $\epsilon^{(k-1)}\simeq \epsilon$ is reasonable, and our estimate Eq.~(\ref{eq:eta-optimal-time-sym}) is fairly tight, if $J$ is small compared to $\beta$. The upper bound Eq.~(\ref{eq:eta-optimal-time-sym}) applies more generally, but it is far from tight if $J$ is much larger than $\beta$, in which case $\epsilon^{(k)} \ll \epsilon$ for $1\le k\le k_{\rm max}$.

\ignore{
Optimal noise strengths for the universal and time-symmetric sequences,
plotted in Fig.~\ref{fig:Concatenated},
are orders of magnitude lower than the noise strengths
achievable without concatenation, shown in Fig.~\ref{fig:plotDDFT}.
Though longer, the time-symmetric sequence performs much better when
$\bar{c}\epsilon\tau_0$
is sufficiently small.

\begin{figure}[!htbp]
\begin{center}
\includegraphics[width=0.48\textwidth]{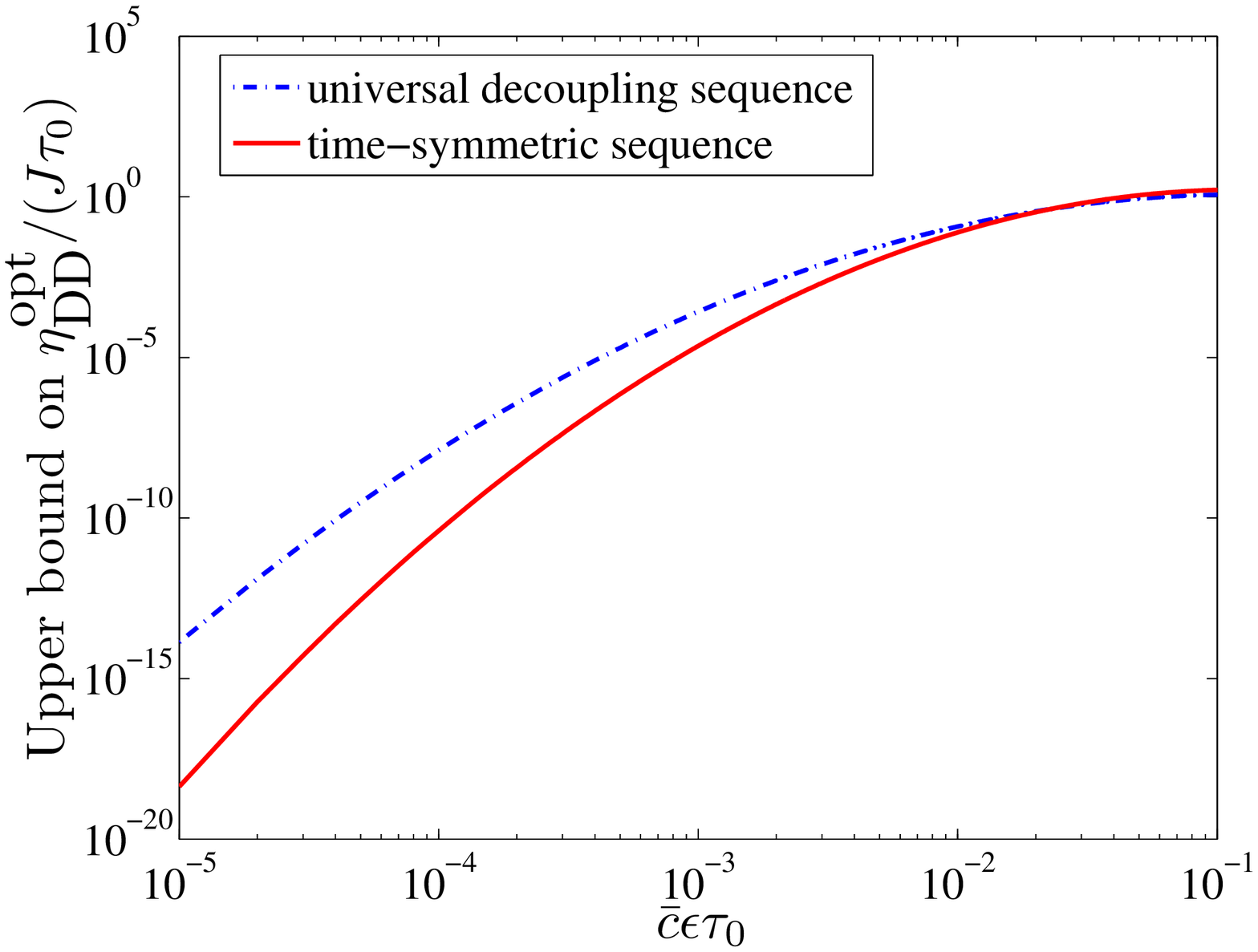}
\end{center}
\caption{\label{fig:Concatenated}
(color online) Upper bounds on the
effective noise strengths achieved by the concatenated universal DD
pulse sequence (blue dashed line, based on Eq.~(\ref{eq:eta-optimal}) with $R=4$) 
and by the
concatenated time-symmetric DD pulse sequence
(red solid line, based on Eq.~(\ref{eq:eta-optimal-time-sym}) with $R=8$),
as a function of $\bar c\epsilon\tau_0$, where $\bar c$ is defined in the text.} 
\end{figure}
}

\subsection{Including pulse errors}
How is this analysis affected if the pulses are imperfect? The answer depends on the degree to which the pulse errors are systematic and reproducible, rather than random. As in our discussion of Eulerian decoupling, let us assume that the errors are systematic. This assumption is reasonable if the pulse errors arise from the time-independent noise Hamiltonian that is ``on'' during the pulses, rather than from variations in the pulse shape. 

In the recursive analysis of the concatenated pulse sequence, the effective Hamiltonian $H^{(k{-}1)}$ incorporates all the damage caused by the pulses errors at level $k-1$ and below. Because the pulse errors are systematic, we may use the same $H^{(k{-}1)}$ to describe the noise in each interval between level-$k$ pulses. Suppose we imagine, at first, that while the pulses at level $k-1$ and below are noisy, the pulses at level $k$ are ideal, and denote by $\hat J^{(k)}$ the upper bound on $\|H_{\rm err}^{(k)}\|$ under this fictitious assumption. Repeating the derivation of Eq.~(\ref{eq:J-level-k}) yields
\begin{eqnarray}
\hat J^{(k)}= c^{(k)}_2 J^{(k{-}1)}\left(\epsilon^{(k{-}1)} T^{(k)}\right).
\end{eqnarray}
But now we must relate $\hat J^{(k)}$ to $J^{(k)}$ by estimating the effects of the pulse errors at the top level. 

The noise in these level-$k$ pulses is governed by the level-0 error
Hamiltonian $H_{\rm err}^{(0)}$ rather than the effective
level-$(k{-}1)$ error Hamiltonian $H_{\rm err}^{(k{-}1)}$. We could
adapt our analysis of the Magnus expansion to this new situation,
using a different upper bound on $H_{\rm err}$ during the pulses than
in the interval between pulses, but then we would face the
complication of revising our estimate of all the higher-order terms in
the expansion. To avoid that complication, we use a different
approach.
As in Sec.~\ref{subsubsec:Omega1-prime}, we assume that
the Hamiltonian describing the sequence of noisy pulses at level $k$
deviates in operator norm from the Hamiltonian describing the sequence of ideal pulses at level
$k$ by at most $2J$ during a total time interval $R\delta$, if there
are $R$ pulses each with width $\delta$.
It then follows from Lemma \ref{lem:error} in Appendix \ref{app:error} that
\begin{equation}
\label{eq:pulse-error-correction}
\left \| e^{\Omega^{(k)}} - e^{\hat \Omega^{(k)}}\right\|\le 2R\delta J,
\end{equation}
where $\Omega^{(k)}$ includes pulse-error corrections at all levels while $\hat \Omega^{(k)}$ includes pulse-error corrections at level $k-1$ and below but not at level $k$. From Eq. \eqref{eq:normAminusB} in Appendix \ref{app:logarithm}, we find that
\begin{equation}
\left \| \Omega^{(k)} - \hat \Omega^{(k)}\right\|\le d^{(k)}\left \| e^{\Omega^{(k)}} - e^{\hat \Omega^{(k)}}\right\|\le 2d^{(k)}R\delta J,
\end{equation}
where the ``constant'' $d^{(k)}$ is close to one if $\| \Omega^{(k)}\|$ and $\| \hat \Omega^{(k)}\|$ are both small;  therefore we obtain an upper bound on $J^{(k)}$:
\begin{eqnarray}
\label{eq:J-level-k-delta}
J^{(k)} &\le& \hat J^{(k)} + 2\left \| \Omega^{(k)} - \hat \Omega^{(k)}\right\|/T^{(k)}\nonumber\\
&\le & c^{(k)}_2 J^{(k{-}1)}\left(\epsilon^{(k{-}1)} T^{(k)}\right) + 4d^{(k)}R\delta J/T^{(k)}.\nonumber\\
\end{eqnarray}

If at each level the second term in Eq.~(\ref{eq:J-level-k-delta}) is small compared to the first term, then our previous analysis of the pulse sequence remains a good approximation, and we conclude that the pulse errors do not compromise the effectiveness of concatenated DD very much. However, the second term imposes a floor on (our upper bound on) the effective noise strength
\begin{eqnarray}
\eta_{\rm DD}^{(k)} = J^{(k)}T^{(k)} \ge  4d^{(k)}R\delta J\ge 4 R\delta J.
\end{eqnarray}
\ignore{
As the level $k$ increases, $\eta_{\rm DD}^{(k)}$ falls as in Eq.~(\ref{eq:eta-level-k}) as long as it remains well above the floor, but reaches a plateau as the floor is approached. Such behavior was observed in the numerical simulations reported in \cite{KL07}. This floor might be substantially suppressed by using Eulerian pulse sequences as in Sec.~\ref{subsec:eulerian}.
}

A noteworthy property of Eq.~(\ref{eq:J-level-k-delta}) is that only the pulse errors at the top level appear explicitly on the right-hand side. The errors at lower levels are included implicitly, through their contributions to $J^{(k{-}1)}$ and $\epsilon^{(k{-}1)}$. Accordingly, Eq.~(\ref{eq:J-level-k-delta}) captures the idea that the cumulative effect of the errors in the $R^k$ pulses is smaller than might have been naively expected, because errors that occur at lower levels in the pulse sequence become suppressed by the upper level pulses. This is an important feature of concatenated DD.

\section{Beyond the local-bath assumption}\label{sec:scaling}

A key element of the noise model formulated in Sec.~\ref{sec:DDFTNoise} is the local-bath assumption: at any given time, the noise Hamiltonians $H_a$ and $H_b$ associated with distinct circuit locations $a$ and $b$ act not only on disjoint sets of qubits but also on disjoint baths. This assumption is important because it allows us to ignore interactions among different circuit locations and thus assign an effective noise strength $\etaDD$ to each DD-protected gate individually. The local-bath assumption may be a reasonable approximation to noise in actual systems, at least in some cases, but it is not strictly satisfied; surely there are bath degrees of freedom that couple to multiple qubits, even while these qubits are participating in distinct gates. Can our analysis be extended to noise models that include correlations that arise because qubits participating in different gates at the same time couple to common bath variables?

Accuracy threshold theorems have been proved for Hamiltonian models of correlated noise in \cite{AGP,AKP,NP}. Perhaps similar methods can be applied to DD-protected circuits, but this seems to be a technically challenging problem which we leave for the future.

However, there is an easier problem that already arises when we consider just a single circuit location, and disregard how the noise at one location is correlated with the noise at another location. How is our analysis affected if the qubits at this location couple not just to a local bath 
\blue{comprising}
nearby bath degrees of freedom but to a global bath that includes bath variables that are far away? Of course, our previous analysis still applies if we replace the norm $\Vert H_{B,a}\Vert$ of the local-bath Hamiltonian by the norm $\Vert H_B\Vert$ of the global-bath Hamiltonian in Eq.~(\ref{eq:beta-define}) and Eq.~(\ref{eq:epsilon-define}), but the trouble with this approach is that $\Vert H_B\Vert$ is a huge number that scales linearly with the volume of the bath, while an accuracy threshold criterion should be stated in terms of intensive quantities that are independent of the size of the system and bath. On the other hand, we expect on physical grounds that the bath has a decomposition into local subsystems, and that the coupling of a given bath subsystem to a system qubit decays as the distance increases between the bath subsystem and the qubit; if in contrast each system qubit were coupled with constant strength to bath subsystems arbitrarily far away, the noise would be unacceptably strong and coherent manipulation of the system would be hopeless. Even though the local-bath assumption formulated  in Sec \ref{sec:DDFTNoise} may not hold exactly, a sensible noise model should be \emph{quasi-local} --- qubits ought to interact only very weakly with bath subsystems that are far away. In this case, can we express the effective noise strength in terms of intensive quantities?

To be concrete, consider a noise model in which a single system qubit is immersed in a bath of 
$N_b$
non-interacting spins in an external magnetic field. The noise Hamiltonian, assuming $H_S=0$, is
\begin{equation}
H=H_B+H_{SB}=\sum_iH_{B,i}+\sum_iH_{SB,i},
\end{equation}
where
\begin{align}
H_{B,i}&\equiv \Id_S\otimes B_i^0,\notag\\
H_{SB,i}&\equiv \sum_\alpha \sigma^\alpha\otimes B_i^\alpha.
\end{align}
Here, the index
$i=1,\ldots,N_b$ labels the bath spins and $\{\sigma_\alpha, ~\alpha=1,2,3\}$ are the Pauli operators acting on the system qubit. We may define the strengths of the individual terms as 
\begin{equation}
\lambda_i\equiv \left\Vert H_{SB,i}\right\Vert \quad\text{and}\quad b_i\equiv \left\Vert H_{B,i}\right\Vert=\left\Vert B_i^0\right\Vert,
\end{equation}
and the strength of the system-bath coupling can be characterized by 
\begin{equation}
J\equiv \sum_i\lambda_i\geq \left\Vert H_{SB}\right\Vert;
\end{equation}
we assume that the sum converges to a (small) finite value in the limit $N_b\to \infty$. On the other hand, the quantity
\begin{equation}
\beta\equiv \sum_i b_i \geq \left\Vert H_B\right\Vert
\end{equation} 
is not expected to remain bounded as $N_b\to \infty$.

Now consider how the bath parameters $\{b_i\}$ enter the Magnus expansion for a DD memory sequence or for a DD-protected gate applied to the system qubit. The Hamiltonian $H_M(t)$ is 
\begin{equation}\label{eq:spinbath}
H_M(t)=H_B+H'(t)=\sum_i H_{B,i}+H'(t),
\end{equation}
where $H'(t)=0$ or $\tilde H_\err(t)(=\tilde H_{SB}(t))$ as in Sec.~\ref{subsec:BoundsMagnus}, so that $\left\Vert H'(t)\right\Vert\leq J$. Furthermore, bath operators acting on different bath spins commute:
\begin{equation}
[B_i^0,B_j^0]=[B_i^0,B_j^\alpha]=[B_i^\alpha,B_j^\alpha]=0,~\forall i\neq j;
\end{equation}
the only nonvanishing commutators of bath operators are $[B_i^0,B_i^\alpha]$ and $[B_i^\alpha,B_i^\gamma]$ (for any spin $i$). 

The bath parameters $\{b_i\}$  do not contribute to $\Omega_1(T)$, so consider $\Omega_2(T)$. To estimate the integral in Eq.~\eqref{eq:O2bound} (taking $\Gamma=0$ so that $H_M(t)=\tilde H(t)$), we need an upper bound on the commutators. We observe that
\begin{align}
&\Vert[ H_B,\tilde H_\err(s_2)]\Vert=\Vert \sum_i[H_{B,i},\tilde H_{SB,i}(s_2)]\Vert\notag\\
\label{eq:2bJ-bound}&\leq \sum_i2\Vert H_{B,i}\Vert\cdot \Vert\tilde H_{SB,i}(s_2)\Vert\leq 2 bJ,
\end{align}
where we have defined the single-spin bath parameter
\begin{equation}
b\equiv \max_i\left\Vert H_{B,i}\right\Vert.
\end{equation}
We also observe that
\begin{align}
\Vert[\tilde H_\err(s_1),\tilde H_\err(s_2)]\Vert&\le \Vert \sum_{i,j}[\tilde H_{SB,i}(s_1),\tilde H_{SB,j}(s_2)]\Vert\notag\\
\label{eq:2J2-bound}&\le \sum_{i,j} 2\lambda_i\lambda_j = 2J^2.
\end{align}
Together, Eq.~(\ref{eq:2bJ-bound}) and Eq.~(\ref{eq:2J2-bound}) imply
\begin{equation}\label{eq:DiscnTwoCom}
\Vert [H_M(s_1),H_M(s_2)]\Vert\leq 4bJ + 2J^2,
\end{equation}
and plugging Eq.~(\ref{eq:DiscnTwoCom}) into Eq.~\eqref{eq:O2bound} yields
\begin{equation}\label{eq:BeyondLocalOmega2}
\Vert\Omega_2(T)\Vert \le \left(4bJ + 2J^2\right)\frac{1}{4}T^2\leq (JT)[(b+J)T].
\end{equation}
Using the local-bath assumption we would conclude $\Vert [H_M(s_1),H_M(s_2)]\Vert\leq 4\beta J + 2J^2$;
The result Eq~(\ref{eq:BeyondLocalOmega2}) matches the conclusion we would reach under the local-bath assumption,  but with $\beta$ now replaced by $b$.

Similarly, upper bounds on the higher-order Magnus terms can be also be expressed in terms of $J$ and $b$, though the ``replace $\beta$ by $b$ rule'' does not quite work beyond second order. Consider, for example, one triple commutator that occurs in $\Omega_3(T)$:
\begin{align}
&\quad\Vert[H_B,[\tilde H_\err(s_2),\tilde H_\err(s_3)]]\Vert\notag\\
&=\Vert\sum_{ijk}[H_{B,k},[\tilde H_{SB,i}(s_2),\tilde H_{SB,j}(s_3)]]\Vert\notag\\
&\leq\sum_{ij}\left(\Vert[H_{B,i},[\tilde H_{SB,i}(s_2),\tilde H_{SB,j}(s_3)]]\Vert\right.\notag\\
&\hspace{0.8cm}\left.+\Vert[H_{B,j},[\tilde H_{SB,i}(s_2),\tilde H_{SB,j}(s_3)]]\Vert\right)\notag\\
&\leq 2(2b)\sum_{i,j}2\lambda_i\lambda_j =8 bJ^2.
\end{align} 
In contrast, in the local-bath model we could upper bound the
corresponding triple commutator by $4 \beta J^2$. Simply replacing
$\beta$ by $b$ gives the wrong answer by a factor of 2, because it
fails to take into account that there are two different bath spins
that do not commute with $[\tilde H_{SB,i}(s_2),\tilde H_{SB,j}(s_3)]$
for $i\ne j$. Similar factors, dependent on $n$, occur in the
higher-order nested commutators contributing to $\Omega_n(T)$, but
these factors do not depend on the total number of bath spins
$N_b$. 

We could also include quasi-local interactions among the bath
spins, and still obtain an upper bound on each Magnus term expressed
in terms of intensive quantities. Suppose for example that we include in the bath Hamiltonian the additional term
\begin{equation}
\frac{1}{2}\sum_{i,j}H_{B,\langle ij\rangle},
\end{equation}
where 
\begin{equation}
H_{B,\langle ij\rangle}= \Id_S\otimes B_{\langle ij\rangle}
\end{equation}
acts trivially on the system qubit but nontrivially on the {\em pair} of bath spins $\langle i j\rangle$. In that case there will be an additional term in our upper bound on $\Vert[ H_B,\tilde H_\err(s_2)]\Vert$:
\begin{align}
&\frac{1}{2}\Vert \sum_{i,j}[H_{B,\langle ij\rangle},\tilde H_{SB,i}(s_2)+\tilde H_{SB,j}(s_2)]\Vert\notag\\
\label{eq:quasi-bath}&\leq 2 \sum_{i,j} \lambda_i \Vert H_{B,\langle ij\rangle}\Vert\leq 2cJ,
\end{align}
where 
\begin{equation}
c = \max_{i}\left(\sum_j \Vert H_{B,\langle ij\rangle}\Vert\right).
\end{equation}
Thus in the modified upper bound on $\Vert \Omega_2(T)\Vert$ we replace $b$ by $b+c$. The expression for $c$ includes a sum over all bath spins, but converges to an intensive quantity if the interaction between bath spins $i$ and $j$ decays sufficiently rapidly with the distance between the spins. Similar convergent sums occur in the upper bounds on higher-order Magnus terms. 

Even when our bounds on the Magnus expansion are intensive, they might
still be useless, if each local bath subsystem has a Hamiltonian with
a large norm. In that case, though, there is another method that might
succeed, which relates the effective noise strength to the frequency
spectrum of bath correlations. We turn to that method next.

\section{Dynamical decoupling and bath correlations}
\label{sec:correlator}
So far, we have described how to analyze the performance of DD using the toggling frame and the Magnus expansion. Another method is to use the interaction picture defined by $H_c(t) +H_B$; that is, to transform away both the control sequence acting on the system and the free bath dynamics. In that case, the interaction-picture Hamiltonian is
\begin{eqnarray}\label{eq:appJham}
\tilde H(t) &=& \left[U_c^\dagger(t)\otimes U_B^\dagger(t)\right] H_{\rm err} \left[U_c(t)\otimes U_B(t)\right] \nonumber\\
&=& \sum_\alpha S_\alpha(t)\otimes B_\alpha(t),
\end{eqnarray}
where
\begin{equation}
S_\alpha(t) = U_c(t)^\dagger S_\alpha U_c(t),\quad B_\alpha(t)=e^{itB_0}B_\alpha e^{-itB_0},
\end{equation}
and we can study the interaction-picture time evolution operator using the Magnus expansion defined by this Hamiltonian. This expansion has the big advantage that the interaction picture sums up the effects of the free bath dynamics to all orders in $\beta$; therefore, higher-order corrections are small provided $J$ is small, even though $\beta$ may be large. But there is also a substantial disadvantage: because the interaction picture bath operator $B_\alpha(t)$ is now time dependent, a pulse sequence that achieves first-order decoupling in the toggling frame may not achieve first-order decoupling in the interaction picture. 

On the other hand, if the bath operator $B_\alpha(t)$ is in some sense slowly varying, then first-order decoupling might be satisfied to a good approximation. Though the rate of change of the operator $B_\alpha(t)$ is actually of order $\beta$, if the {\em state} of the bath has suitable properties, then the expectation value of $B_\alpha(t)$ in that state may vary slowly; then DD may work well because the typical {\em frequencies} of the bath are sufficiently small, even though $\beta$ may be large. 

When estimating $\eta_{\rm DD}$ using the Magnus expansion, we did not make any assumption about the state of the bath. The new estimates we derive in this Section depend on the bath's frequency spectrum and hence implicitly on the bath's state. In order to obtain a simple formula for $\eta_{\rm DD}$ we will impose a further limitation on the noise model that was not needed in the Magnus expansion analysis --- we assume that the state of the bath is discarded at the end of each circuit location, and replaced by a fresh bath state at the beginning of the next location. Thus we will include the effects of the bath's memory in analyzing the effectiveness of the DD pulse sequence at each circuit location, but we assume that noise correlations between consecutive circuit locations can be neglected. We recognize the artificiality of this noise model, but we adopt it anyway because it allows us to derive an explicit expression for $\eta_{\rm DD}$. See Appendix \ref{app:Dyson} for further discussion.

\subsection{Dyson expansion}
\label{sec:Dyson}

In the toggling frame, it is convenient to analyze DD using the Magnus
expansion because for a well chosen sequence of ideal pulses
$\Omega_1$ is a pure bath term, and the remaining noise acting on the
system resides in the higher order terms. But if we use the
interaction picture instead, so that first-order decoupling is not
exact even for ideal pulses, it is simpler to estimate the effective
noise strength $\eta_{\rm DD}$ using the Dyson expansion rather than
the Magnus expansion. The interaction-picture time evolution operator
$\tilde U(t) = \left[U_c^\dagger(t)\otimes U_B^\dagger(t)\right]
U(t,0)$ is
\begin{equation}
\tilde U(t) = {\cal T}\exp\left(-i \int_0^t dt' \tilde H(t')\right)
\end{equation}
where ${\cal T}$ denotes time ordering.
For the local-bath noise model, augmented by the assumption that the bath is refreshed at the beginning at each circuit location, the arguments in Appendix \ref{app:Dyson}
show that the noise strength $\bar \eta$ can be expressed as 
\begin{eqnarray}\label{eq:eta-dyson}
\bar\eta^2 &=& \max_{a,|\Psi\rangle} \left\langle \left( \tilde U^\dagger(T) - \Id\right)\left( \tilde U(T) - \Id\right)\right\rangle \nonumber\\
&=& \max_{a,|\Psi\rangle} \left\langle 2\Id- \tilde U(T) - \tilde U^\dagger(T)\right\rangle;
\end{eqnarray}
here $T$ is the duration of the location, the expectation value $\langle \cdot \rangle$ is evaluated in the pure state $|\Psi\rangle\otimes |\Phi_a\rangle $ where $|\Phi_a\rangle$ is (a purification of) the initial state of the bath at the beginning of location $a$, and the maximum is with respect to all circuit locations and all system states. 
As is also shown in Appendix \ref{app:Dyson},
\begin{eqnarray}\label{eq:eta-dyson2}
\bar \eta^2 &\le& \max \int_0^T dt_1 \int_0^T dt_2 \left\langle\tilde H(t_1)\tilde H(t_2) \right\rangle \nonumber\\
&+& 2\left( e^{JT} - 1 - JT - \frac{1}{2}(JT)^2\right).
\end{eqnarray} 
For each term in the expansion Eq.~(\ref{eq:appJham}) the expectation value in the product state factorizes and we have
\begin{eqnarray}\label{eq:eta-dyson3}
\bar \eta^2 &\le&\int_0^T dt_1  dt_2 \sum_{\alpha,\beta}\left\langle S_\alpha(t_1)S_\beta(t_2) \right\rangle_S \left\langle B_\alpha(t_1) B_\beta(t_2) \right\rangle_B\nonumber\\
&+& 2\left( e^{JT} - 1 - JT - \frac{1}{2}(JT)^2\right),
\end{eqnarray} 
where the maximum over circuit locations and system states is implicit.

Now suppose that the bath's time correlations are stationary, {\em i.e.}, that the expectation value $\langle B_\alpha(t_1) B_\beta(t_2)\rangle_B$ is a function of the time difference $t_1-t_2$; this will be true if the initial state of the bath commutes with $H_B$, for example if the state is a mixture of energy eigenstates such as a thermal state. Then the bath correlation function may be expressed as
\begin{equation}
\langle B_\alpha(t_1) B_\beta(t_2)\rangle_B = \int_{-\infty}^\infty \frac {d\omega}{2\pi} e^{-i\omega(t_1-t_2)}~K_{\alpha\beta}(\omega),
\end{equation}
and Eq.~(\ref{eq:eta-dyson3}) becomes 
\begin{equation}
\bar \eta^2\le \max\int_{-\infty}^\infty \frac {d\omega}{2\pi} \sum_{\alpha,\beta}\langle \tilde S_\alpha(\omega)\tilde S_\beta(-\omega) \rangle_S ~K_{\alpha\beta}(\omega)+\dots,
\end{equation}
where
\begin{equation}
\tilde S_\alpha(\omega) = \int_0^T dt ~e^{-i\omega t}~ S_\alpha(t),
\end{equation}
and the ellipsis indicates the terms higher order in $J$. Defining the bath's spectral function $J^2_{\alpha\beta,i}$ by
\begin{equation}
K_{\alpha\beta}(\omega)= 2\pi\sum_i J^2_{\alpha\beta,i}\delta(\omega-\omega_i),
\end{equation}
our expression for (the square of) the noise strength is
\begin{equation}\label{eq:eta-S-FT}
\bar \eta^2 \le \max \sum_{i,\alpha,\beta} J^2_{\alpha\beta,i} \langle \tilde S_\alpha(\omega_i)\tilde S_\beta(-\omega_i) \rangle_S+\cdots.
\end{equation}
Thus, speaking loosely, DD is effective if $\tilde S_\alpha(\omega)$ is suppressed when $\omega$ is a ``typical frequency'' where the bath spectral function has support. We use the symbol $J^2$ advisedly, because $\sqrt{J^2_{\alpha\beta,i}}$, like $J=\max \|H_{\rm err}\|$, scales linearly with the strength of the system-bath coupling.

The operator $\tilde S_\alpha(\omega)$ can be written as $T\bar S_\alpha(\omega T)$, where $\bar S_\alpha$ is dimensionless. Adapting our terminology to this correlation function analysis, let us say that a pulse sequence achieves $n$th-order decoupling if the first $n$ terms in the Taylor expansion of $\bar S_\alpha(\omega T)$ vanish, so that
\begin{equation}\label{eq:FT-nth-order}
\tilde S_\alpha(\omega) = T\left(\bar{S}_{\alpha,n}
(\omega T)^n  + O[(\omega T)^{n+1}]\right).
\end{equation} 
Equivalently, the pulse sequence achieves $n$th-order decoupling provided
\begin{equation}
\int_0^T dt ~t^m S_\alpha(t)=0
\end{equation} 
for all $\alpha$ and for $m=0,1,2,\dots, n-1$.
Denoting the norm of the operator 
$\bar{S}_{\alpha,n}$ by $C_{\alpha,n}$, we find that
for a pulse sequence achieving $n$th-order decoupling, the noise strength is
\begin{equation}
\eta_{\rm DD} \le \left[\sum_{i,\alpha,\beta} C_{\alpha,n}C_{\beta,n}J^2_{\alpha\beta,i}T^2\left(\omega_i T\right)^{2n}\right]^{1/2}+\dots,
\end{equation}
where now the ellipsis includes corrections both higher order in $\omega T$ and higher order in the Dyson expansion. Therefore, ignoring the $O[(JT)^3]$ corrections higher order in the Dyson expansion, $n$th-order decoupling implies that DD suppresses the effective noise strength by $n$ powers of $\omega T$ where $\omega$ is a characteristic bath frequency, rather than $n$ powers of $\epsilon T$ as in our previous analysis using the Magnus expansion.

\subsection{Universal decoupling}

To be concrete, consider the case of a single qubit with noise Hamiltonian
\begin{equation}
H= \Id\otimes B_0 + \sum_{\alpha=x,y,z} \sigma_\alpha\otimes B_\alpha.
\end{equation}
For a sequence of ideal zero-width Pauli operator pulses, the time-dependent system operator in the interaction-picture Hamiltonian becomes
\begin{equation}
\sigma_\alpha(t) = F_\alpha(t)\sigma_\alpha,
\end{equation}
where $F_\alpha(t) = \pm 1$ (${+}1$ if $\sigma_\alpha$ commutes with $U_c(t)$ and ${-}1$ if $\sigma_\alpha$ anticommutes with $U_c(t)$). For the universal decoupling sequence
\begin{equation}
U_c(t_{\rm DD})=Z\Id X\Id Z\Id X\Id,
\end{equation}
these functions are
\begin{eqnarray}
F_x &=& (++--),\nonumber\\
F_y &=& (+-+-),\nonumber\\
F_z &=& (+--+);
\end{eqnarray}
here, for example, $F_x= (++--)$ means that $F_x$ has the value ${+}1$ in the intervals $[0,\tau_0]$ and $[\tau_0,2\tau_0]$ and has the value ${-}1$ in the intervals $[2\tau_0,3\tau_0]$ and $[3\tau_0,4\tau_0]$. All three functions integrate to zero over the interval $[0,4\tau_0]$ and hence achieve first-order decoupling. Evaluating the Fourier transform
\begin{equation}
\tilde F_\alpha(\omega)= \int_0^{4\tau_0} dt~e^{-i\omega t}F_\alpha(t),
\end{equation}
we find
\begin{widetext}
\begin{eqnarray}\label{eq:f-alpha-integrals}
(-i\omega)\tilde F_x(\omega) &=& (x-1)(1+x-x^2-x^3)=-(x^2-1)^2=4e^{-2i\omega\tau_0}\sin^2(\omega\tau_0)=4(\omega\tau_0)^2 + \dots,\nonumber\\
(-i\omega)\tilde F_y(\omega) &=& (x-1)(1-x+x^2-x^3)=-(x-1)(x^4-1)/(x+1)\nonumber\\
&=&2e^{-2i\omega\tau_0}\tan(\omega\tau_0/2)\sin(2\omega\tau_0)=2(\omega\tau_0)^2+\dots,\nonumber\\
(-i\omega)\tilde F_z(\omega) &=& (x-1)(1-x-x^2+x^3)=(x-1)(x^2-1)^2/(x+1)\nonumber\\
&=&4ie^{-2i\omega\tau_0}\tan(\omega\tau_0/2)\sin^2(\omega\tau_0)=2i(\omega\tau_0)^3+\dots,\nonumber\\
\end{eqnarray}
\end{widetext}
where $x=e^{-i\omega \tau_0}$. The low-frequency suppression of $\tilde F_y(\omega)$ is stronger by a factor of 2 than the suppression of $F_x(\omega)$ because the period of $F_y(t)$ is shorter than the period of $F_x(t)$. The function $\tilde F_z(\omega)$ is suppressed by a further power of $\omega\tau_0$ because $F_z(t)$ is time-symmetric: $F_z(4\tau_0 -t)= F_z(t)$. Indeed, for any function $F(t)$ satisfying $F(T-t)=F(t)$, we have
\begin{eqnarray}
\tilde F(\omega) &=& \int_0^T dt~e^{-i\omega t}~F(t)=\int_0^T dt~e^{-i\omega t}~F(T-t)\nonumber\\
&=& \int_0^T dt~e^{-i\omega (T-t)}~F(t)= e^{-i\omega T} \tilde F(-\omega);
\end{eqnarray}
thus $\tilde F(\omega)=e^{-i\omega T/2}\tilde F_\text{even}(\omega)$, where $\tilde F_\text{even}(\omega)$ is an even function of $\omega$, and $\tilde F(\omega)= O(\omega^2)$ if $\tilde F(0)$ vanishes.

The time-symmetric pulse sequence
\begin{equation}
U_c(\tDD)=\Id X\Id Z\Id X\Id \Id X\Id Z\Id X\Id,
\end{equation}
achieves second-order decoupling because all three functions obey $F(t)=F(T-t)$:
\begin{eqnarray}
F_x &=& (++----++),\nonumber\\
F_y &=& (+-+--+-+),\nonumber\\
F_z &=& (+--++--+).
\end{eqnarray}
Compared to the four-pulse sequence, the functions $F_x$ and $F_y$ are repeated twice, but with a sign flip, so the Fourier transform is suppressed by an additional factor of $1-x^4=2ie^{-2i\omega\tau_0}\sin(2\omega\tau_0)\approx 4i(\omega\tau_0)$. The function $F_z$ is repeated without the sign flip, so its Fourier transform is multiplied by $1+x^4=2e^{-2i\omega\tau_0}\cos(2\omega\tau_0)\approx 2$. Therefore we have
\begin{eqnarray}
\tilde F_x(\omega) &=&-16\tau_0(\omega\tau_0)^2 + \dots,\nonumber\\
\tilde F_y(\omega) &=&-8\tau_0(\omega\tau_0)^2+\dots,\nonumber\\
\tilde F_z(\omega) &=&-4\tau_0(\omega\tau_0)^2+\dots.
\end{eqnarray}
Again, different types of low-frequency Pauli noise are suppressed by different (constant) factors, with the heaviest suppression for phase ({\em i.e.}, $\sigma_z$) noise. By altering the pulse sequence, the stronger suppression could be applied to $\sigma_x$ or $\sigma_y$ noise instead. 

\subsection{Finite-width pulses}
If the pulses are not ideal, then first-order decoupling will not be exact. For example if the pulses have nonzero width, then there is a contribution to $\bar\eta^2$ of the form
\begin{eqnarray}
&&\int \left(dt_1  dt_2\right)_{\rm PW} \sum_{\alpha,\beta}\left\langle S_\alpha(t_1)S_\beta(t_2) \right\rangle_S \left\langle B_\alpha(t_1) B_\beta(t_2) \right\rangle_B\nonumber\\
&=& \sum_{i,\alpha,\beta} J^2_{\alpha\beta,i} \langle \tilde S_{\alpha,{\rm PW}}(\omega_i)\tilde S_{\beta,{\rm PW}}(-\omega_i) \rangle_S;
\end{eqnarray} 
here $\int (dt)_{\rm PW}$ denotes integration over the nonzero-width pulses, and
\begin{equation}\label{eq:correlation-pulse-error}
\tilde S_{\alpha, {\rm PW}}(\omega) = \int(dt)_{\rm PW} ~e^{-i\omega t}~ S_\alpha(t).
\end{equation}
For a sequence of $N$ pulses, each with duration $\delta$, we expect 
$\tilde{S}_{\alpha,{\rm PW}}(\omega)\approx N\delta\|S_\alpha\|$
for $\omega t \ll 1$. Comparing with Eq.~(\ref{eq:FT-nth-order}), we conclude that for a pulse sequence that achieves $n$th-order decoupling in the ideal case, pulse-width corrections are small provided 
\begin{equation}
N\delta/T \ll (\omega T)^n
\end{equation}
where $\omega$ is a typical bath frequency. This is similar to the criterion we found using the Magnus expansion, except with the frequency $\omega$ now replacing the operator norm $\epsilon$.

For an Eulerian sequence with reproducible pulse errors, $\tilde S_{\alpha}(\omega)$ vanishes in the limit $\omega\to 0$ (by the same reasoning as in Sec.~\ref{subsec:eulerian}); therefore first-order decoupling is exact. Furthermore $\tilde S_{\alpha}(\omega)$ is an even function of $\omega$ for any time-symmetric pulse sequence, and therefore a time-symmetric Eulerian sequence achieves second-order decoupling.

\subsection{Gaussian noise}

We have seen that, while in our previous analysis we required $\beta = \max \|H_B\|$ to be small compared to $1/\tau_0$ in order to get a useful estimate of $\eta_{\rm DD}$, the analysis based on bath correlation functions can provide a useful estimate even if $\beta$ is large. However we still require that $J = \max \| H_{\err}\|$ is small to justify neglecting the higher-order corrections in the Dyson expansion in Eq.~(\ref{eq:eta-dyson2}). In some cases it is possible to go further and express these higher-order corrections in terms of correlation functions as well, thereby obtaining an estimate that makes sense even if the system qubits are coupled to bath operators with large norm ({\em e.g.}, the quadrature amplitudes of a bath of harmonic oscillators). 

Consider, for example, a single qubit coupled to bath operators whose correlators obey Gaussian statistics in the interaction picture: the interaction-picture Hamiltonian is 
\begin{equation}
\tilde H(t) = \sum_\alpha \sigma_\alpha(t)\otimes B_\alpha(t), 
\end{equation}
where the expectation value of an odd number of bath operators vanishes, and the expectation of an even number of bath operators is
\begin{eqnarray}\label{eq:gaussian-def}
&& \left\langle B(1)B(2) \cdots B(2n)\right\rangle \nonumber\\
&=& \sum_{\rm contractions} K(i_1,i_2)K(i_3,i_4)\cdots
  K(i_{2n-1},i_{2n}).\notag \\
\end{eqnarray}
Here the sum is over the $(2n)!/2^n n!$ ways to divide the labels $1,2, \dots 2n$ into $n$ unordered pairs, and we use the shorthand $B(i)=B_{\alpha_i}(t_i)$, $K(i,j) = \langle B(i)B(j)\rangle_B$. Thus terms of odd order in the Dyson expansion for $\bar\eta^2$ vanish, and 
 we may bound the ($2n$)-th order term as
\begin{widetext}
\begin{eqnarray}
\label{eq:gaussian-bound}
&&\left| \left\langle \frac{1}{(2n)!}\int_0^T dt_1\dots dt_{2n}{\cal T}\left(\tilde H(t_1) \cdots \tilde H(t_{2n})\right)\right\rangle\right| \notag \\
&&\le \frac{1}{(2n)!}\int_0^T dt_1\dots dt_{2n}\sum_{\alpha_1, \dots \alpha_{2n}}\sum_{\rm contractions}\left|K(i_1,i_2)\cdots K(i_{2n-1},i_{2n})\right|\nonumber\\
&=& \frac{1}{(2n)!}\sum_{\rm contractions} \left(2K\right)^{n} = \frac{K^n}{n!},\quad {\rm where}\quad K = \frac{1}{2}\int_0^T dtds \sum_{\alpha,\beta}\left|\left\langle B_\alpha(t)B_\beta(s)\right\rangle_B\right|.
\end{eqnarray}
\end{widetext}
To derive Eq.~(\ref{eq:gaussian-bound}), we use Eq.~(\ref{eq:gaussian-def}) and $\|\sigma_\alpha(t)\|=1$, and we note that the value of $\left|K(i,j)\right|$ does not depend on the time ordering of $t_i$ and $t_j$. We conclude that, in the case of Gaussian noise, the sum of all corrections higher than quadratic order in the Dyson expansion can be bounded above by
\begin{equation}
\sum_{n=2}^\infty K^n/n!= e^K -1-K,
\end{equation}
and that the quadratic term provides a good approximation to the effective noise strength for $K$ sufficiently small. 

\blue{\subsection{Nonuniformly spaced pulses}}

Another approach to analyzing DD is to use the Dyson expansion and to also expand $B_\alpha(t)$ in powers of $B_0 t$, thus obtaining a double expansion in powers of $JT$ and $\beta T$. In that case we might say that ``$n$th-order decoupling'' is achieved if, in the expression for the interaction-picture evolution operator $\tilde U(T)$, all terms of order $T^m$ are pure-bath terms for $m=1,2, \dots, n$. For the case of a qubit subject to pure dephasing noise ($B_1=B_2=0$), it is shown in \cite{Uhrig,Yang} that in this sense $n$th-order decoupling can be achieved by a sequence of $X$ pulses with $n+1$ pulse intervals, where the pulses are nonuniformly spaced in time. For general 
\green{single-qubit}
noise, $n$th-order decoupling can be achieved by a sequence of nonuniformly spaced $X$ and $Z$ pulses with altogether $(n+1)^2$ pulse intervals \cite{WFL:09}%
\blue{, and for general $m$-qubit noise, $(n+1)^{2m}$ pulse intervals suffice \cite{NUDD}.}

The corrections higher-order in $T$ are not necessarily small unless both $\beta T\ll 1$ and $JT\ll 1$ are satisfied. However, the ideal pulse sequence constructed in \cite{WFL:09} has the property
\begin{equation}
\int_0^T dt ~t^m F_\alpha(t) =0, \quad m=0,1,\dots n-1,\quad \alpha=x,y,z.
\end{equation}
(The sequence in \cite{Uhrig,Yang} has this property only for $\alpha=z$.) Therefore, even if $\beta T$ is not small, we can use the correlation function analysis to show that same sequence also achieves $n$th-order decoupling in the sense of Eq.~(\ref{eq:FT-nth-order}). Therefore DD works effectively if $\omega T \ll 1$, where $\omega$ is a typical bath frequency, provided that either $JT \ll 1$ or (in the case of Gaussian noise) $K \ll 1$. The same remark applies to pure dephasing noise for the pulse sequence in \cite{Uhrig,Yang}.
\newline
\newline

\subsection{Concatenated dynamical decoupling}
\label{subsec:dyson-concatenated}

Instead of using the Magnus expansion, we can analyze the performance
of concatenated DD sequences using the Dyson expansion and bath
correlation functions. As in Sec.~\ref{sec:Dyson}, we will
suppose that the higher-order terms in the Dyson expansion can be
neglected, and will focus on the lowest-order term
Eq.~(\ref{eq:eta-S-FT}). The objective is to show that, by
concatenating $k$ times a pulse sequence that achieves first-order
decoupling, $k$-th order decoupling can be achieved, in the sense that
$\tilde{S}_\alpha(\omega)=O[(\omega T)^k]$.

To illustrate the idea, consider the simple pulse sequence that decouples pure-dephasing noise for a single qubit:
\begin{equation}
U_c(t_{\rm DD})= X\Id X \Id,
\end{equation}
so that the ``level-1'' function multiplying $\sigma_z$ in the interaction picture can be represented as
\begin{equation}
F_z^{(1)}= (+-).
\end{equation}
When we concatenate the pulse sequence, $F_z^{(1)}$ is replaced by $F_z^{(2)}$, in which $F_z^{(1)}$ is repeated twice, but with a sign flip in the second repetition:
\begin{equation}
F_z^{(2)}= (+--+),
\end{equation}
and for higher-level sequences we have
\begin{eqnarray}
F_z^{(3)}&=& (+--+-++-),\nonumber\\
F_z^{(4)}&=& (+--+-++--++-+--+),\nonumber\\
\end{eqnarray}
etc. Evaluating the Fourier transforms of these functions, 
\begin{equation}
\tilde F_z^{(k)}(\omega) = \int_0^{2^k\tau_0}dt~e^{-i\omega t}F_z^{(k)}(t),
\end{equation}
we see that
\begin{eqnarray}
\tilde F_z^{(1)}(\omega) &=& (-i\omega)^{-1}(x-1)(1-x)
\end{eqnarray}
and 
\begin{eqnarray}
\tilde F_z^{(k)}(\omega) &=& \left(1- x^{2^{k{-}1}}\right)\tilde F_z^{(k{-}1)}(\omega),
\end{eqnarray}
where $x=e^{-i\omega\tau_0}$, and hence
\begin{eqnarray}
\tilde F_z^{(n)}(\omega) &=& (-i\omega)^{-1}x^{1/2}\left(x^{1/2}-x^{-1/2}\right)\nonumber\\
&&\times\prod_{k=1}^n x^{2^{k{-}2}}\left(x^{-2^{k{-}2}}-x^{2^{k{-}2}}\right)\nonumber\\
&=& 2\omega^{-1}x^{1/2}\sin(\omega\tau_0/2)\nonumber\\
&&\times\prod_{k=1}^n x^{2^{k{-}2}}\left(2i\sin\left(2^{k{-}2}\omega\tau_0\right)\right).
\end{eqnarray}
The leading behavior of this function for small $\omega\tau_0$ is
\begin{equation}
\tilde F_z^{(n)}(\omega) = \tau_0(i)^n2^{n(n-1)/2}\left(\omega\tau_0\right)^n+ \cdots,
\end{equation}
and therefore Eq.~(\ref{eq:eta-S-FT}) becomes
\begin{equation}
\eta_{\rm DD}^{(n)} \le 2^{n(n-1)/2}\left(\sum_{i} \left(J^2_{33,i}\tau_0^2\right) \left(\omega_i\tau_0\right)^{2n} \right)^{1/2}+\cdots,
\end{equation}
where we neglect corrections both higher order in the Dyson expansion and higher order in frequency. Naively, this expression for the effective noise strength $\eta_{\rm DD}$ is optimized by choosing the level of concatenation $n$ to be the largest integer such that 
$2^{n-1}\left(\omega\tau_0\right) < 1$
where $\omega$ is a ``typical'' bath frequency. Note, however, that for $2^{n}\left(\omega\tau_0\right) \approx 1$ the  higher-order corrections in $(\omega\tau_0)$ modify $\eta_{\rm DD}$ by an $O(1)$ multiplicative factor. Note also that $2^n\tau_0=T^{(n)}$ is the duration of the level-$n$ pulse sequence, and thus the optimal pulse sequence has duration comparable to a typical inverse frequency of the bath.

Other concatenated pulse sequences can be studied similarly. Consider for example the universal DD sequence. We have seen in Eq.~(\ref{eq:f-alpha-integrals}) that this sequence suppresses noise asymmetrically (the best suppression for $\sigma_z$, the worst for $\sigma_x$), so we might choose to alter the sequence at higher levels to provide more balanced noise suppression. But if we do not do that, the functions $\tilde F_\alpha^{(k)}(\omega)$ can be specified by augmenting Eq.~(\ref{eq:f-alpha-integrals}) with
\begin{widetext}
\begin{eqnarray}\label{eq:f-alpha-level-k}
\tilde F_x^{(k)}(\omega) &=& \left(1+x^{4^{k-1}}-\left(x^{4^{k-1}}\right)^2-\left(x^{4^{k-1}}\right)^3\right)\tilde F_x^{(k-1)}(\omega)\nonumber\\
&=&\left(x^{4^{k-1}}\right)^{3/2}(4i)\cos\left(4^{k-1}(\omega\tau_0)/2\right)\sin\left(4^{k-1}(\omega\tau_0)\right)\tilde F_x^{(k-1)}(\omega),\nonumber\\
\tilde F_y^{(k)}(\omega) &=& \left(1-x^{4^{k-1}}+\left(x^{4^{k-1}}\right)^2-\left(x^{4^{k-1}}\right)^3\right)\tilde F_y^{(k-1)}(\omega)=\left(x^{4^{k-1}}\right)^{3/2}\frac{(i)\sin\left(2\cdot 4^{k-1}(\omega\tau_0)\right)}{\cos\left(4^{k-1}(\omega\tau_0)/2\right)}\tilde F_x^{(k-1)}(\omega),\nonumber\\
\tilde F_z^{(k)}(\omega) &=& \left(1-x^{4^{k-1}}-\left(x^{4^{k-1}}\right)^2+\left(x^{4^{k-1}}\right)^3\right)\tilde F_z^{(k-1)}(\omega)=\left(x^{4^{k-1}}\right)^{3/2}\frac{(-2)\sin^2\left(4^{k-1}(\omega\tau_0)\right)}{\cos\left(4^{k-1}(\omega\tau_0)/2\right)}\tilde F_x^{(k-1)}(\omega),\nonumber\\
\end{eqnarray}
\end{widetext}
The weakest suppression of low-frequency noise occurs for $\tilde F_x^{(k)}(\omega)$, where
\begin{equation}
\tilde F_x^{(k)}(\omega) =\left( 4^k \left(i \omega\tau_0\right) + \cdots\right) \tilde F_x^{(k{-}1)}(\omega),
\end{equation}
and hence
\begin{eqnarray}
\tilde F_x^{(k)}(\omega) &=&\tau_0 \prod_{k=1}^n \left(4^k i \omega\tau_0\right) +\cdots\nonumber\\
&=&\tau_0(i)^n 4^{n(n+1)/2}\left(\omega\tau_0\right)^n +\cdots,
\end{eqnarray}
where we neglect the terms higher order in $\omega\tau_0$. From Eq.~(\ref{eq:eta-S-FT}) we obtain the estimate of the noise strength
\begin{eqnarray}\label{eta-level-n-correlator}
\eta_{\rm DD}^{(n)}\approx 4^{n(n+1)/2}\left(\sum_{i,\alpha,\beta} \left(J^2_{\alpha\beta,i}\tau_0^2\right)\left(\omega_i\tau_0\right)^{2n}\right)^{1/2}+\cdots.\nonumber\\
\end{eqnarray}
Noting that the universal DD sequence has length $R=4$, we see that Eq.~(\ref{eta-level-n-correlator}) resembles Eq.~(\ref{eq:eta-level-k}), but with the operator norm $\epsilon$ replaced by a bath frequency.

\section{Conclusions}\label{sec:DDFTConc}

We have derived upper bounds on the effective noise strength $\etaDD$
for DD-protected quantum gates, in terms of the parameters of a
Hamiltonian noise model. From the upper bounds on the noise strength
we can extract a noise suppression threshold condition, a sufficient
condition for DD-protected gates to outperform unprotected gates. We
can also derive an accuracy threshold condition; when the noise
parameters obey this condition, scalable quantum computing is
possible. Our results show that DD, and in particular concatenated DD, can improve the
gate accuracy and overhead cost of fault-tolerant quantum computing. 

\green{
Dynamical decoupling works when the noise varies slowly on a time scale determined by the pulse sequence. Therefore, estimates of the achievable effective noise strength depend on parameters quantifying the speed of the bath dynamics.
We have used two different methods to quantify the accuracy of DD-protected gates, appropriate for two different ways of characterizing the time variation of the noise. From the Magnus expansion in the toggling frame we derived an expression for $\etaDD$ in terms of the operator norm of the noise Hamiltonian; an advantage of this method is that $\etaDD$ does not depend on the state of the bath. From the Dyson expansion in the interaction picture we derived an expression for $\etaDD$ in terms of the frequency spectrum of bath correlations. While the bath frequency spectrum does depend on the state of the bath, the second method sometimes yields useful result when the first method fails, because the norm $\beta = \|H_{B,a}\|$ of the local bath Hamiltonian is too large. Our correlation function analysis can remain applicable even in the formal limit $\beta\rightarrow \infty$.
}
%



Our analysis of fault-tolerant circuits built from DD-protected gates applies only to Hamiltonian noise models satisfying suitable assumptions. For the Magnus expansion analysis we used the local-bath model; this allows us to study each DD-protected gate individually, ignoring noise correlations among distinct gates being executed in parallel at the same time. For the correlation function analysis we used an even more artificial model, in which the state of the bath is refreshed after each DD-protected gate. This assumption allows us to include non-Markovian effects during the DD pulse sequence at each protected gate, but to ignore these effects when the DD-protected gates are composed in a quantum circuit. It is clearly desirable to extend our analysis to models with more general noise correlations.

Here we have proposed to combine DD with fault-tolerant quantum computing straightforwardly, by replacing each gate in a fault-tolerant circuit by the corresponding DD-protected gate. 
\green{
We have not studied systematically the improvements in fault tolerance that might be achieved using Eulerian dynamically corrected gates \cite{KV09,KV09a,KLV09} which are robust against pulse imperfections. Nor have we considered the potential advantages of qubit encodings that allow gates and DD pulses to commute, so that both can be applied simultaneously.}
\blue{This \green{latter} strategy has been shown numerically to lead to robust gates for a spin bath model \cite{West:10}.}
Perhaps 
\green{other} ways to combine DD with fault tolerance can be found, leading to further gains in efficiency and accuracy. 
%

\acknowledgments
Research of HKN and JP is supported by NSF under Grant No. PHY-0803371. JP's research is also supported by DOE under Grant No. DE-FG03-92-ER40701, and by NSA/ARO under Grant No. W911NF-09-1-0442. DAL thanks the Institute for Quantum Information at Caltech, where this work was done, and acknowledges funding from the US Department of Defense, NSF PHY-803304, NSF PHY-802678, and NSF CCF-726439. We thank Kurt Litsch for doing a numerical analysis of the recursion relations in Sec.~\ref{subsec:magnus-concatenated}, and for suggesting ways to improve some of our arguments.

\appendix

\section{Review of the Magnus expansion}
\label{app:magnus}
Here we will briefly review some properties of the Magnus expansion that are used in our arguments. For a more detailed discussion, see \cite{Blanes08}.

The foundation of the Magnus expansion is this theorem:

\begin{theorem}
Suppose
\begin{equation}
\frac{d}{dt}e^{\Omega(t)}= M(t) e^{\Omega(t)}.
\end{equation}
Then 
\begin{equation}\label{eq:magnus-theorem}
\frac{d}{dt}\Omega(t)= \sum_{n=0}^\infty \frac{B_n}{n!} ~{\rm ad}^n_{\Omega(t)}[M(t)].
\end{equation}
Here the $\{B_n\}$ are the Bernoulli numbers defined by 
\begin{equation}
\frac{x}{e^x-1}=\sum_{n=0}^\infty\frac{B_n}{n!}x^n,
\end{equation}
and ${\rm ad}^n_B[A]$ is defined by
\begin{equation}
{\rm ad}^n_B[A]\equiv [B,[B,[\cdots [B,[B,A]]\cdots ]]]
\end{equation}
(${\rm ad}^0_B[A]=A$, and ${\rm ad}^n_B[A]$ for $n\ge 1$ contains $n$
nested commutators). The series converges provided $\|\Omega(t)\| <
\pi$.
\end{theorem}

\begin{proof}
To obtain a useful expression for $M(t)= \left(\frac{d}{dt} e^{\Omega(t)}\right)e^{-\Omega(t)}$, we first evaluate
\begin{eqnarray}
&&\frac{d}{d \lambda}\left[\frac{d}{d t}\left(e^{\lambda\Omega(t)}\right)e^{-\lambda\Omega(t)}\right]\nonumber\\
&=&\left[\frac{d}{dt}\left(\frac{d}{d\lambda}e^{\lambda \Omega}\right)e^{-\lambda\Omega(t)}\right]
-\left(\frac{d}{dt}e^{\lambda\Omega(t)}\right)\Omega(t)e^{-\lambda \Omega(t)}\nonumber\\
&=&\left[\frac{d}{dt}\left(e^{\lambda \Omega(t)}\Omega(t)\right)e^{-\lambda\Omega(t)}\right]
-\left(\frac{d}{dt}e^{\lambda\Omega(t)}\right)\Omega(t)e^{-\lambda \Omega(t)}\nonumber\\
&=& e^{\lambda \Omega(t)}\left(\frac{d}{dt}\Omega(t)\right) e^{-\lambda\Omega(t)}\nonumber\\
&=& \sum_{n=0}^\infty \frac{\lambda^n}{n!} {\rm ad}^n_{\Omega(t)}\left[\frac{d}{dt}\Omega(t)\right].
\end{eqnarray}
In the last line we have used the identity
\begin{eqnarray}
e^{\lambda B} A e^{-\lambda B} = \sum_{n=0}^{\infty} \frac{\lambda^n}{n!} ~{\rm ad}^n_B[A],
\end{eqnarray}
which can be verified by differentiating both sides $k$ times with respect to $\lambda$ and then setting $\lambda=0$.
Expressing $M(t)$ as the integral of its derivative, we find
\begin{eqnarray}
M(t)&=& \int_0^1 d\lambda \frac{d}{d \lambda}\left[\frac{d}{d t}\left(e^{\lambda\Omega(t)}\right)e^{-\lambda\Omega(t)}\right]\nonumber\\
&=& \int_0^1 d\lambda \sum_{n=0}^\infty \frac{\lambda^n}{n!} {\rm ad}^n_{\Omega(t)}\left[\frac{d}{dt}\Omega(t)\right]\nonumber\\
&=&\sum_{n=0}^\infty \frac{1}{(n+1)!} {\rm ad}^n_{\Omega(t)}\left[\frac{d}{dt}\Omega(t)\right].
\end{eqnarray}

Thus we have shown that $M(t)={\cal O}_{{\rm ad}_{\Omega(t)}}\left[\frac{d}{dt}\Omega(t)\right]$, where
\begin{equation}
{\cal O}_A= \sum_{n=0}^\infty \frac{1}{(n+1)!}A^n= \frac{e^A-1}{A}, 
\end{equation}
which is inverted by
\begin{equation}
{\cal O}_A^{-1}= \left(\frac{e^A-1}{A}\right)^{-1}= \sum_{n=0}^\infty \frac{B_n}{n!} A^n. 
\end{equation}
Therefore, 
\begin{equation}
\frac{d}{dt}\Omega(t) = {\cal O}_{{\rm ad}_{\Omega(t)}}^{-1}\left[M(t)\right], 
\end{equation}
from which Eq.~(\ref{eq:magnus-theorem}) follows.

Regarding the convergence of the expansion, we note that $\|{\rm ad}_B^n\| \le \left(2 \| B\|\right)^n$, and that the series expansion of $x/(e^x-1)$ converges for $|x| < 2\pi$, because the nearest poles to the origin in the complex $x$-plane are at $x=\pm 2\pi i$. Therefore the expansion in Eq.~(\ref{eq:magnus-theorem}) converges for $\|2\Omega(t)\|< 2\pi$.

\end{proof}

In the Magnus expansion, we express $\Omega(t)=\sum_{n=1}^\infty \Omega_n(t)$, where $\Omega_n(t)$ is $n$th-order in $M$. Using this expansion, Eq.~(\ref{eq:magnus-theorem}) becomes
\begin{eqnarray}\label{eq:omega-dt-S}
\frac{d}{dt}\Omega_1(t) &=& M(t),\nonumber\\
\frac{d}{dt}\Omega_n(t) &=& \sum_{j=1}^{n-1}\frac{B_j}{j!} S_n^{(j)}(t),\quad n\ge 2,
\end{eqnarray}
where
\begin{eqnarray}
S_n^{(j)}(t)=\sum_{i_1,i_2,\dots, i_j}^{(n-1)}{\rm ad}_{\Omega_{i_{1}}(t)}{\rm ad}_{\Omega_{i_{2}}(t)}\cdots{\rm ad}_{\Omega_{i_{j}}(t)}\left[M(t)\right];\nonumber\\
\end{eqnarray}
here the sum is over nonnegative integers $\{i_1,i_2,\dots ,i_j\}$ satisfying $i_1+i_2+\cdots +i_j=n-1$. We see that
\begin{equation}\label{eq:S_n^1}
S_n^{(1)}(t)=[\Omega_{n-1}(t),M(t)],
\end{equation}
and that $S_n^{(j)}$ for $j>1$ can be expressed as
\begin{equation}\label{eq:S_n^j}
S_n^{(j)}=\sum_{m=1}^{n-j}\left[\Omega_m(t),S_{n-m}^{(j-1)}(t)\right], \quad 2\le j\le n-1.
\end{equation}
The relations Eq.~(\ref{eq:omega-dt-S}), (\ref{eq:S_n^1}), (\ref{eq:S_n^j}) provide an algorithm for generating the terms in the Magnus expansion recursively, and we use these recursion relations to derive our upper bounds on the higher-order terms.

\section{Even Magnus terms vanish for a time-symmetric Hamiltonian}
\label{app:Sym}

Here we prove the fact that, if $H_M(t)$ is time-symmetric, all even Magnus terms vanish. This was previously known in the NMR literature, at least for the case of a piecewise constant Hamiltonian \cite{Wang72}.
\begin{lemma}
If $H_M(T-t)=H_M(t)$, then $\Omega_n(T)=0$ for all even $n$.
\end{lemma}
\begin{proof}
First we show that $\Omega(T)$ is an odd function in $A(t)=-iH_M(t)$ when $H_M(t)$ (or correspondingly $A(t)$) is time-symmetric about $T/2$. Defining $\Delta_N\equiv T/2N$ for $N$ a positive integer, the evolution operator from $t=0$ to $t=T$ can be written as
\begin{align}
U(T,0)&=\lim_{N\rightarrow\infty}e^{A(T)\Delta_N}e^{A(T-\Delta_N)\Delta_N}\cdots e^{A(\frac{T}{2}+\Delta_N)\Delta_N}\notag\\
&\qquad\times e^{A(\frac{T}{2}-\Delta_N)\Delta_N}\ldots e^{A(\Delta_N)\Delta_N}e^{A(0)\Delta_N}\nonumber\\
&=\lim_{N\rightarrow\infty}e^{A(0)\Delta_N}e^{A(\Delta_N)\Delta_N}\cdots e^{A(\frac{T}{2}-\Delta_N)\Delta_N}\notag\\
\label{eq:U-limit}&\qquad\times e^{A(\frac{T}{2}-\Delta_N)\Delta_N}\cdots e^{A(\Delta_N)\Delta_N}e^{A(0)\Delta_N},
\end{align}
where in the second equality, we have used the time-symmetry $A(T-t)=A(t)$. Taking the adjoint of Eq.~(\ref{eq:U-limit}), and noting that $A(t)^\dagger = -A(t)$, we find
\begin{align}
U^\dagger(T,0)&=\lim_{N\rightarrow\infty}e^{-A(0)\Delta_N}e^{-A(\Delta_N)\Delta_N}\cdots e^{-A(\frac{T}{2}-\Delta_N)\Delta_N}\nonumber\\
&~\qquad\times e^{-A(\frac{T}{2}-\Delta_N)\Delta_N}\cdots e^{-A(\Delta_N)\Delta_N}e^{-A(0)\Delta_N}.
\end{align}
Thus $U^\dagger(T,0)$ has the same form as $U(T,0)$, except for the replacement $A(t)\rightarrow -A(t)$. 

Since $U(T,0)=\exp\left(\Omega(T)\right)$ and $U^\dagger(T,0) = \exp\left(-\Omega(T)\right)$, we conclude that under the replacement $A(t)\rightarrow -A(t)$, $\Omega(T)$ transforms as $\Omega(T)\rightarrow -\Omega(T)+i2\pi \ell$, for some integer $\ell$. In fact, since the integer $\ell$ cannot jump discontinuously when $A(t)$ is smoothly deformed, $\ell$ must be a constant independent of $A(t)$, and by taking the limit $A(t) \to 0$ we see that $\ell=0$; thus $\Omega(T)$ changes sign under $A(t)\rightarrow -A(t)$, i.e., is an odd function of $A(t)$.

In general, $\Omega_n(T)$ is an integral of an expression containing $n$ factors of $A(t)$. Thus, $\Omega_n(T)$ is invariant under the replacement $A(t)\rightarrow -A(t)$ for $n$ even, and changes sign under this replacement for $n$ odd. Since in the time-symmetric case $\Omega(T)$ changes sign under $A(t)\rightarrow -A(t)$, we conclude that $\Omega_n(T)$ vanishes for $n$ even.
\end{proof}

\section{Error estimate for time evolution}
\label{app:error}

Here we prove:

\begin{lemma}\label{lem:error}
Suppose that the time evolution operator $U(t)$ satisfies the differential equation
\begin{equation}
\frac{d}{dt} U(t) = -i H(t) U(t)
\end{equation}
with the initial condition $U(t_0)=U_0$, while $\tilde U(t)$ satisfies
\begin{equation}
\frac{d}{dt} \tilde U(t) = -i \tilde H(t) \tilde U(t)
\end{equation}
with the same initial condition, where both $H(t)$ and $\tilde H(t)$ are Hermitian. Then
\begin{eqnarray}
\| \tilde U(t) - U(t)\| \le \int_{t_0}^t ds \| \tilde H(s) - H(s)\|.
\end{eqnarray}
\end{lemma}

\begin{proof}
\begin{eqnarray}
&&\| \tilde U(t) - U(t)\|= \| \tilde U(t)U(t)^{-1}-\Id\|\nonumber\\
&=& \left\| \int_{t_0}^t  ds \frac{d}{ds}\left( \tilde U(s)U(s)^{-1}\right)\right\|\nonumber\\
&=& \left\| -i\int_{t_0}^t  ds  ~\tilde U(s)\left(\tilde H(s)-H(s)\right)U(s)^{-1}\right\|\nonumber\\
&\le&  \int_{t_0}^t  ds  ~\left\|\tilde U(s)\left(\tilde H(s)-H(s)\right)U(s)^{-1}\right\|\nonumber\\
&\le&  \int_{t_0}^t  ds  ~\left\|\tilde H(s)-H(s)\right\|.
\end{eqnarray}
\end{proof}

In this paper, we use Lemma \ref{lem:error} in three ways. In one application, we consider the case where both Hamiltonians are time independent, and conclude that (compare with Eq. \eqref{eq:Udiff})
\begin{equation}
\Vert\tilde U(t)-U(t)\Vert\leq(t-t_0)\Vert\tilde H-H\Vert.
\end{equation}
This inequality allows us to relate the effective noise strength $\eta_{\rm DD}$ achieved by dynamical decoupling to our bounds on the terms in the Magnus expansion.

In another application, we consider $H$ to be $H_S+H_B$, where $H_S$ governs the ideal system dynamics and $H_B$ governs the bath dynamics, while the Hamiltonian for the noisy joint evolution of system and bath is $\tilde H = H + H_{SB}$, where $H_{SB}$ is responsible for the noise. Then in the local-bath model, if $\|H_{SB}\| \le J$ and a gate is executed in time $\tau_0$, Lemma \ref{lem:error} implies that the norm of the ``bad'' part of the gate is bounded above by $J\tau_0$. Thus we may estimate the effective noise strength in the absence of DD as $\eta=J\tau_0$, as in Eq.~(\ref{eq:eta-unprotected}).

In the third application, we use Lemma \ref{lem:error} to estimate the
error arising from pulses with nonzero width. We consider $\tilde
H(t)$ to be the Hamiltonian describing the actual DD sequence with
realistic pulses, and $H(t)$ to be the idealized evolution for
zero-width pulses, where both Hamiltonians are expressed in the toggling frame determined by the ideal sequence. Suppose that there are $R$ pulses, and that each
realistic pulse has support in a time interval of width $\delta$. Both $\tilde H(t)$ and $H(t)$ can be expressed as a sum of a bath Hamiltonian and an error Hamiltonian; the bath Hamiltonian cancels in the difference $\tilde H(t) - H(t)$, and we suppose that for both the realistic and ideal sequences the norm of the error Hamiltonian is bounded above by $J$ during the  pulses. Thus $\|\tilde H(t) - H(t)\| \le 2J$ during the pulses (a total duration of $R\delta$), while $\tilde H(t)=H(t)$ outside the pulses; thus Lemma \ref{lem:error} implies $\|\tilde U(T)-U(T)\| \le 2R\delta J$, as in Eq.~(\ref{eq:pulse-error-correction}).

\section{Bounds for even Magnus terms in the time-symmetric case}
\label{app:Even}

We want to generalize the argument used to compute the bound for $\Omega_2(T)$ in the case where $H_M(t)$ is time-symmetric except for $t\in \Delta$. To do this for higher-order terms requires a formula for the Magnus terms for which all the multiple time-integrals are explicit. Such a formula can be found in \cite{SS01} (for $n\geq 2$):
\begin{align}
\Omega_n(T)&=\frac{1}{n}\int_0^Tdt_1\ldots\int_0^Tdt_nL_n\notag\\
&\hspace{1cm}\times[[\ldots [A(t_1),A(t_2)],\ldots],A(t_n)]
\end{align}
where
\begin{align}
L_n&\equiv \sum_{l=1}^{n-1}\frac{1}{l}(-1)^{l+1}\sum_{1\leq j_1<\ldots<j_{n-l}<n}\prod_{m=1}^{n-l}\Theta(j_m,j_m+1).
\end{align}
The $L_n$ coefficients take care of the time-ordering and relabeling of the integration variables. For $n$ even, following what we did in the $\Omega_2(T)$ case, we split up the $n$ time-integrals into $n$ different cases: (1) none of $t_i, i=1,\ldots n$ are in $\Delta$, (2) exactly one of $t_i\in\Delta$, (3) exactly two of $t_i\in\Delta$, $\ldots$, (n) exactly $n$ of $t_i\in\Delta$. Case (1) is zero from the time symmetry of $H_M(t)$ for $t\notin\Delta$; the remaining cases we bound by first bounding the nested commutator and $L_n$, and then doing the time-integral.

The $(n-1)$-nested commutator can be bounded as
\begin{align}
&\quad\left\Vert [[\ldots [A(t_1),A(t_2)],\ldots],A(t_n)]\right\Vert\notag\\
&\leq 2^{n-2}\left\Vert[A(t_1),A(t_2)]\right\Vert\Vert A(t_3)\Vert\ldots \Vert A(t_n)\Vert\nonumber\\
&\leq 2^{n-2}(4J\epsilon)\epsilon^{n-2}\nonumber\\
\label{eq:nestedCom}&=2^n J\epsilon^{n-1}.
\end{align}
The $2^{n-2}$ factor in the first line comes from opening up $(n-2)$-nested commutators using submultiplicativity of the operator norm. The $(4J\epsilon)$ factor in the second line is an upper bound on $\left\Vert [A(t_1),A(t_2)]\right\Vert$. The coefficient $L_n$ can be bounded by ignoring the step function (i.e. ignoring the time-ordering, since we do not have the details of $\Delta$ anyway):
\begin{align}
\vert L_n\vert &\leq \sum_{l=1}^{n-1}\frac{1}{l}\sum_{1\leq j_1<j_2<\ldots<j_{n-l}<n}1\notag\\
&=\sum_{l=1}^{n-1}\frac{1}{l}\binom{n-1}{n-l}=\sum_{l=1}^{n-1}\frac{1}{n-l}\binom{n-1}{l}.
\end{align}
The binomial factor arises from counting the number of terms in the sum over $j_i$: we pick $n-l$ elements from the numbers $1$ to $n-1$, and arranging them in ascending order gives a single choice of $(j_1,j_2,\ldots,j_{n-l})$ and hence a single term in the sum. The number of ways of choosing $n-l$ elements from $n-1$ distinct numbers is given by the binomial factor. To bound the remaining sum, consider
\begin{align}
\int_0^1dx (1+x)^{n-1}&=\sum_{l=0}^{n-1}\binom{n-1}{l}\left.\frac{1}{n-l}x^{n-l}\right\vert_{x=0}^{x=1}\notag\\
&=\sum_{l=0}^{n-1}\binom{n-1}{l}\frac{1}{n-l}.
\end{align}
Therefore, we have that
\begin{align}
\vert L_n\vert&\leq \int_0^1dx(1+x)^{n-1}-\binom{n-1}{0}\frac{1}{n}\notag\\
&=\frac{2}{n}\left(2^{n-1}-1\right).
\end{align}

Putting these back in $\Omega_n(T)$ ($n$ even) and doing the time-integrals, we find that
\begin{align}
&\quad \left\Vert\Omega_n(T)\right\Vert\nonumber\\
&\leq \frac{2}{n^2}\left(2^{n-1}-1\right)\left(2^nJ\epsilon^{n-1}\right)\left[\binom{n}{1}\Delta (T-\Delta)^{n-1}\right.\notag\\
&\hspace{1cm}\left.+\binom{n}{2}\Delta^2(T-\Delta)^{n-2}+\ldots+\binom{n}{n}\Delta^n\right]\nonumber\\
\label{eq:Even}&=\frac{2^{n+1}J\epsilon^{n-1}}{n^2}\left(2^{n-1}-1\right)\left[T^n-(T-\Delta)^n\right].
\end{align}
In the first inequality above, the terms in the brackets are the $n-1$ cases for choosing the times $t_1,\ldots,t_n$, with at least one being in $\Delta$. 

One can check that Eq.~(\ref{eq:Even}) agrees with Eq.~(\ref{eq:omega-4-even-bound}) in the $n=4$ case. We also see that, for each $n$, $\Omega_n(T)$ is of order $\Delta T^{n-1}$, and thus vanishes in the limit $\Delta\rightarrow 0$.

\section{\texorpdfstring{$S_n^{(j)}$}{Snj} coefficients}
\label{app:snj}

Here we derive bounds on the $\snj$ coefficients found in the recursive formulas (Eqs.~\eqref{eq:A} -- \eqref{eq:snj}) for the Magnus terms.
\begin{lemma}\label{prop:snj}
For all $n\geq 2$, $1\leq j\leq n-1$, 
\begin{equation}
\Vert \snj(t)\Vert\leq \fnj J\left(2\epsilon t\right)^{n-1},
\end{equation}
where the coefficients are defined recursively:
\begin{subequations}\label{eq:fnj}
\begin{align}
f_1^{(0)}&=1,\quad f_n^{(0)}=0,\quad n>1,\\
\fnj&=2\sum_{m=1}^{n-j}\sum_{p=0}^{m-1}\frac{|B_p|}{p!m}f_m^{(p)}f_{n-m}^{(j-1)},\quad n\geq 2;
\end{align}
\end{subequations}
here the $\{B_p\}$ are the Bernoulli numbers, defined by
\begin{equation}\label{eq:Bernoulli}
\frac{x}{e^x-1}=\sum_{p=0}^\infty \frac{B_p}{p!} ~x^p.
\end{equation}
\end{lemma}
\begin{proof}
We will prove the lemma by induction. We begin with the smallest case where $n=2, j=1$:
\begin{align}
\Vert S_2^{(1)}(t)\Vert&=\Vert[\Omega_1(t),-iH_M(t)]\Vert\notag\\
&\leq \int_0^tds~\Vert\left[H_M(s),H_M(t)\right]\Vert.
\end{align}
The commutator can be bounded as $\left\Vert\left[H_M(s),H_M(t)\right]\right\Vert\leq 4J\epsilon$. This thus gives $\Vert S_2^{(1)}(t)\Vert\leq 4J\epsilon t$. Since $f_2^{(1)}=2$, this can be rewritten as $\Vert S_2^{(1)}\Vert\leq 4J\epsilon t=f_2^{(1)}J(2\epsilon t)$.

For a given $n\geq 3$, suppose that the lemma holds for all $S_m^{(p)}$ for $m<n,1\leq p\leq m-1$. There are three different types of $\snj$:
\begin{subequations}
\begin{align}
S_n^{(1)}(t)&=\left[\Omega_{n-1}(t),-iH_M(t)\right];\\
S_n^{(n-1)}(t)&=\left[\Omega_1(t),S_{n-1}^{(n-2)}(t)\right];\\
S_n^{(j)}(t)&=\left[\Omega_1(t),S_{n-1}^{(j-1)}(t)\right]+\sum_{m=2}^{n-j}\left[\Omega_m(t),S_{n-m}^{(j-1)}(t)\right],\notag\\
&\qquad\text{for }2\leq j\leq n-2.
\end{align}
\end{subequations}
Note that the last case occurs only for $n\geq 4$. We will bound each case separately. First, for $S_n^{(1)}$,
\begin{align}
\Vert S_n^{(1)}(t)\Vert&\leq 2\Vert\Omega_{n-1}(t)\Vert\Vert H_M(t)\Vert\nonumber\\
&\leq 2\epsilon\sum_{p=1}^{n-2}\frac{\vert B_p\vert}{p!}\int_0^t ds\Vert S_{n-1}^{(p)}\Vert\nonumber\\
&\leq J(2\epsilon t)^{n-1}\sum_{p=1}^{n-2}\frac{\vert B_p\vert}{p! (n-1)}f_{n-1}^{(p)}.
\end{align}
Eq.~(\ref{eq:fnj}) becomes $f_n^{(1)}=2\sum_{p=1}^{n-2}\frac{\vert B_p\vert}{p! (n-1)}f_{n-1}^{(p)}$ when $j=1$; therefore $\Vert S_n^{(1)}(t)\Vert\leq f_n^{(1)} J(2\epsilon t)^{n-1}$. 

Next we bound $S_n^{(n-1)}$:
\begin{align}
\Vert S_n^{(n-1)}(t)\Vert&\leq 2\Vert\Omega_1(t)\Vert \Vert S_{n-1}^{(n-2)}(t)\Vert\notag\\
&\leq f_{n-1}^{(n-2)} J (2\epsilon t)^{n-1}
\end{align}
Eq.~(\ref{eq:fnj}) becomes $f_n^{(n-1)}=2f_{n-1}^{(n-2)}$ when $j=n-1$; therefore $\Vert S_n^{(n-1)}(t)\Vert\leq f_n^{(n-1)}J(2\epsilon t)^{n-1}$.

Lastly, the $2\leq j\leq n-2$ cases:
\begin{align}
&\quad~\Vert S_n^{(j)}(t)\Vert\notag\\
&\leq 2\Vert\Omega_1(t)\Vert\Vert S_{n-1}^{(j-1)}(t)\Vert\notag\\
&\quad+2\sum_{m=2}^{n-j}\sum_{p=1}^{m-1}\frac{\vert B_p\vert}{p!}\left(\int_0^t ds\Vert S_m^{(p)}(t)\Vert\right)\Vert S_{n-m}^{(j-1)}(t)\Vert\nonumber\\
&\leq f_{n-1}^{(j-1)} J (2\epsilon t)^{n-1}\notag\\
&\quad+J^2t (2\epsilon t)^{n-2}\left[2\sum_{m=2}^{n-j}\sum_{p=1}^{m-1}\frac{\vert B_p\vert}{p!m}f_m^{(p)}f_{n-m}^{(j-1)}\right].
\end{align}
The expression within the brackets in the last line looks like $f_n^{(j)}$ in Eq.~\eqref{eq:fnj}, except we need to add in the $m=1$ terms, as well as the $p=0$ terms. In fact,
\begin{eqnarray}
&&2\sum_{m=2}^{n-j}\sum_{p=1}^{m-1}\frac{\vert B_p\vert}{p!m}f_m^{(p)}f_{n-m}^{(j-1)}\notag\\
&=& f_n^{(j)}-2\frac{\vert B_0\vert}{0! 1}f_1^{(0)}f_{n-1}^{j-1}-2\sum_{m=2}^{n-j}\frac{\vert B_0\vert}{0! m} f_m^{(0)}f_{n-m}^{(j-1)}\nonumber\\
&=&f_n^{(j)}-2f_{n-1}^{(j-1)},
\end{eqnarray}
where in the last line, we have used the fact that $f_{m>1}^{(0)}=0$. Putting this into $\Vert S_n^{(j)}(t)\Vert$, and using the fact that $J\leq \epsilon$, we get
\begin{align}
\Vert S_n^{(j)}(t)\Vert&\leq f_{n-1}^{(j-1)} J (2\epsilon t)^{n-1}\notag\\
&\quad+J (\epsilon t) (2\epsilon t)^{n-2}\left[f_n^{(j)}-2f_{n-1}^{(j-1)}\right]\notag\\
&\leq f_n^{(j)}J(2\epsilon t)^{n-1}.
\end{align}
This completes the induction.
\end{proof}

\section{\texorpdfstring{$f_n$}{fn} coefficients}
\label{app:fn}

In \cite{Moan}, the $\{f_n\}$ were shown to be coefficients in the power series expansion of of $G^{-1}(y)=\sum_{n=1}^\infty f_n y^n$, the inverse function of
\begin{equation}\label{eq:yfunc}
y=G(s)=\int_0^s dx\left[2+\frac{x}{2}\left(1-\cot\frac{x}{2}\right)\right]^{-1}.
\end{equation}
Here, we will provide a self-contained proof of the above claim. It suffices to show that the coefficients of $G^{-1}$ can be written in the form Eq.~\eqref{eq:fn}, with $\fnj$ defined via the recursion relations \eqref{eq:fnj}.

First, we prove a lemma that applies to a general function $y(s)$:
\begin{lemma}\label{lem:fn}
Suppose the smooth function $y\equiv G(s)$ is monotonic on its domain and satisfies $y(0)=0$. Then $G^{-1}(y)$ can be written as $\sum_{n=1}^\infty f_ny^n$ where
\begin{equation}
\label{eq:fnD}
f_n=\left.\frac{1}{n!}\left[\left(g(s)\frac{d}{ds}\right)^{n-1}g(s)\right]\right\vert_{s=0},
\end{equation}
and $\frac{dy}{ds}= \frac{1}{g(s)}$. 
\end{lemma}
\begin{proof}
Since $y$ is monotonic on its domain, the inverse function $G^{-1}(y)(=s)$ exists and has derivatives
\begin{equation}\label{eq:dGinv}
\frac{d^n}{dy^n}G^{-1}(y)=\left(g(s)\frac{d}{ds}\right)^n s=\left(g(s)\frac{d}{ds}\right)^{n-1}g(s),
\end{equation}
where in the first equality, we have used the chain rule of differentiation: $\frac{d}{dy}=\frac{ds}{dy} \frac{d}{ds}=\left(\frac{dy}{ds}\right)^{-1}\frac{d}{ds}=g(s)\frac{d}{ds}$. Since $y$ is a smooth function on its domain, so is $g(s)$ and hence all derivatives of $G^{-1}(y)$ exist. We can then expand $G^{-1}(y)$ as a Taylor series about $y=0$ and write $G^{-1}(y)=\sum_{n=0}^\infty f_n y^n$ for some coefficients $f_n$. We see that $f_0=0$ since $G^{-1}(0)=0$. For $n\geq 1$, the Taylor coefficients are given by
\begin{equation}
f_n=\left.\frac{1}{n!}\frac{d^n}{dy^n}G^{-1}(y)\right\vert_{y=0},
\end{equation}
which, upon inserting Eq.~\eqref{eq:dGinv} and noting that $y(0)=0$, immediately gives Eq.~\eqref{eq:fnD}.
\end{proof}

For our purposes, the function $y(s)$ is given in Eq.~\eqref{eq:yfunc}, i.e. $y(s)=G(s)$ which is smooth and monotonic over the domain $s\in[-2\pi,2\pi]$. It is also clear that $y(0)=0$. Lemma \ref{lem:fn} thus tells us that we can write $G^{-1}(y)=\sum_{n=1}^\infty f_ny^n$, where $f_n$ is given in Eq.~\eqref{eq:fnD} with
\begin{equation}
g(s)\equiv\left(\frac{dy}{ds}\right)^{-1}=2+\frac{s}{2}\left(1-\cot\frac{s}{2}\right).
\end{equation}

\begin{lemma}
The coefficients $f_n$ in $G^{-1}(y)=\sum_{n=1}^\infty f_ny^n$ can be written in the form Eq.~\eqref{eq:fn}, with $\fnj$ defined according to Eq.~\eqref{eq:fnj}.
\end{lemma}
\begin{proof}
For $n=1$, the index $j$ in Eq.~\eqref{eq:fn} can only take value $0$, so $f_1$ can be written in the form Eq.~\eqref{eq:fn} if we set $f_1^{(0)}=1$.  To handle the case $n\ge 2$, we use Eq.~(\ref{eq:Bernoulli}) to  expand $\cot(s/2)$ in terms of Bernoulli numbers, finding 
\begin{equation}\label{eq:cot-Bernoulli}
\frac{s}{2}\cot\left(\frac{s}{2}\right)= B_0 + \left(B_1 + \frac{1}{2}\right)(is) +\sum_{j=2}^\infty \frac{B_j}{j!}(is)^j.
\end{equation}
Noting that $B_0=1$, $B_1=-1/2$, $B_{2j+1}=0$ for $j\ge 1$, $B_{4j}< 0$ for $j\ge 1$ and $B_{4j +2} > 0$ for $j \ge 0$, Eq.~(\ref{eq:cot-Bernoulli}) becomes
\begin{equation}\label{eq:g}
g(s)=2+\frac{s}{2}\left(1-\cot\frac{s}{2}\right)=\sum_{j=0}^\infty\frac{|B_j|}{j!}s^j.
\end{equation}
Using this series expansion of $g(s)$, we can rewrite \eqref{eq:fnD} for $f_{n\geq 2}$ as:
\begin{equation}\label{eq:fn-sum-j}
f_n=\left.\frac{1}{n 2^{n-1}}\sum_{j=1}^{n-1}\frac{|B_j|}{j!}\frac{2^{n-1}}{(n-1)!}\left[\left(g\frac{d}{ds}\right)^{n-1} s^j\right]\right\vert_{s=0}.
\end{equation}
We omit the $j=0$ term in Eq.~(\ref{eq:fn-sum-j}) because the derivative of a constant vanishes, and the sum over $j$ terminates at $j=n-1$ because higher-order terms vanish when we set $s=0$. Thus $f^{(0)}_{n\ge2}=0$, and by comparing with Eq.~\eqref{eq:fn} we define $\fnj$ as
\begin{equation}\label{eq:deffnj}
\fnj=\frac{2^{n-1}}{(n-1)!}\left.\left[\left(g\frac{d}{ds}\right)^{n-1} s^j\right]\right\vert_{s=0}.
\end{equation}

Now we need to show that the $\{\fnj\}$ obey the recursive relation \eqref{eq:fnj}. Using our definition of $\fnj$ from \eqref{eq:deffnj}, the right-hand side of Eq.~\eqref{eq:fnj} can be rewritten as
\begin{eqnarray}
&&2\sum_{m=1}^{n-j}\sum_{p=0}^{m-1}\frac{|B_p|}{p!m}f_m^{(p)}f_{n-m}^{(j-1)}\nonumber\\
&=&2\sum_{m=1}^{n-j}\sum_{p=0}^{m-1}\frac{|B_p|}{p!m}\frac{2^{m-1}}{(m-1)!}\left.\left[\left(g\frac{d}{ds}\right)^{m-1}s^p\right]\right\vert_{s=0}\notag\\
&&\qquad\times\frac{2^{n-m-1}}{(n-m-1)!}\left.\left[\left(g\frac{d}{ds}\right)^{n-m-1}s^{j-1}\right]\right\vert_{s=0}\nonumber\\
&=&\frac{2^{n-1}}{(n-1)!}\sum_{m=1}^{n-j}\binom{n-1}{m}\left.\left[\left(g\frac{d}{ds}\right)^{m-1}\sum_{p=0}^{m-1}\frac{|B_p|}{p!}s^p\right]\right\vert_{s=0}\notag\\
&&\qquad\times\left.\left[\left(g\frac{d}{ds}\right)^{n-m-1}s^{j-1}\right]\right\vert_{s=0}
\end{eqnarray}
The expression $\left(\sum_{p=0}^{m-1}\frac{|B_p|}{p!}s^p\right)$ is just $g(s)$ if we can extend the upper limit of the sum to infinity. We can indeed do this, because in the equation above, the expression is differentiated $m-1$ times and $s$ is set to 0. Hence, higher-order terms in the power series expansion of $g(s)$ with $p\geq m$ do not contribute. Therefore,
\begin{eqnarray}
&&2\sum_{m=1}^{n-j}\sum_{p=0}^{m-1}\frac{|B_p|}{p!m}f_m^{(p)}f_{n-m}^{(j-1)}\nonumber\\
&=&\frac{2^{n-1}}{(n-1)!}\sum_{m=1}^{n-j}\binom{n-1}{m}\left.\left[\left(g\frac{d}{ds}\right)^{m-1}g \right]\right\vert_{s=0}\notag\\
&&\hspace{2cm}\times\left.\left[\left(g\frac{d}{ds}\right)^{n-m-1}s^{j-1}\right]\right\vert_{s=0}\nonumber\\
&=&\frac{2^{n-1}}{(n-1)!}\sum_{m=1}^{n-1}\binom{n-1}{m}\left.\left[\left(g\frac{d}{ds}\right)^m s\right]\right\vert_{s=0}\notag\\
&&\hspace{2cm}\times\left.\left[\left(g\frac{d}{ds}\right)^{n-1-m} s^{j-1}\right]\right\vert_{s=0}\label{eq:fnj2}
\end{eqnarray}
Now, for any differential operator $\cD$ satisfying the product rule, i.e. $\cD(xy)=\cD(x)y+x\cD(y)$ (where $x$ and $y$ commute), $\cD^n$ has the binomial expansion
\begin{equation}
\cD^n(xy)=\sum_{m=0}^n\binom{n}{m}\left[\cD^m(x)\right]\left[\cD^{n-m}(y)\right].
\end{equation}
Take $\cD=g\frac{d}{ds}$, $x=s$ and $y=s^{j-1}$. Then (note that the $m=0$ term is zero),
\begin{align}
\left.\left(g\frac{d}{ds}\right)^{n-1}s^j\right\vert_{s=0}&=\sum_{m=0}^{n-1}\binom{n-1}{m}\left.\left[\left(g\frac{d}{ds}\right)^m s\right]\right\vert_{s=0}\notag\\
&\quad\times\left.\left[\left(g\frac{d}{ds}\right)^{n-1-m} s^{j-1}\right]\right\vert_{s=0}.
\end{align}
Putting this into \eqref{eq:fnj2} gives exactly the expression for $\fnj$ in \eqref{eq:deffnj}.
\end{proof}

\section{Pure-bath term in second order of the Magnus expansion}\label{app:omega2-bath}
Here we consider the case where the toggling-frame Hamiltonian has a decomposition $H(t)=H_B+H_{\rm err}(t)$ such that
\begin{eqnarray}
H_B=B_0\otimes \Id,\quad H_{\rm err}(t)= \sum_{\alpha} B_\alpha\otimes S_\alpha(t),
\end{eqnarray}
where $S_\alpha(t)= U_c^\dagger(t) S_\alpha U_c(t)$ and the operators $\{S_\alpha\}$ are a Hermitian basis for traceless operators acting on the system such that
\begin{equation}
{\rm tr}\left(S_\alpha S_\beta\right)=0
\end{equation}
for all $\alpha\ne \beta$. 
We further assume that each pulse either commutes or anticommutes with each $S_\alpha$, so that
\begin{equation}
S_\alpha(t)= U_c^\dagger(t) S_\alpha U_c(t) = \pm S_\alpha,
\end{equation}
and hence
\begin{equation}
H_{\rm err}(t)= \sum_{\alpha} \zeta_\alpha(t) B_\alpha\otimes S_\alpha
\end{equation}
where $\zeta_\alpha(t)=\pm 1$. These assumptions are true, in particular, for an $n$-qubit system if each $S_\alpha$ and each pulse is a traceless $n$-qubit Pauli operator. We will show that under these assumptions the second-order term in the Magnus expansion
\begin{equation}
\Omega_2(T) = -\frac{1}{2}\int_0^T dt_1\int_0^{t_1} dt_2 ~[H(t_1),H(t,_2)]
\end{equation}
contains no pure bath term; that is, ${\rm tr}_S(\Omega_2(T))=0$.

Because $[H_B,H_B]=0$, it suffices to show that the system trace vanishes for $[H_B,H_{\rm err}(t_1)]$ and $[H_{\rm err}(t_1),H_{\rm err}(t_2)]$ for any $t_1,t_2\in[0,T]$. First we observe that
\begin{eqnarray}
[H_B,H_{\rm err}(t)]&=& \sum_\alpha \zeta_\alpha(t)[B_0\otimes \Id ,B_\alpha\otimes S_\alpha]\nonumber\\
&=& \sum_{\alpha} \zeta_\alpha(t)[B_0,B_\alpha]\otimes S_\alpha 
\end{eqnarray}
has vanishing system trace. Next we note that if the product $S_\alpha S_\beta$ is traceless it can be expanded in the basis $\{S_\alpha\}$, so that
\begin{eqnarray}
S_\alpha S_\beta = \delta_{\alpha\beta} T_\alpha + \sum_\gamma g_{\alpha\beta\gamma} S_\gamma
\end{eqnarray}
(where $T_\alpha$ might have a nonvanishing trace). Therefore
\begin{eqnarray}
H_{\rm err}(t_1)H_{\rm err}(t_2) = \sum_\alpha \zeta_\alpha(t_1)\zeta_\alpha(t_2) B_\alpha B_\alpha\otimes T_\alpha + \cdots\nonumber\\
H_{\rm err}(t_2)H_{\rm err}(t_1) = \sum_\alpha \zeta_\alpha(t_2)\zeta_\alpha(t_1) B_\alpha B_\alpha\otimes T_\alpha + \cdots\nonumber\\
\end{eqnarray}
where the ellipsis represents terms with vanishing system trace. Thus in the commutator $[H_{\rm err}(t_1),H_{\rm err}(t_2)]$ the terms proportional to $T_\alpha$ cancel, and what remains has vanishing system trace, as we wished to show.

\section{Noise parameters for concatenated dynamical decoupling}
\label{app:CDD}
For the analysis of concatenated DD in Sec.~\ref{sec:concatenated}, we considered dividing the third-order term in the Magnus expansion $\Omega_3$ into a pure bath term and a remainder. For that purpose we may use:

\begin{lemma}\label{lem:CDD}
Suppose an operator ${\cal O}$ has a decomposition
\begin{equation}
{\cal O} = \Id\otimes B_0 + \sum_\alpha S_\alpha\otimes B_\alpha,
\end{equation}
where both terms are Hermitian and ${\rm tr}\left(S_\alpha\right)=0$ for each $\alpha$. Then
\begin{equation}\label{eq:B0-lemma}
\left\| B_0\right\| \le \|{\cal O}\|,
\end{equation}
and
\begin{equation}\label{eq:B-alpha-lemma}
\| \sum_\alpha S_\alpha\otimes B_\alpha\|\le 2\|{\cal O}\|.
\end{equation}
\end{lemma}

\begin{proof}
To derive Eq.~(\ref{eq:B0-lemma}), suppose that $|\psi\rangle$ is a normalized pure state such that $|\langle \psi|B_0|\psi\rangle|=\|B_0\|$, and consider the expectation value
\begin{eqnarray}\label{eq:chi-exp-value}
&&\langle \chi|\otimes \langle \psi|\left(\sum_\alpha S_\alpha\otimes B_\alpha\right)|\chi\rangle\otimes|\psi\rangle\nonumber\\
&=&\langle \chi|\left(\sum_\alpha S_\alpha\langle \psi|B_\alpha|\psi\rangle\right)|\chi\rangle,
\end{eqnarray}
where $|\chi\rangle$ is also a normalized pure state
\blue{(of the system)}. 
The right-hand side of Eq.~(\ref{eq:chi-exp-value}) is the expectation value in the state $|\chi\rangle$ of a traceless Hermitian operator. Unless this operator is zero, the expectation value can be either positive or negative depending on how $|\chi\rangle$ is chosen. By choosing $|\chi\rangle$ so that the expectation value $\langle \sum_\alpha S_\alpha\otimes B_\alpha\rangle$ in the state $|\chi\rangle\otimes|\psi\rangle$ is either zero or has the same sign as $\langle \Id\otimes B_0\rangle$, we have
\begin{equation}
|\langle {\cal O}\rangle| \ge |\langle\Id\otimes B_0\rangle|,
\end{equation}
and
\blue{using $\| \Id \otimes B_0\| = \| B_0 \|$,}
Eq.~(\ref{eq:B0-lemma}) follows. From the triangle inequality,
\begin{equation}
\| \sum_\alpha S_\alpha\otimes B_\alpha\| = \|{\cal O} -\Id\otimes B_0\| \le \|{\cal O}\|+ \|\Id\otimes B_0\|\le 2\|{\cal O}\|,
\end{equation}
which proves Eq.~(\ref{eq:B-alpha-lemma}).
\end{proof}

The inequality Eq.~(\ref{eq:B-alpha-lemma}) is tight if we do not restrict the dimension of the system, but if the system is a qubit (two dimensional), it can be improved to
\begin{equation}\label{eq:B-alpha-lemma-qubit}
\| \sum_\alpha \sigma_\alpha\otimes B_\alpha\|\le \|{\cal O}\|.
\end{equation}
For a qubit, there is an anti-unitary time-reversal operator $T:|\psi\rangle\to \sigma_y|\psi\rangle^*$ such that $T^\dagger \sigma_\alpha T=-\sigma_\alpha$. Suppose  $|\psi\rangle$ is a normalized pure state such that $|\langle \psi|\sum_\alpha \sigma_\alpha\otimes B_\alpha|\psi\rangle|=\|\sum_\alpha \sigma_\alpha\otimes B_\alpha\|$. By applying $T\otimes \Id$ if necessary, we can choose $|\psi\rangle$ so that $\langle \sum_\alpha \sigma_\alpha\otimes B_\alpha\rangle$ and $\langle \Id\otimes B_0\rangle$ have the same sign (unless $\langle \Id\otimes B_0\rangle=0$). Therefore 
\begin{equation}
|\langle {\cal O}\rangle| \ge 
\left|\left\langle \sum_\alpha \sigma_\alpha\otimes B_\alpha\right\rangle\right|,
\end{equation}
and Eq.~(\ref{eq:B-alpha-lemma-qubit}) follows.

\section{Relating distance between operators to distance between their exponentials}
\label{app:logarithm}
Here we prove:

\begin{lemma}\label{lem:logarithm}
\begin{eqnarray}\label{eq:log-lemma}
&&\left\| e^A - e^B\right\| \ge 2\| A- B\| \nonumber\\
&&- 2\exp\left(\frac{1}{2}\|A+B\|\right)\sinh \left(\frac{1}{2}\|A-B\|\right).
\end{eqnarray}
\end{lemma}

\begin{proof}
Expanding the exponentials, we obtain
\begin{eqnarray}
e^A-e^B =A-B +\sum_{n=2}^\infty \frac{1}{n!}\left[ A^n-B^n\right],
\end{eqnarray}
and therefore 
\begin{eqnarray}\label{eq:exp-bound-minus}
\|e^A-e^B \|\ge \|A-B\| -\sum_{n=2}^\infty \frac{1}{n!}\|A^n-B^n\|.
\end{eqnarray}

Defining
\begin{equation}
N=\frac{1}{2}\left(A+B\right),\quad M=\frac{1}{2}\left(A-B\right),
\end{equation}
we have
\begin{eqnarray}
 A^n-B^n=(N+M)^n-(N-M)^n,
\end{eqnarray}
and when we apply the binomial expansion to $(N+M)^n-(N-M)^n$ the
terms even order in $M$ cancel. There are
$2\binom{n}{m}$
terms of order $m$ in $M$ for $m$ odd, each with an operator norm bounded above
by $\|M\|^m \|N\|^{n-m}$; therefore
\begin{eqnarray}
\|A^n-B^n\| &\le& 2\sum_{{\rm odd}~m}
\binom{n}{m}
\|M\|^m \|N\|^{n-m}\nonumber\\
&=& \left(\|N\| +\| M\|\right)^n - \left(\|N\| -\| M\|\right)^n.\nonumber\\
\end{eqnarray}
Thus we find
\begin{eqnarray}
&&\sum_{n=2}^\infty \frac{1}{n!}\|A^n-B^n\| \le \exp\left(\|N\|+\| M\|\right)\nonumber\\
&&\quad - \exp\left(\|N\| -\| M\|\right) -2\|M\|\nonumber\\
&&\quad =\exp\left(\|N\|\right)\cdot 2\sinh\left(\|M\|\right)-2\|M\|,
\end{eqnarray}
and substituting into Eq.~(\ref{eq:exp-bound-minus}) yields Eq.~(\ref{eq:log-lemma}).
\end{proof}

If the norm of the sum $A+B$ is not too large, we can use Lemma \ref{lem:logarithm} to show that $A$ is close to $B$ when $e^A$ is close to $e^B$. For example, suppose that 
\begin{eqnarray}
\|A+B\| \le \epsilon_+,\quad \|A-B\| \le \epsilon_-.
\end{eqnarray}
Then Lemma \ref{lem:logarithm} implies
\begin{eqnarray}\label{eq:normAminusB}
\| A-B\|   \le c(\epsilon_+,\epsilon_-)\left\|e^A-e^B\right\|,
\end{eqnarray}
where 
\begin{equation}
c(\epsilon_+,\epsilon_-)=\left( 2- e^{\epsilon_+/2}\frac{\sinh \epsilon_-/2}{\epsilon_-/2}\right)^{-1}.
\end{equation}
For example, if $\epsilon_+ = \epsilon_- = 0.3$, we find $c(\epsilon_+,\epsilon_-)=1.20$.

\section{Bath-state-dependent noise strength and Dyson expansion}
\label{app:Dyson}
In the local-bath model, the noisy operation applied at the circuit location $a$ is a unitary transformation $\Gb_a$ acting jointly on the system and bath. We may express $\Gb_a$ as the sum of a ``good'' part $\cG_a=G_a\otimes B_a$ (where $G_a$ is the ideal gate), and a ``bad'' part $\cB_a=\Gb_a-\cG_a$. 

The accuracy threshold theorem proved in \cite{AGP,NP} establishes that quantum computing is scalable provided the noise strength $\bar \eta$ is smaller than a critical value $\eta_0$. For this purpose, the noise strength may be defined in the following way. Recall that we model a noisy preparation of a qubit as an ideal preparation followed by noisy Hamiltonian evolution for a prescribed period. Therefore, we may assume that the initial state of the system at the very beginning of a quantum computation is ideal, and that the initial state of the system and bath is a product state 
\begin{equation}
|\Psi_S^0\rangle\langle \Psi_S^0|\otimes \rho_B^0.
\end{equation} 
It is convenient to introduce a reference system $R$ that purifies the initial state of the bath; then the initial state of system, bath, and reference system is a pure state
\begin{equation}
|\Phi_{SBR}^0\rangle=|\Psi_S^0\rangle \otimes|\Upsilon_{BR}^0\rangle,
\end{equation}
where 
\begin{equation}
\rho_B^0 = {\rm tr}_R\Big(|\Upsilon_{BR}^0\rangle\langle \Upsilon_{BR}^0|\Big).
\end{equation}
Now consider a quantum circuit acting on the initial state $|\Phi_{SBR}^0\rangle$, and let ${\cal I}_r$ denote a set of $r$ locations in the circuit. Let $U^{\rm bad}({\cal I}_r)$ denote the transformation that results if we place the noisy gate $\Gb_a$ at each location $a \notin {\cal I}_r$ and place the bad part $\cB_a$ at each location $a\in {\cal I}_r$; $U^{\rm bad}({\cal I}_r)$ acts trivially on $R$. We may say that the noise strength is $\bar \eta$ if
\begin{equation}\label{eq:eta-bar-r}
\|U^{\rm bad}({\cal I}_r)|\Phi^0\rangle\| \le \bar \eta^r
\end{equation}
for any set ${\cal I}_r$ of $r$ locations \cite{NP}.

Since each $\Gb_a$ is unitary and therefore has operator norm 1, the submultiplicative property of the norm implies 
\begin{equation}
\|U^{\rm bad}({\cal I}_r)|\Phi^0\rangle\|\le \|U^{\rm bad}({\cal I}_r)\|\le \prod_{a\in{\cal I}_r}\| \cB_a \|.
\end{equation}
Therefore, we may choose the noise strength to be 
\begin{equation}\label{appJ-norm-strength}
\bar \eta = \max_a \| \cB_a\|,
\end{equation}
We used this definition for the analysis in Sec.~\ref{sec:DDFTDDgates}-\ref{sec:concatenated}, based on the Magnus expansion, of the effective noise strength achieved by dynamical decoupling.

The threshold theorem can be formulated in a more general way \cite{AGP,NP}, so that the local-bath assumption is not really needed to define the noise strength or prove the theorem. We adopt the local-bath model in this paper so that we can study the efficacy of the DD pulse sequence for each circuit location individually; otherwise we would need to include noise correlations among distinct gates that are executed simultaneously, which would greatly complicate the analysis. 

The expression \cite{AGP,NP} for the noise strength does not depend on the initial state of the bath, but for the analysis of the effective noise strength in Sec.~\ref{sec:correlator}, based on bath correlation functions and the Dyson expansion, we use a different definition of $\bar \eta$ that {\em does} depend on the initial state of the bath. To state the new definition simply, it is convenient to put a further limitation on the noise model that was not needed in the Magnus expansion analysis --- we assume that the state of the bath is discarded at the end of each circuit location, and replaced by a fresh bath state at the beginning of the next location. We admit that this new more restricted noise model is even more artificial than the local-bath model we analyzed previously using the Magnus expansion. In a rather perverse compromise, we include the effects of the bath's memory  in our analysis of the DD pulse sequence at each circuit location, but assume such effects are negligible when we stitch the DD-protected gates together in a quantum circuit. 

Under this assumption, the noisy operation at location $a$ is applied to a product state, where the initial state $\rho_{B,a}$ of the local bath for location $a$ does not depend on the noisy operations applied at earlier circuit locations. Thus Eq.~(\ref{eq:eta-bar-r}) is satisfied if we define
\begin{equation}
\bar \eta = \max_{a,|\Psi\rangle} \| \cB_a\left(|\Psi\rangle\otimes |\Phi_a\rangle\right)\|,
\end{equation}
where $|\Phi_a\rangle$ is a purification of $\rho_{B,a}$, and the maximum is over all circuit locations  and over all pure states of the system. In terms of the interaction-picture operator applied at location $a$,
\begin{equation}
\tilde U_a = \cG_a^\dagger\Gb_a= \cG_a^\dagger\left(\cG_a+\cB_a\right)= \Id_a + \cG_a^\dagger \cB_a,
\end{equation}
we may write $\bar\eta$ as
\begin{equation}
\bar \eta = \max_{a,|\Psi\rangle} \left\| \left(\tilde U_a - \Id_a \right)\left(|\Psi\rangle\otimes |\Phi_a\rangle\right)\right\|,
\end{equation}
or equivalently
\begin{eqnarray}\label{eq:bar-eta-sqared-bath}
\bar\eta^2 &=& \max_{a,|\Psi\rangle} \left\langle \left( \tilde U_a^\dagger - \Id_a\right)\left( \tilde U_a - \Id_a\right)\right\rangle,\nonumber\\
&=& \max_{a,|\Psi\rangle}\left\langle 2\Id_a- \tilde U_a - \tilde U^\dagger_a\right\rangle,
\end{eqnarray}
where $\langle \cdot\rangle$ denotes the expectation value in the state $|\Psi\rangle\otimes|\Phi_a\rangle$. This is the formula used in \red{Eq. \eqref{eq:eta-dyson} in} Sec.~\ref{sec:correlator}.

Now we can explain how the analysis would need to be modified if we relaxed the assumption that the bath is refreshed at the beginning of each circuit location. In the proof of the threshold theorem, we need to derive an upper bound not on the amplitude for a fault at a single circuit location, but instead on the amplitude for faults occurring at each of the $r$ specified locations in the set $\mathcal{I}_r$, as in Eq.~(\ref{eq:eta-bar-r}). Therefore, in our expression for $\bar\eta^2$ in Eq.~(\ref{eq:bar-eta-sqared-bath}), we should consider the state $|\Phi_a\rangle$ of the bath to be not the actual bath state at the beginning of location $a$, but rather the \emph{conditional} state of the bath, given that faults have already occurred at a specified set of previous locations. In \cite{NP} we obtained an upper bound on $\bar \eta^2$ for the case of Gaussian noise by doing a global analysis of the whole quantum circuit --- generalizing that analysis to DD-improved gates seems difficult. On the other hand, we may still express $\bar \eta^2$ as in Eq.~(\ref{eq:bar-eta-sqared-bath}) in the more general setting (without assuming the bath is refreshed), with the proviso that $\bar\eta^2$ is maximized over all such \emph{conditional} bath states. This is not a very useful criterion as it stands, since this value of $\bar\eta^2$ cannot be easily extracted from any feasible experiment. But it could become more useful were we able to infer properties of the bath correlations in the conditional state from weaker assumptions about the noise model. 

\red{To derive Eq. \eqref{eq:eta-dyson2}, consider a location with duration $T$. The} interaction-picture time-evolution operator is given by Dyson's formula 
\begin{equation}
\tilde U(T) = {\cal T}\exp\left(-i \int_0^T dt~ \tilde H(t)\right)
\end{equation}
where ${\cal T}$ denotes time-ordering and $\tilde H(t)$ is the interaction-picture Hamiltonian, which obeys $\| \tilde H(t)\|= \|H_{\rm err}\| \le J$. Expanding the exponential, we find \begin{equation}
\tilde U(T) = \Id + \sum_{n=1}^\infty \tilde U_n(T),
\end{equation}
where
\begin{equation}
\tilde U_n(T) = \frac{(-i)^n}{n!}\int_0^T dt_1  \cdots dt_n {\cal T} \left(\tilde H(t_1) \cdots \tilde H(t_n)\right),
\end{equation}
and hence
\begin{equation}
\| \tilde U_n(T)\| \le \frac{1}{n!} T^n\|\tilde H(t_1) \cdots \tilde H(t_n)\| \le \frac{(JT)^n}{n!}.
\end{equation}
Similarly, $U^\dagger(T)$ has the expansion 
\begin{equation}
\tilde U^\dagger(T)=\Id +\sum_{n=1}^\infty \tilde U_n^\dagger(T),
\end{equation}
where
\begin{equation}
\tilde U^\dagger_n(T) = \frac{(i)^n}{n!}\int_0^T dt_1  \cdots dt_n {\cal T}' \left(\tilde H(t_1) \cdots \tilde H(t_n)\right);
\end{equation}
here ${\cal T}'$ denotes reverse-time ordering, and again 
\begin{equation}
\| \tilde U^\dagger_n(T)\| \le \frac{(JT)^n}{n!}.
\end{equation}

Noting that $\tilde U_1(T) + \tilde U_1^\dagger(T)=0$, we find
\begin{eqnarray}
&&~~\left\langle 2\Id- \tilde U(T) - \tilde U^\dagger(T)\right\rangle\nonumber\\
&&\le -\left\langle\tilde U_2(T) +\tilde U_2^\dagger(T)\right\rangle+\sum_{n=3}^\infty\|\tilde U_n(T) + \tilde U^\dagger_n(T)\|\nonumber\\
&&\le - \left\langle\tilde U_2(T) +\tilde U_2^\dagger(T)\right\rangle+2\sum_{n=3}^\infty\frac{(JT)^n}{n!}\nonumber\\
&&\le -\left\langle\tilde U_2(T) +\tilde U_2^\dagger(T)\right\rangle+ 2\left(e^{JT}-1-JT-\frac{1}{2}(JT)^2\right).\nonumber\\
\end{eqnarray}
To evaluate the expectation value of $\tilde U_2(T) +\tilde U_2^\dagger(T)$, we observe that
\begin{eqnarray}
&&{\cal T}\left(\tilde H(t_1) \tilde H(t_2)\right) + {\cal T}'\left(\tilde H(t_1) \tilde H(t_2)\right) \nonumber\\
&&= \tilde H(t_1) \tilde H(t_2)+\tilde H(t_2) \tilde H(t_1),
\end{eqnarray}
so that
\begin{eqnarray}
&&\tilde U_2(T)+\tilde U_2^\dagger(T) \nonumber\\
&&= -\frac{1}{2} \int_0^T dt_1dt_2\left( \tilde H(t_1) \tilde H(t_2)+\tilde H(t_2) \tilde H(t_1)\right)\nonumber\\
&&= -\int_0^T dt_1dt_2 \tilde H(t_1) \tilde H(t_2).
\end{eqnarray}
Finally we may express the noise strength as
\begin{eqnarray}
\bar\eta^2 &=&  \max \int_0^T dt_1dt_2\left\langle\tilde H(t_1)\tilde H(t_2) \right\rangle \nonumber\\
&+& 2\left( e^{JT} - 1 - JT - \frac{1}{2}(JT)^2\right),
\end{eqnarray}
as in Eq.~(\ref{eq:eta-dyson2}).


\begin{thebibliography}{74}
\expandafter\ifx\csname natexlab\endcsname\relax\def\natexlab#1{#1}\fi
\expandafter\ifx\csname bibnamefont\endcsname\relax
  \def\bibnamefont#1{#1}\fi
\expandafter\ifx\csname bibfnamefont\endcsname\relax
  \def\bibfnamefont#1{#1}\fi
\expandafter\ifx\csname citenamefont\endcsname\relax
  \def\citenamefont#1{#1}\fi
\expandafter\ifx\csname url\endcsname\relax
  \def\url#1{\texttt{#1}}\fi
\expandafter\ifx\csname urlprefix\endcsname\relax\def\urlprefix{URL }\fi
\providecommand{\bibinfo}[2]{#2}
\providecommand{\eprint}[2][]{\url{#2}}

\bibitem[{\citenamefont{Haeberlen}(1976)}]{NMRBook}
\bibinfo{author}{\bibfnamefont{U.}~\bibnamefont{Haeberlen}},
  \emph{\bibinfo{title}{High Resolution NMR in Solids}}, Advances in Magnetic
  Resonance Series, Supplement 1 (\bibinfo{publisher}{Academic Press},
  \bibinfo{address}{New York}, \bibinfo{year}{1976}).

\bibitem[{\citenamefont{Viola and Lloyd}(1998)}]{VL98}
\bibinfo{author}{\bibfnamefont{L.}~\bibnamefont{Viola}} \bibnamefont{and}
  \bibinfo{author}{\bibfnamefont{S.}~\bibnamefont{Lloyd}},
  \bibinfo{journal}{Phys. Rev. A} \textbf{\bibinfo{volume}{58}},
  \bibinfo{pages}{2733} (\bibinfo{year}{1998}).

\bibitem[{\citenamefont{Zanardi}(1999)}]{Zanardi99}
\bibinfo{author}{\bibfnamefont{P.}~\bibnamefont{Zanardi}},
  \bibinfo{journal}{Phys. Lett. A} \textbf{\bibinfo{volume}{258}},
  \bibinfo{pages}{77} (\bibinfo{year}{1999}).

\bibitem[{\citenamefont{Duan and Guo}(1999)}]{Duan99}
\bibinfo{author}{\bibfnamefont{L.-M.} \bibnamefont{Duan}} \bibnamefont{and}
  \bibinfo{author}{\bibfnamefont{G.}~\bibnamefont{Guo}},
  \bibinfo{journal}{Phys. Lett. A} \textbf{\bibinfo{volume}{261}},
  \bibinfo{pages}{139} (\bibinfo{year}{1999}).

\bibitem[{\citenamefont{Vitali and Tombesi}(1999)}]{Vitali99}
\bibinfo{author}{\bibfnamefont{D.}~\bibnamefont{Vitali}} \bibnamefont{and}
  \bibinfo{author}{\bibfnamefont{P.}~\bibnamefont{Tombesi}},
  \bibinfo{journal}{Phys. Rev. A} \textbf{\bibinfo{volume}{59}},
  \bibinfo{pages}{4178} (\bibinfo{year}{1999}).

\bibitem[{\citenamefont{Viola et~al.}(1999{\natexlab{a}})\citenamefont{Viola,
  Knill, and Lloyd}}]{Viola99}
\bibinfo{author}{\bibfnamefont{L.}~\bibnamefont{Viola}},
  \bibinfo{author}{\bibfnamefont{E.}~\bibnamefont{Knill}}, \bibnamefont{and}
  \bibinfo{author}{\bibfnamefont{S.}~\bibnamefont{Lloyd}},
  \bibinfo{journal}{Phys. Rev. Lett.} \textbf{\bibinfo{volume}{82}},
  \bibinfo{pages}{2417} (\bibinfo{year}{1999}{\natexlab{a}}).

\bibitem[{\citenamefont{Viola and Knill}(2003)}]{Viola03}
\bibinfo{author}{\bibfnamefont{L.}~\bibnamefont{Viola}} \bibnamefont{and}
  \bibinfo{author}{\bibfnamefont{E.}~\bibnamefont{Knill}},
  \bibinfo{journal}{Phys. Rev. Lett.} \textbf{\bibinfo{volume}{90}},
  \bibinfo{pages}{037901} (\bibinfo{year}{2003}).

\bibitem[{\citenamefont{Byrd and Lidar}(2003)}]{BL03}
\bibinfo{author}{\bibfnamefont{M.~S.} \bibnamefont{Byrd}} \bibnamefont{and}
  \bibinfo{author}{\bibfnamefont{D.~A.} \bibnamefont{Lidar}},
  \bibinfo{journal}{Phys. Rev. A} \textbf{\bibinfo{volume}{67}},
  \bibinfo{pages}{012324} (\bibinfo{year}{2003}).

\bibitem[{\citenamefont{Khodjasteh and Lidar}(2005)}]{KL05}
\bibinfo{author}{\bibfnamefont{K.}~\bibnamefont{Khodjasteh}} \bibnamefont{and}
  \bibinfo{author}{\bibfnamefont{D.~A.} \bibnamefont{Lidar}},
  \bibinfo{journal}{Phys. Rev. Lett.} \textbf{\bibinfo{volume}{95}},
  \bibinfo{pages}{180501} (\bibinfo{year}{2005}).

\bibitem[{\citenamefont{Khodjasteh and Lidar}(2007)}]{KL07}
\bibinfo{author}{\bibfnamefont{K.}~\bibnamefont{Khodjasteh}} \bibnamefont{and}
  \bibinfo{author}{\bibfnamefont{D.~A.} \bibnamefont{Lidar}},
  \bibinfo{journal}{Phys. Rev. A} \textbf{\bibinfo{volume}{75}},
  \bibinfo{pages}{062310} (\bibinfo{year}{2007}).

\bibitem[{\citenamefont{Viola and Knill}(2005)}]{Viola05}
\bibinfo{author}{\bibfnamefont{L.}~\bibnamefont{Viola}} \bibnamefont{and}
  \bibinfo{author}{\bibfnamefont{E.}~\bibnamefont{Knill}},
  \bibinfo{journal}{Phys. Rev. Lett.} \textbf{\bibinfo{volume}{94}},
  \bibinfo{pages}{060502} (\bibinfo{year}{2005}).

\bibitem[{\citenamefont{Kern and Alber}(2009)}]{Kern05}
\bibinfo{author}{\bibfnamefont{O.}~\bibnamefont{Kern}} \bibnamefont{and}
  \bibinfo{author}{\bibfnamefont{G.}~\bibnamefont{Alber}},
  \bibinfo{journal}{Phys. Rev. Lett.} \textbf{\bibinfo{volume}{95}},
  \bibinfo{pages}{250501} (\bibinfo{year}{2009}).

\bibitem[{\citenamefont{Uhrig}(2007)}]{Uhrig}
\bibinfo{author}{\bibfnamefont{G.~S.} \bibnamefont{Uhrig}},
  \bibinfo{journal}{Phys. Rev. Lett.} \textbf{\bibinfo{volume}{98}},
  \bibinfo{pages}{100504} (\bibinfo{year}{2007}).

\bibitem[{\citenamefont{Berglund}(2000)}]{Berglund01}
\bibinfo{author}{\bibfnamefont{A.~J.} \bibnamefont{Berglund}}
  (\bibinfo{year}{2000}), \eprint{quant-ph/0010001}.

\bibitem[{\citenamefont{Fortunato et~al.}(2002)\citenamefont{Fortunato, Viola,
  Hodges, Teklemariam, and Cory}}]{Fortunato02}
\bibinfo{author}{\bibfnamefont{E.~M.} \bibnamefont{Fortunato}},
  \bibinfo{author}{\bibfnamefont{L.}~\bibnamefont{Viola}},
  \bibinfo{author}{\bibfnamefont{J.}~\bibnamefont{Hodges}},
  \bibinfo{author}{\bibfnamefont{G.}~\bibnamefont{Teklemariam}},
  \bibnamefont{and} \bibinfo{author}{\bibfnamefont{D.~G.} \bibnamefont{Cory}},
  \bibinfo{journal}{New J. Phys.} \textbf{\bibinfo{volume}{4}},
  \bibinfo{pages}{5.1} (\bibinfo{year}{2002}).

\bibitem[{\citenamefont{Fraval et~al.}(2009)\citenamefont{Fraval, Sellars, and
  Longdell}}]{Fraval05}
\bibinfo{author}{\bibfnamefont{E.}~\bibnamefont{Fraval}},
  \bibinfo{author}{\bibfnamefont{M.~J.} \bibnamefont{Sellars}},
  \bibnamefont{and} \bibinfo{author}{\bibfnamefont{J.~J.}
  \bibnamefont{Longdell}}, \bibinfo{journal}{Phys. Rev. Lett.}
  \textbf{\bibinfo{volume}{95}}, \bibinfo{pages}{030506}
  (\bibinfo{year}{2009}).

\bibitem[{\citenamefont{Petta et~al.}(2005)\citenamefont{Petta, Johnson,
  Taylor, Laird, Yacoby, Lukin, Marcus, Hanson, and Gossard}}]{Petta05}
\bibinfo{author}{\bibfnamefont{J.}~\bibnamefont{Petta}},
  \bibinfo{author}{\bibfnamefont{A.}~\bibnamefont{Johnson}},
  \bibinfo{author}{\bibfnamefont{J.}~\bibnamefont{Taylor}},
  \bibinfo{author}{\bibfnamefont{E.}~\bibnamefont{Laird}},
  \bibinfo{author}{\bibfnamefont{A.}~\bibnamefont{Yacoby}},
  \bibinfo{author}{\bibfnamefont{M.~D.} \bibnamefont{Lukin}},
  \bibinfo{author}{\bibfnamefont{C.}~\bibnamefont{Marcus}},
  \bibinfo{author}{\bibfnamefont{M.}~\bibnamefont{Hanson}}, \bibnamefont{and}
  \bibinfo{author}{\bibfnamefont{A.~C.} \bibnamefont{Gossard}},
  \bibinfo{journal}{Science} \textbf{\bibinfo{volume}{309}},
  \bibinfo{pages}{2180} (\bibinfo{year}{2005}).

\bibitem[{\citenamefont{Morton et~al.}(2006)\citenamefont{Morton, Tyryshkin,
  Ardavan, Benjamin, Porfyrakis, Lyon, and Briggs}}]{Morton06}
\bibinfo{author}{\bibfnamefont{J.}~\bibnamefont{Morton}},
  \bibinfo{author}{\bibfnamefont{A.}~\bibnamefont{Tyryshkin}},
  \bibinfo{author}{\bibfnamefont{A.}~\bibnamefont{Ardavan}},
  \bibinfo{author}{\bibfnamefont{S.}~\bibnamefont{Benjamin}},
  \bibinfo{author}{\bibfnamefont{K.}~\bibnamefont{Porfyrakis}},
  \bibinfo{author}{\bibfnamefont{S.}~\bibnamefont{Lyon}}, \bibnamefont{and}
  \bibinfo{author}{\bibfnamefont{G.}~\bibnamefont{Briggs}},
  \bibinfo{journal}{Nature Phys.} \textbf{\bibinfo{volume}{2}},
  \bibinfo{pages}{40} (\bibinfo{year}{2006}).

\bibitem[{\citenamefont{Morton et~al.}(2008)\citenamefont{Morton, Tyryshkin,
  Brown, Shankar, Lovett, Ardavan, Schenkel, Haller, Ager, and
  Lyon}}]{Morton08}
\bibinfo{author}{\bibfnamefont{J.~J.~L.} \bibnamefont{Morton}},
  \bibinfo{author}{\bibfnamefont{A.~M.} \bibnamefont{Tyryshkin}},
  \bibinfo{author}{\bibfnamefont{R.~M.} \bibnamefont{Brown}},
  \bibinfo{author}{\bibfnamefont{S.}~\bibnamefont{Shankar}},
  \bibinfo{author}{\bibfnamefont{B.~W.} \bibnamefont{Lovett}},
  \bibinfo{author}{\bibfnamefont{A.}~\bibnamefont{Ardavan}},
  \bibinfo{author}{\bibfnamefont{T.}~\bibnamefont{Schenkel}},
  \bibinfo{author}{\bibfnamefont{E.~E.} \bibnamefont{Haller}},
  \bibinfo{author}{\bibfnamefont{J.~W.} \bibnamefont{Ager}}, \bibnamefont{and}
  \bibinfo{author}{\bibfnamefont{S.~A.} \bibnamefont{Lyon}},
  \bibinfo{journal}{Nature} \textbf{\bibinfo{volume}{455}},
  \bibinfo{pages}{1085} (\bibinfo{year}{2008}).

\bibitem[{\citenamefont{Biercuk
  et~al.}(2009{\natexlab{a}})\citenamefont{Biercuk, Uys, VanDevender, Shiga,
  Itano, and Bollinger}}]{Biercuk09}
\bibinfo{author}{\bibfnamefont{M.~J.} \bibnamefont{Biercuk}},
  \bibinfo{author}{\bibfnamefont{H.}~\bibnamefont{Uys}},
  \bibinfo{author}{\bibfnamefont{A.~P.} \bibnamefont{VanDevender}},
  \bibinfo{author}{\bibfnamefont{N.}~\bibnamefont{Shiga}},
  \bibinfo{author}{\bibfnamefont{W.~M.} \bibnamefont{Itano}}, \bibnamefont{and}
  \bibinfo{author}{\bibfnamefont{J.~J.} \bibnamefont{Bollinger}},
  \bibinfo{journal}{Nature} \textbf{\bibinfo{volume}{458}},
  \bibinfo{pages}{996} (\bibinfo{year}{2009}{\natexlab{a}}).

\bibitem[{\citenamefont{Uys et~al.}(2009)\citenamefont{Uys, Biercuk, and
  Bollinger}}]{Uys09}
\bibinfo{author}{\bibfnamefont{H.}~\bibnamefont{Uys}},
  \bibinfo{author}{\bibfnamefont{M.~J.} \bibnamefont{Biercuk}},
  \bibnamefont{and} \bibinfo{author}{\bibfnamefont{J.~J.}
  \bibnamefont{Bollinger}}, \bibinfo{journal}{Phys. Rev. Lett.}
  \textbf{\bibinfo{volume}{103}}, \bibinfo{pages}{040501}
  (\bibinfo{year}{2009}).

\bibitem[{\citenamefont{Biercuk
  et~al.}(2009{\natexlab{b}})\citenamefont{Biercuk, Uys, VanDevender, Shiga,
  Itano, and Bollinger}}]{Biercuk09a}
\bibinfo{author}{\bibfnamefont{M.~J.} \bibnamefont{Biercuk}},
  \bibinfo{author}{\bibfnamefont{H.}~\bibnamefont{Uys}},
  \bibinfo{author}{\bibfnamefont{A.~P.} \bibnamefont{VanDevender}},
  \bibinfo{author}{\bibfnamefont{N.}~\bibnamefont{Shiga}},
  \bibinfo{author}{\bibfnamefont{W.~M.} \bibnamefont{Itano}}, \bibnamefont{and}
  \bibinfo{author}{\bibfnamefont{J.~J.} \bibnamefont{Bollinger}},
  \bibinfo{journal}{Phys. Rev. A} \textbf{\bibinfo{volume}{79}},
  \bibinfo{pages}{062324} (\bibinfo{year}{2009}{\natexlab{b}}).

\bibitem[{\citenamefont{Damodarakurup et~al.}(2009)\citenamefont{Damodarakurup,
  Lucamarini, Giuseppe, Vitali, and Tombesi}}]{Damo08}
\bibinfo{author}{\bibfnamefont{S.}~\bibnamefont{Damodarakurup}},
  \bibinfo{author}{\bibfnamefont{M.}~\bibnamefont{Lucamarini}},
  \bibinfo{author}{\bibfnamefont{G.~D.} \bibnamefont{Giuseppe}},
  \bibinfo{author}{\bibfnamefont{D.}~\bibnamefont{Vitali}}, \bibnamefont{and}
  \bibinfo{author}{\bibfnamefont{P.}~\bibnamefont{Tombesi}},
  \bibinfo{journal}{Phys. Rev. Lett.} \textbf{\bibinfo{volume}{103}},
  \bibinfo{pages}{040502} (\bibinfo{year}{2009}).

\bibitem[{\citenamefont{Beavan et~al.}(2009)\citenamefont{Beavan, Fraval,
  Sellars, and Longdell}}]{beavan:032308}
\bibinfo{author}{\bibfnamefont{S.~E.} \bibnamefont{Beavan}},
  \bibinfo{author}{\bibfnamefont{E.}~\bibnamefont{Fraval}},
  \bibinfo{author}{\bibfnamefont{M.~J.} \bibnamefont{Sellars}},
  \bibnamefont{and} \bibinfo{author}{\bibfnamefont{J.~J.}
  \bibnamefont{Longdell}}, \bibinfo{journal}{Phys. Rev. A}
  \textbf{\bibinfo{volume}{80}}, \bibinfo{pages}{032308}
  (\bibinfo{year}{2009}).

\bibitem[{\citenamefont{Sagi et~al.}(2010)\citenamefont{Sagi, Almog, and
  Davidson}}]{Sagi09}
\bibinfo{author}{\bibfnamefont{Y.}~\bibnamefont{Sagi}},
  \bibinfo{author}{\bibfnamefont{I.}~\bibnamefont{Almog}}, \bibnamefont{and}
  \bibinfo{author}{\bibfnamefont{N.}~\bibnamefont{Davidson}},
  \bibinfo{journal}{Phys. Rev. Lett.} \textbf{\bibinfo{volume}{105}},
  \bibinfo{pages}{053201} (\bibinfo{year}{2010}).

\bibitem[{\citenamefont{Shor}(1996)}]{Shor96}
\bibinfo{author}{\bibfnamefont{P.}~\bibnamefont{Shor}}, in
  \emph{\bibinfo{booktitle}{Proceedings of the 37th Symposium on Foundations of
  Computing}} (\bibinfo{publisher}{IEEE Computer Society Press},
  \bibinfo{address}{Los Alamitos, CA}, \bibinfo{year}{1996}),
  p.~\bibinfo{pages}{56}.


\bibitem[{\citenamefont{Aharonov and Ben-Or}(2008)}]{AB99}
\bibinfo{author}{\bibfnamefont{D.}~\bibnamefont{Aharonov}} \bibnamefont{and}
  \bibinfo{author}{\bibfnamefont{M.}~\bibnamefont{Ben-Or}},
  \bibinfo{journal}{SIAM J. Comput.} \textbf{\bibinfo{volume}{38}},
  \bibinfo{pages}{1207} (\bibinfo{year}{2008}).

\bibitem[{\citenamefont{Kitaev}(1997)}]{Kitaev97}
\bibinfo{author}{\bibfnamefont{A.}~\bibnamefont{Kitaev}},
  \bibinfo{journal}{Russ. Math. Surveys} \textbf{\bibinfo{volume}{52}},
  \bibinfo{pages}{1191} (\bibinfo{year}{1997}).

\bibitem[{\citenamefont{Knill et~al.}(1998)\citenamefont{Knill, Laflamme, and
  Zurek}}]{Knill98}
\bibinfo{author}{\bibfnamefont{E.}~\bibnamefont{Knill}},
  \bibinfo{author}{\bibfnamefont{R.}~\bibnamefont{Laflamme}}, \bibnamefont{and}
  \bibinfo{author}{\bibfnamefont{W.~H.} \bibnamefont{Zurek}},
  \bibinfo{journal}{Proc. R. Soc. London, Ser. A}
  \textbf{\bibinfo{volume}{454}}, \bibinfo{pages}{365} (\bibinfo{year}{1998}).

\bibitem[{\citenamefont{Terhal and Burkard}(2005)}]{TB}
\bibinfo{author}{\bibfnamefont{B.~M.} \bibnamefont{Terhal}} \bibnamefont{and}
  \bibinfo{author}{\bibfnamefont{G.}~\bibnamefont{Burkard}},
  \bibinfo{journal}{Phys. Rev. A} \textbf{\bibinfo{volume}{71}},
  \bibinfo{pages}{012336} (\bibinfo{year}{2005}).

\bibitem[{\citenamefont{Aliferis et~al.}(2008)\citenamefont{Aliferis,
  Gottesman, and Preskill}}]{AGP}
\bibinfo{author}{\bibfnamefont{P.}~\bibnamefont{Aliferis}},
  \bibinfo{author}{\bibfnamefont{D.}~\bibnamefont{Gottesman}},
  \bibnamefont{and} \bibinfo{author}{\bibfnamefont{J.}~\bibnamefont{Preskill}},
  \bibinfo{journal}{Quant. Inf. Comp.} \textbf{\bibinfo{volume}{8}},
  \bibinfo{pages}{181} (\bibinfo{year}{2008}).

\bibitem[{\citenamefont{Aharonov et~al.}(2006)\citenamefont{Aharonov, Kitaev,
  and Preskill}}]{AKP}
\bibinfo{author}{\bibfnamefont{D.}~\bibnamefont{Aharonov}},
  \bibinfo{author}{\bibfnamefont{A.}~\bibnamefont{Kitaev}}, \bibnamefont{and}
  \bibinfo{author}{\bibfnamefont{J.}~\bibnamefont{Preskill}},
  \bibinfo{journal}{Phys. Rev. Lett.} \textbf{\bibinfo{volume}{96}},
  \bibinfo{pages}{050504} (\bibinfo{year}{2006}).

\bibitem[{\citenamefont{Ng and Preskill}(2009)}]{NP}
\bibinfo{author}{\bibfnamefont{H.~K.} \bibnamefont{Ng}} \bibnamefont{and}
  \bibinfo{author}{\bibfnamefont{J.}~\bibnamefont{Preskill}},
  \bibinfo{journal}{Phys. Rev. A} \textbf{\bibinfo{volume}{79}},
  \bibinfo{pages}{032318} (\bibinfo{year}{2009}).

\bibitem[{\citenamefont{Viola et~al.}(1999{\natexlab{b}})\citenamefont{Viola,
  Lloyd, and Knill}}]{Viola:99a}
\bibinfo{author}{\bibfnamefont{L.}~\bibnamefont{Viola}},
  \bibinfo{author}{\bibfnamefont{S.}~\bibnamefont{Lloyd}}, \bibnamefont{and}
  \bibinfo{author}{\bibfnamefont{E.}~\bibnamefont{Knill}},
  \bibinfo{journal}{Phys. Rev. Lett.} \textbf{\bibinfo{volume}{83}},
  \bibinfo{pages}{4888} (\bibinfo{year}{1999}{\natexlab{b}}).

\bibitem[{\citenamefont{Byrd and Lidar}(2002)}]{ByrdLidar:01a}
\bibinfo{author}{\bibfnamefont{M.~S.} \bibnamefont{Byrd}} \bibnamefont{and}
  \bibinfo{author}{\bibfnamefont{D.~A.} \bibnamefont{Lidar}},
  \bibinfo{journal}{Phys. Rev. Lett.} \textbf{\bibinfo{volume}{89}},
  \bibinfo{pages}{047901} (\bibinfo{year}{2002}).

\bibitem[{\citenamefont{Khodjasteh and Lidar}(2003)}]{KhodjastehLidar:03}
\bibinfo{author}{\bibfnamefont{K.}~\bibnamefont{Khodjasteh}} \bibnamefont{and}
  \bibinfo{author}{\bibfnamefont{D.~A.} \bibnamefont{Lidar}},
  \bibinfo{journal}{Phys. Rev. A} \textbf{\bibinfo{volume}{68}},
  \bibinfo{pages}{022322} (\bibinfo{year}{2003}), \bibinfo{note}{erratum: {\it
  ibid}, Phys. Rev. A {\bf 72}, 029905 (2005).}

\bibitem[{\citenamefont{Boulant et~al.}(2002)\citenamefont{Boulant, Pravia,
  Fortunato, Havel, and Cory}}]{Boulant02}
\bibinfo{author}{\bibfnamefont{N.}~\bibnamefont{Boulant}},
  \bibinfo{author}{\bibfnamefont{M.~A.} \bibnamefont{Pravia}},
  \bibinfo{author}{\bibfnamefont{E.~M.} \bibnamefont{Fortunato}},
  \bibinfo{author}{\bibfnamefont{T.~F.} \bibnamefont{Havel}}, \bibnamefont{and}
  \bibinfo{author}{\bibfnamefont{D.~G.} \bibnamefont{Cory}},
  \bibinfo{journal}{Quant. Inf. Proc.} \textbf{\bibinfo{volume}{1}},
  \bibinfo{pages}{135} (\bibinfo{year}{2002}).

\bibitem[{\citenamefont{{R. Bhatia}}(1997)}]{Bhatia}
\bibinfo{author}{\bibnamefont{{R. Bhatia}}}, \emph{\bibinfo{title}{{Matrix
  Analysis}}}, no. \bibinfo{number}{169} in \bibinfo{series}{{Graduate Texts in
  Mathematics}} (\bibinfo{publisher}{Springer-Verlag}, \bibinfo{address}{{New
  York}}, \bibinfo{year}{1997}).

\bibitem[{\citenamefont{Yang and Liu}(2008)}]{Yang}
\bibinfo{author}{\bibfnamefont{W.}~\bibnamefont{Yang}} \bibnamefont{and}
  \bibinfo{author}{\bibfnamefont{R.-B.} \bibnamefont{Liu}},
  \bibinfo{journal}{Phys. Rev. Lett.} \textbf{\bibinfo{volume}{101}},
  \bibinfo{pages}{180403} (\bibinfo{year}{2008}).

\bibitem[{\citenamefont{Lee et~al.}(2008)\citenamefont{Lee, Witzel, and
  Sarma}}]{Lee08}
\bibinfo{author}{\bibfnamefont{B.}~\bibnamefont{Lee}},
  \bibinfo{author}{\bibfnamefont{W.~M.} \bibnamefont{Witzel}},
  \bibnamefont{and} \bibinfo{author}{\bibfnamefont{S.~D.} \bibnamefont{Sarma}},
  \bibinfo{journal}{Phys. Rev. Lett.} \textbf{\bibinfo{volume}{100}},
  \bibinfo{pages}{160505} (\bibinfo{year}{2008}).

\bibitem[{\citenamefont{Uhrig}(2009)}]{Uhrig09}
\bibinfo{author}{\bibfnamefont{G.~S.} \bibnamefont{Uhrig}},
  \bibinfo{journal}{Phys. Rev. Lett.} \textbf{\bibinfo{volume}{102}},
  \bibinfo{pages}{120502} (\bibinfo{year}{2009}).

\bibitem[{\citenamefont{Uhrig and Pasini}()}]{Uhrig09a}
\bibinfo{author}{\bibfnamefont{G.~S.} \bibnamefont{Uhrig}} \bibnamefont{and}
  \bibinfo{author}{\bibfnamefont{S.}~\bibnamefont{Pasini}},
  \eprint{arXiv:0906.3605}.

\bibitem[{\citenamefont{Magnus}(1954)}]{Magnus}
\bibinfo{author}{\bibfnamefont{W.}~\bibnamefont{Magnus}},
  \bibinfo{journal}{Comm. Pure Appl. Math.} \textbf{\bibinfo{volume}{7}},
  \bibinfo{pages}{649} (\bibinfo{year}{1954}).

\bibitem[{\citenamefont{Blanes et~al.}(2009)\citenamefont{Blanes, Casas, Oteo,
  and Ros}}]{Blanes08}
\bibinfo{author}{\bibfnamefont{S.}~\bibnamefont{Blanes}},
  \bibinfo{author}{\bibfnamefont{F.}~\bibnamefont{Casas}},
  \bibinfo{author}{\bibfnamefont{J.~A.} \bibnamefont{Oteo}}, \bibnamefont{and}
  \bibinfo{author}{\bibfnamefont{J.}~\bibnamefont{Ros}},
  \bibinfo{journal}{Phys. Reports} \textbf{\bibinfo{volume}{470}},
  \bibinfo{pages}{151} (\bibinfo{year}{2009}).

\bibitem[{\citenamefont{Moan et~al.}(1999)\citenamefont{Moan, Oteo, and
  Ros}}]{Moan99}
\bibinfo{author}{\bibfnamefont{P.~C.} \bibnamefont{Moan}},
  \bibinfo{author}{\bibfnamefont{J.~A.} \bibnamefont{Oteo}}, \bibnamefont{and}
  \bibinfo{author}{\bibfnamefont{J.}~\bibnamefont{Ros}}, \bibinfo{journal}{J.
  Phys. A} \textbf{\bibinfo{volume}{32}}, \bibinfo{pages}{5133}
  (\bibinfo{year}{1999}).

\bibitem[{\citenamefont{Pasini et~al.}(2008)\citenamefont{Pasini, Fischer,
  Karbach, and Uhrig}}]{Pasini08}
\bibinfo{author}{\bibfnamefont{S.}~\bibnamefont{Pasini}},
  \bibinfo{author}{\bibfnamefont{T.}~\bibnamefont{Fischer}},
  \bibinfo{author}{\bibfnamefont{P.}~\bibnamefont{Karbach}}, \bibnamefont{and}
  \bibinfo{author}{\bibfnamefont{G.~S.} \bibnamefont{Uhrig}},
  \bibinfo{journal}{Phys. Rev. A} \textbf{\bibinfo{volume}{77}},
  \bibinfo{pages}{032315} (\bibinfo{year}{2008}).

\bibitem[{\citenamefont{Wang and Ramshaw}(1972)}]{Wang72}
\bibinfo{author}{\bibfnamefont{C.~H.} \bibnamefont{Wang}} \bibnamefont{and}
  \bibinfo{author}{\bibfnamefont{J.~D.} \bibnamefont{Ramshaw}},
  \bibinfo{journal}{Phys. Rev. B} \textbf{\bibinfo{volume}{6}},
  \bibinfo{pages}{3253} (\bibinfo{year}{1972}).

\bibitem[{\citenamefont{Aliferis and Cross}(2007)}]{AC}
\bibinfo{author}{\bibfnamefont{P.}~\bibnamefont{Aliferis}} \bibnamefont{and}
  \bibinfo{author}{\bibfnamefont{A.~W.} \bibnamefont{Cross}},
  \bibinfo{journal}{Phys. Rev. Lett.} \textbf{\bibinfo{volume}{98}},
  \bibinfo{pages}{220502} (\bibinfo{year}{2007}).

\bibitem[{\citenamefont{DiVincenzo and Aliferis}(2007)}]{DiVincenzo}
\bibinfo{author}{\bibfnamefont{D.~P.} \bibnamefont{DiVincenzo}}
  \bibnamefont{and} \bibinfo{author}{\bibfnamefont{P.}~\bibnamefont{Aliferis}},
  \bibinfo{journal}{Phys. Rev. Lett.} \textbf{\bibinfo{volume}{98}},
  \bibinfo{pages}{020501} (\bibinfo{year}{2007}).

\bibitem[{\citenamefont{Wocjan et~al.}(2002)\citenamefont{Wocjan, R{\"o}tteler,
  Janzing, and Beth}}]{Wocjan}
\bibinfo{author}{\bibfnamefont{P.}~\bibnamefont{Wocjan}},
  \bibinfo{author}{\bibfnamefont{M.}~\bibnamefont{R{\"o}tteler}},
  \bibinfo{author}{\bibfnamefont{D.}~\bibnamefont{Janzing}}, \bibnamefont{and}
  \bibinfo{author}{\bibfnamefont{T.}~\bibnamefont{Beth}},
  \bibinfo{journal}{Phys. Rev. A} \textbf{\bibinfo{volume}{65}},
  \bibinfo{pages}{042309} (\bibinfo{year}{2002}).

\bibitem[{\citenamefont{R{\"o}tteler and Wocjan}(2006)}]{RW06}
\bibinfo{author}{\bibfnamefont{M.}~\bibnamefont{R{\"o}tteler}}
  \bibnamefont{and} \bibinfo{author}{\bibfnamefont{P.}~\bibnamefont{Wocjan}},
  \bibinfo{journal}{IEEE Trans. Inform. Theory} \textbf{\bibinfo{volume}{52}},
  \bibinfo{pages}{4171} (\bibinfo{year}{2006}).

\bibitem[{\citenamefont{Gullion et~al.}(1990)\citenamefont{Gullion, Baker, and
  Conradi}}]{Gullion:1990:479}
\bibinfo{author}{\bibfnamefont{T.}~\bibnamefont{Gullion}},
  \bibinfo{author}{\bibfnamefont{D.~B.} \bibnamefont{Baker}}, \bibnamefont{and}
  \bibinfo{author}{\bibfnamefont{M.~S.} \bibnamefont{Conradi}},
  \bibinfo{journal}{J. Magn. Res.} \textbf{\bibinfo{volume}{89}},
  \bibinfo{pages}{479} (\bibinfo{year}{1990}).

\bibitem[{\citenamefont{Lizak et~al.}(1991)\citenamefont{Lizak, Gullion, and
  Conradi}}]{Lizak91}
\bibinfo{author}{\bibfnamefont{M.}~\bibnamefont{Lizak}},
  \bibinfo{author}{\bibfnamefont{T.}~\bibnamefont{Gullion}}, \bibnamefont{and}
  \bibinfo{author}{\bibfnamefont{M.}~\bibnamefont{Conradi}},
  \bibinfo{journal}{J. Magn. Res.} \textbf{\bibinfo{volume}{91}},
  \bibinfo{pages}{254} (\bibinfo{year}{1991}).

\bibitem[{\citenamefont{Khodjasteh and Viola}(2009{\natexlab{a}})}]{KV09}
\bibinfo{author}{\bibfnamefont{K.}~\bibnamefont{Khodjasteh}} \bibnamefont{and}
  \bibinfo{author}{\bibfnamefont{L.}~\bibnamefont{Viola}},
  \bibinfo{journal}{Phys. Rev. Lett.} \textbf{\bibinfo{volume}{102}},
  \bibinfo{pages}{080501} (\bibinfo{year}{2009}{\natexlab{a}}).

\bibitem[{\citenamefont{Khodjasteh and Viola}(2009{\natexlab{b}})}]{KV09a}
\bibinfo{author}{\bibfnamefont{K.}~\bibnamefont{Khodjasteh}} \bibnamefont{and}
  \bibinfo{author}{\bibfnamefont{L.}~\bibnamefont{Viola}},
  \bibinfo{journal}{Phys. Rev. A} \textbf{\bibinfo{volume}{80}},
  \bibinfo{pages}{032314} (\bibinfo{year}{2009}{\natexlab{b}}).

\bibitem[{\citenamefont{Khodjasteh et~al.}(2010)\citenamefont{Khodjasteh,
  Lidar, and Viola}}]{KLV09}
\bibinfo{author}{\bibfnamefont{K.}~\bibnamefont{Khodjasteh}},
  \bibinfo{author}{\bibfnamefont{D.}~\bibnamefont{Lidar}}, \bibnamefont{and}
  \bibinfo{author}{\bibfnamefont{L.}~\bibnamefont{Viola}},
  \bibinfo{journal}{Phys. Rev. Lett.} \textbf{\bibinfo{volume}{104}},
  \bibinfo{pages}{090501} (\bibinfo{year}{2010}).

\bibitem[{\citenamefont{Klarsfeld and Oteo}(1989)}]{Klarsfeld}
\bibinfo{author}{\bibfnamefont{S.}~\bibnamefont{Klarsfeld}} \bibnamefont{and}
  \bibinfo{author}{\bibfnamefont{J.~A.} \bibnamefont{Oteo}},
  \bibinfo{journal}{Phys. Rev. A} \textbf{\bibinfo{volume}{39}},
  \bibinfo{pages}{3270} (\bibinfo{year}{1989}).

\bibitem[{\citenamefont{Blanes et~al.}(1998)\citenamefont{Blanes, Casas, Oteo,
  and Ros}}]{Blanes98}
\bibinfo{author}{\bibfnamefont{S.}~\bibnamefont{Blanes}},
  \bibinfo{author}{\bibfnamefont{F.}~\bibnamefont{Casas}},
  \bibinfo{author}{\bibfnamefont{J.~A.} \bibnamefont{Oteo}}, \bibnamefont{and}
  \bibinfo{author}{\bibfnamefont{J.}~\bibnamefont{Ros}}, \bibinfo{journal}{J.
  Phys. A} \textbf{\bibinfo{volume}{31}}, \bibinfo{pages}{259}
  (\bibinfo{year}{1998}).

\bibitem[{\citenamefont{Moan}(2002)}]{Moan}
\bibinfo{author}{\bibfnamefont{P.~C.} \bibnamefont{Moan}}, Ph.D. thesis,
  \bibinfo{school}{University of Cambridge} (\bibinfo{year}{2002}).

\bibitem[{\citenamefont{Uhrig and Lidar}(2010)}]{Lidar-Uhrig}
\bibinfo{author}{\bibfnamefont{G.}~\bibnamefont{Uhrig}} \bibnamefont{and}
  \bibinfo{author}{\bibfnamefont{D.}~\bibnamefont{Lidar}},
  \bibinfo{journal}{Phys. Rev. A} \textbf{\bibinfo{volume}{82}},
  \bibinfo{pages}{012301} (\bibinfo{year}{2010}).

\bibitem[{\citenamefont{Litsch}(2010)}]{Litsch}
\bibinfo{author}{\bibfnamefont{K.}~\bibnamefont{Litsch}}
  (\bibinfo{year}{2010}), \bibinfo{note}{unpublished}.

\bibitem[{\citenamefont{{Witzel} and {Das Sarma}}(2007)}]{Witzel:07a}
\bibinfo{author}{\bibfnamefont{W.~M.} \bibnamefont{{Witzel}}} \bibnamefont{and}
  \bibinfo{author}{\bibfnamefont{S.}~\bibnamefont{{Das Sarma}}},
  \bibinfo{journal}{Phys. Rev. B} \textbf{\bibinfo{volume}{76}},
  \bibinfo{pages}{241303(R)} (\bibinfo{year}{2007}).

\bibitem[{\citenamefont{Zhang et~al.}(2007)\citenamefont{Zhang, Dobrovitski,
  Santos, Viola, and Harmon}}]{PhysRevB.75.201302}
\bibinfo{author}{\bibfnamefont{W.}~\bibnamefont{Zhang}},
  \bibinfo{author}{\bibfnamefont{V.~V.} \bibnamefont{Dobrovitski}},
  \bibinfo{author}{\bibfnamefont{L.~F.} \bibnamefont{Santos}},
  \bibinfo{author}{\bibfnamefont{L.}~\bibnamefont{Viola}}, \bibnamefont{and}
  \bibinfo{author}{\bibfnamefont{B.~N.} \bibnamefont{Harmon}},
  \bibinfo{journal}{Phys. Rev. B} \textbf{\bibinfo{volume}{75}},
  \bibinfo{pages}{201302} (\bibinfo{year}{2007}).

\bibitem[{\citenamefont{Zhang et~al.}(2008)\citenamefont{Zhang, Konstantinidis,
  Dobrovitski, Harmon, Santos, and Viola}}]{Zhang:08}
\bibinfo{author}{\bibfnamefont{W.}~\bibnamefont{Zhang}},
  \bibinfo{author}{\bibfnamefont{N.~P.} \bibnamefont{Konstantinidis}},
  \bibinfo{author}{\bibfnamefont{V.~V.} \bibnamefont{Dobrovitski}},
  \bibinfo{author}{\bibfnamefont{B.~N.} \bibnamefont{Harmon}},
  \bibinfo{author}{\bibfnamefont{L.~F.} \bibnamefont{Santos}},
  \bibnamefont{and} \bibinfo{author}{\bibfnamefont{L.}~\bibnamefont{Viola}},
  \bibinfo{journal}{Phys. Rev. B} \textbf{\bibinfo{volume}{77}},
  \bibinfo{pages}{125336} (\bibinfo{year}{2008}).

\bibitem[{\citenamefont{West et~al.}(2010{\natexlab{a}})\citenamefont{West,
  Lidar, Fong, and Gyure}}]{West:10}
\bibinfo{author}{\bibfnamefont{J.~R.} \bibnamefont{West}},
  \bibinfo{author}{\bibfnamefont{D.~A.} \bibnamefont{Lidar}},
  \bibinfo{author}{\bibfnamefont{B.~H.} \bibnamefont{Fong}}, \bibnamefont{and}
  \bibinfo{author}{\bibfnamefont{M.~F.} \bibnamefont{Gyure}},
  \bibinfo{journal}{Phys. Rev. Lett.} \textbf{\bibinfo{volume}{105}},
  \bibinfo{pages}{230503} (\bibinfo{year}{2010}{\natexlab{a}}).

\bibitem[{\citenamefont{\'Alvarez et~al.}(2010)\citenamefont{\'Alvarez, Ajoy,
  Peng, and Suter}}]{Alvarez:10}
\bibinfo{author}{\bibfnamefont{G.~A.} \bibnamefont{\'Alvarez}},
  \bibinfo{author}{\bibfnamefont{A.}~\bibnamefont{Ajoy}},
  \bibinfo{author}{\bibfnamefont{X.}~\bibnamefont{Peng}}, \bibnamefont{and}
  \bibinfo{author}{\bibfnamefont{D.}~\bibnamefont{Suter}},
  \bibinfo{journal}{Phys. Rev. A} \textbf{\bibinfo{volume}{82}},
  \bibinfo{pages}{042306} (\bibinfo{year}{2010}).

\bibitem[{\citenamefont{{Tyryshkin} et~al.}(2010)\citenamefont{{Tyryshkin},
  {Wang}, {Zhang}, {Haller}, {Ager}, {Dobrovitski}, and
  {Lyon}}}]{2010arXiv1011.1903T}
\bibinfo{author}{\bibfnamefont{A.~M.} \bibnamefont{{Tyryshkin}}},
  \bibinfo{author}{\bibfnamefont{Z.}~\bibnamefont{{Wang}}},
  \bibinfo{author}{\bibfnamefont{W.}~\bibnamefont{{Zhang}}},
  \bibinfo{author}{\bibfnamefont{E.~E.} \bibnamefont{{Haller}}},
  \bibinfo{author}{\bibfnamefont{J.~W.} \bibnamefont{{Ager}}},
  \bibinfo{author}{\bibfnamefont{V.~V.} \bibnamefont{{Dobrovitski}}},
  \bibnamefont{and} \bibinfo{author}{\bibfnamefont{S.~A.} \bibnamefont{{Lyon}}}
  (\bibinfo{year}{2010}), \eprint{arxiv:1011.1903}.

\bibitem[{\citenamefont{{Wang} et~al.}(2010)\citenamefont{{Wang}, {Zhang},
  {Tyryshkin}, {Lyon}, {Ager}, {Haller}, and
  {Dobrovitski}}}]{2010arXiv1011.6417W}
\bibinfo{author}{\bibfnamefont{Z.}~\bibnamefont{{Wang}}},
  \bibinfo{author}{\bibfnamefont{W.}~\bibnamefont{{Zhang}}},
  \bibinfo{author}{\bibfnamefont{A.~M.} \bibnamefont{{Tyryshkin}}},
  \bibinfo{author}{\bibfnamefont{S.~A.} \bibnamefont{{Lyon}}},
  \bibinfo{author}{\bibfnamefont{J.~W.} \bibnamefont{{Ager}}},
  \bibinfo{author}{\bibfnamefont{E.~E.} \bibnamefont{{Haller}}},
  \bibnamefont{and} \bibinfo{author}{\bibfnamefont{V.~V.}
  \bibnamefont{{Dobrovitski}}} (\bibinfo{year}{2010}),
  \eprint{arxiv:1011.6417}.

\bibitem[{\citenamefont{{Barthel} et~al.}(2010)\citenamefont{{Barthel},
  {Medford}, {Marcus}, {Hanson}, and {Gossard}}}]{2010arXiv1007.4255B}
\bibinfo{author}{\bibfnamefont{C.}~\bibnamefont{{Barthel}}},
  \bibinfo{author}{\bibfnamefont{J.}~\bibnamefont{{Medford}}},
  \bibinfo{author}{\bibfnamefont{C.~M.} \bibnamefont{{Marcus}}},
  \bibinfo{author}{\bibfnamefont{M.~P.} \bibnamefont{{Hanson}}},
  \bibnamefont{and} \bibinfo{author}{\bibfnamefont{A.~C.}
  \bibnamefont{{Gossard}}} (\bibinfo{year}{2010}), \eprint{arxiv:1007.4255}.

\bibitem[{\citenamefont{West et~al.}(2010{\natexlab{b}})\citenamefont{West,
  Fong, and Lidar}}]{WFL:09}
\bibinfo{author}{\bibfnamefont{J.~R.} \bibnamefont{West}},
  \bibinfo{author}{\bibfnamefont{B.~H.} \bibnamefont{Fong}}, \bibnamefont{and}
  \bibinfo{author}{\bibfnamefont{D.~A.} \bibnamefont{Lidar}},
  \bibinfo{journal}{Phys. Rev. Lett.} \textbf{\bibinfo{volume}{104}},
  \bibinfo{pages}{130501} (\bibinfo{year}{2010}{\natexlab{b}}).

\bibitem[{\citenamefont{Mukhtar et~al.}(2010)\citenamefont{Mukhtar, Soh, Saw,
  and Gong}}]{Mukhtar:10}
\bibinfo{author}{\bibfnamefont{M.}~\bibnamefont{Mukhtar}},
  \bibinfo{author}{\bibfnamefont{W.~T.} \bibnamefont{Soh}},
  \bibinfo{author}{\bibfnamefont{T.~B.} \bibnamefont{Saw}}, \bibnamefont{and}
  \bibinfo{author}{\bibfnamefont{J.}~\bibnamefont{Gong}},
  \bibinfo{journal}{Phys. Rev. A} \textbf{\bibinfo{volume}{82}},
  \bibinfo{pages}{052338} (\bibinfo{year}{2010}).

\bibitem[{\citenamefont{Wang and Liu}(2011)}]{NUDD}
\bibinfo{author}{\bibfnamefont{Z.-Y.} \bibnamefont{Wang}} \bibnamefont{and}
  \bibinfo{author}{\bibfnamefont{R.-B.} \bibnamefont{Liu}},
  \bibinfo{journal}{Phys. Rev. A} \textbf{\bibinfo{volume}{83}},
  \bibinfo{pages}{022306} (\bibinfo{year}{2011}).

\bibitem[{pri()}]{prior-CDD-comment}
\bibinfo{note}{A result similar to Eq.~(\protect\ref{J-level-k-solution}) was
  found in \protect\cite{KL07}; in our notation, Eq.~(51) of
  \protect\cite{KL07} reads $J^{(k)} = (\beta \tau_0)^k R^{k(k+2)/2}J$, which
  was obtained under the assumption $\beta > J$, and with $\epsilon$
  approximated by $\beta$.}

\bibitem[{\citenamefont{Su{\'a}rez and S{\'a}enz}(2001)}]{SS01}
\bibinfo{author}{\bibfnamefont{R.}~\bibnamefont{Su{\'a}rez}} \bibnamefont{and}
  \bibinfo{author}{\bibfnamefont{L.}~\bibnamefont{S{\'a}enz}},
  \bibinfo{journal}{J. Math. Phys.} \textbf{\bibinfo{volume}{42}},
  \bibinfo{pages}{4582} (\bibinfo{year}{2001}).

\end{thebibliography}


\end{document}